\documentclass[printer]{aa}  

\usepackage{graphicx}
\usepackage{txfonts}
\usepackage[version=3]{mhchem} 
\usepackage{siunitx} 
\usepackage{mathabx} 
\usepackage{amsmath}
\usepackage{multirow}
\usepackage{placeins}
 \usepackage{hyperref}

\newcommand{\vplanet}{\texttt{\footnotesize{VPLanet}}}
\newcommand{\magmoc}{\texttt{\footnotesize{MagmOc}}}

\begin{document}

   \title{From CO$_2$- to H$_2$O-dominated atmospheres and back}

   \subtitle{How mixed outgassing changes the volatile distribution in magma oceans around M dwarf stars}

   \author{L. Carone
          \inst{1,2,3}
          \and
          R. Barnes\inst{4}
          \and L. Noack\inst{5}
          \and K. Chubb\inst{2,6}
          \and P. Barth\inst{1,2,7,8,9}
          \and B. Bitsch\inst{10,3}
          \and A. Thamm\inst{5}
          \and A. Balduin\inst{5}
          \and R. Garcia\inst{4}
          \and Ch. Helling\inst{1,11}
          }

   \institute{Space Research Institute, Austrian Academy of Sciences, Schmiedlstrasse 6, A-8042 Graz, Austria\\
              \email{ludmila.carone@oeaw.ac.at}
              \and
            Centre for Exoplanet Science, University of St Andrews, North Haugh, St Andrews, KY16 9SS, UK
             \and
             Max-Planck-Institut f\"ur Astronomie, K\"onigstuhl 17, Heidelberg, D-69117, Germany
         \and
            Department of Astronomy, University of Washington, Seattle, WA 98105, USA
            \and 
            Institute of Geological Sciences, Freie Universität Berlin, Malteserstr. 74-100, D-12249 Berlin, Germany
            \and
            University of Bristol, School of Physics, Tyndall Avenue, Bristol, BS8 1TL, UK
            \and
             School of Earth \& Environmental Sciences, University of St Andrews, Bute Building, Queen’s Terrace, St Andrews, KY16 9TS, UK
             \and
             SUPA, School of Physics \& Astronomy, University of St Andrews, North Haugh, St Andrews, KY16 9SS, UK
             \and
             Stuttgart Center for Simulation Science, University of Stuttgart, Pfaffenwaldring 5a, 70569 Stuttgart, Germany
            \and
            Department of Physics, University College Cork, Cork, T12 R229,Ireland
\and
Fakult\"at für Mathematik, Physik und Geod\"asie, TU Graz, Petersgasse 16, Graz, A-8010, Austria
             }

   \date{Received April 12, 2024; accepted: December 9, 2024}

 
  \abstract
   {}
   {We investigate the impact of \ce{CO2} on the distribution of water on TRAPPIST-1 e, f and g  during the magma ocean stage. These potentially habitable rocky planets are currently the most accessible for astronomical observations. A constraint on the volatile budget during the magma ocean stage is a key link to planet formation and also to judge their habitability.}
   {We expand the \magmoc{} module of the \vplanet{} environment to perform simulations with 1-100 terrestrial oceans (TO) of \ce{H2O} with and without \ce{CO2} and for albedos 0 and 0.75. The \ce{CO2} mass is scaled with initial \ce{H2O} by a constant factor between 0.1 and 1.}
   {The magma ocean state of rocky planets begins with a \ce{CO2}-dominated atmosphere but can evolve into a \ce{H2O} dominated state, depending on initial conditions. For less than 10~TO initial \ce{H2O}, the atmosphere tends to desiccate and the evolution may end with a \ce{CO2} dominated atmosphere. Otherwise, the final state is a thick ($>1000$~bar) \ce{H2O}-\ce{CO2} atmosphere.
   Complete atmosphere desiccation with less than 10~TO initial \ce{H2O} can be significantly delayed for TRAPPIST-1e and f, when \ce{H2O} has to diffuse through a \ce{CO2} atmosphere to reach the upper atmosphere, where XUV photolysis occurs. As a consequence of \ce{CO2} diffusion-limited water loss, the time of mantle solidification for TRAPPIST-1 e, f, and g can be significantly extended compared to a pure \ce{H2O} evolution by up to 40~Myrs for albedo 0.75 and by up to 200~Mrys for albedo 0.
   The addition of \ce{CO2} further results in a higher water content in the melt during the magma ocean stage. Thus, more water can be sequestered in the solid mantle. However, only up to 6\% of the initial water mass can be stored in the mantle at the end of the magma ocean stage.
Our compositional model adjusted for the measured metallicity of TRAPPIST-1 yields for the dry inner planets (b, c, d) an iron fraction of 27 wt\%. For TRAPPIST-1 e, this iron fraction would be compatible with a (partly) desiccated evolution scenario and a \ce{CO2} atmosphere with surface pressures of a few $100$~bar. 

   }
   {A comparative study between TRAPPIST-1 e and the inner planets may yield the most insights about formation and evolution scenarios by confronting, respectively, a scenario with a desiccated evolution due to volatile-poor formation to a volatile-rich scenario with extended atmospheric erosion. }

   \keywords{Planetary systems --
                Planets and satellites: atmospheres --
                Planets and satellites: terrestrial planets --
                Planets and satellites: physical evolutions
               }

   \maketitle
%

\section{Introduction} \label{sec:intro}
After planet formation, rocky planets\footnote{The transition between bona-fide rocky planets and volatile-rich planets is predicted to occur at 2 \citep{ChenKipping2017} or 4 Earth masses \citep{Baron2024}. In this work we discuss planets well below 2 Earth masses.} may start their evolution with a global magma ocean with outgassing from its interior that replaces the primary hydrogen-helium envelope inherited from the protoplanetary disk with a secondary atmosphere \citep[e.g.][]{Lammer2018, Stueken2020}. The composition of the latter depends on the redox state of the mantle \citep{Deng2020,Ortenzi2020}. 

It is generally recognized that the magma ocean represents not only a direct link between atmospheric properties and the
rocky planet's mantle, it is further crucial for determining the abundances of important volatiles such as water and carbon-dioxide in the mantle and the atmosphere \citep{Chaso2021,Barth2021,Moore2023, Krissansen2022}. Therefore,  despite its relatively short duration (1-100 Myrs), the magma ocean phase ``sets the stage'' for the long-term planetary evolution over billion of years \citep[e.g][]{Chaso2021,Lammer2018,Stueken2020}. 

For Earth-sized planets like Venus, Earth and the potentially habitable exoplanets TRAPPIST-1 e, f and g,  predominantly \ce{H2O}-\ce{CO2} outgassing is expected \citep{Deng2020}.  However, a water vapor dominated atmosphere is subject to XUV photolysis with subsequent escape of \ce{H2} and thus prone to significant mass loss, even in the Solar System with its relatively calm host star. \citet{Hamano2013} demonstrated that atmospheric erosion in the inner Solar System could deplete all of Earth's oceans within 100 Myrs. Venus experienced a relatively long magma ocean stage of 100~Myrs, during which most of its water was lost, whereas the shorter magma ocean stage on Earth, lasting just a few Myrs, prevented significant water loss. For Earth-sized planets, the magma ocean stage thus plays a particularly critical role in determining their surface water content and thus their habitability \citep[e.g.][]{Barth2021,Hamano2013,Tian2015}. The interaction between \ce{H2O} outgassing and atmospheric escape during the magma ocean stage is even more crucial for rocky planets orbiting active M dwarf stars, which emit intense XUV flux during a prolonged pre-main sequence stage \citep[e.g.][]{Johnstone2021,Tian2015}.  

It was shown by \citet{Raymond2022} that the TRAPPIST-1 planets - in contrast to Earth - were not strongly modified by large impactors during late accretion, which would alter the composition of the outgassed atmosphere after the magma ocean stage \citep[e.g.][]{Zahnle2020,Zahnle2015}. Thus, any \ce{CO2} and \ce{H2O} observed in the atmospheres today must be primordial \citep[see e.g.][]{Krissansen2022}. Hence, atmospheric constraints on the rocky planets in the TRAPPIST-1 system may shed light on the volatile budget of the magma ocean, planet formation and also its potential for habitability.
 
Initial observations with JWST suggest that TRAPPIST-1~b \citep{Greene2023} and c \citep{Zieba2023} do not possess dense \ce{CO2} atmospheres and even indicate the absence of a substantial atmosphere with $p_{\rm surf}\geq 0.1$~bar. However, recent analyses suggest that the MIRI data for TRAPPIST-1~c are consistent with a variety of atmosphere compositions: water vapor \citep{Acuna2023}, \ce{O2} with \ce{CO2} or \ce{H2O} \citep{Lincowski2023} with surface pressures up to 40~bar. These results are, however, based on only five secondary eclipses for TRAPPIST-1b and four for TRAPPIST-1c, which may also be affected by stellar contamination \citep{Lim2023}.
 
 The combination of ultra-precise density constraints \citep{Agol2021} and interior models \citep{Noack2016,Unterborn2018,Dorn2018} suggest that at least TRAPPIST-1 g currently incorporates $>10$~wt\% of water in its interior structure, which would correspond in mass to more than 100 terrestrial oceans (TO) of water\footnote{1 TO of water is $1.39\times 10^{21}$~kg of \ce{H2O}.} \citep{Unterborn2018,Barth2021,Raymond2022}. The high inferred volatile content of TRAPPIST-1 g is supported by planet formation models \citep{Miguel2020,Unterborn2018,Schoonenberg2019} that suggest a volatile-rich formation scenario for the outer TRAPPIST-1 planets  with initial water mass fractions of up to 50~wt\%. Determining the magma ocean solidification time for rocky planets and the resulting distribution of volatiles are thus pressing science questions of current and future missions like PLATO \citep{Turbet2019,Schlecker2024} and mission concepts like the Large Interferometer for Exoplanets (LIFE, \citet{Bonati2019}).

There are, however, still several open questions in the modelling magma oceans on diverse rocky planets. Many previous magma ocean models, including the open-source \magmoc{1.0} \citep{Barth2021} as part of the \texttt{VPLANET} modeling suite \citep{Barnes2020}, assumed a pure \ce{H2O} steam atmosphere \citep{Hamano2013,Goldblatt2013,Lichtenberg2021}. Not only water is a critical molecule in terrestrial planet evolution \citet{Elkins-Tanton2008,Niko2019,Krissansen2022}. Mixed \ce{H2O}-\ce{CO2} outgassing can also modify the volatile distribution as the atmosphere composition changes from an initial \ce{CO2}-dominated to \ce{H2O}-dominated atmosphere for the Earth \citep{Bower2019}. The impact of simultaneous \ce{H2O} and \ce{CO2} outgassing during the magma ocean stage on the potentially habitable TRAPPIST-1 planets e, f and g together with atmospheric escape has not yet been investigated.

We tackle for the first time the \ce{H2O}-\ce{CO2} outgassing feedback between \ce{H2O} and \ce{CO2} outgassing on TRAPPIST-1~e, f and g with an upgraded version of the \citet{Barth2021} model, \vplanet/\magmoc{2.0}. In Section~\ref{sec: outgas}, we first introduce a new solution for coupled outgassing of two volatiles. We next present an update of the atmospheric evolution model, where we now take into account the vertical extent of a mixed \ce{CO2}-\ce{H2O} atmosphere (Sect.~\ref{sec: PT}).

We employ full radiative transfer (RT) calculations with \texttt{petitRADTRANS} \citep{Molliere2019} for two different atmosphere treatments: In the RT grid model, we compile the outgoing longwave radiation (OLR) in a 3D emission grid for different surface pressures, surface temperatures, and water volume mixing ratios and interpolate during simulation time (Sect.~\ref{sec: RT}). In the corrected gray atmosphere model, we use the  radiative transfer grid calculation to formulate an empirical approximation that reproduces the results of the RT grid atmosphere model generally within 10\% accuracy (Sect.~\ref{sec: grey}). 

We also investigate how the runaway greenhouse limit changes with different surface gravities (7.5~m/s${}^2$-22.5~m/s${}^2$, Sect.~\ref{sec: Grey_exo}). After validation for various Earth scenarios and outgassing laws (Sect.~\ref{sec:Earth}), we apply \magmoc{2.0} to the potentially habitable TRAPPIST-1 planets e, f, and g for which we investigate initial volatile contents between 1-100~TO \ce{H2O} (Sect.~\ref{sec: Simulations}). For each initial water scenario, we calculate evolutionary tracks with no \ce{CO2} and initial \ce{CO2} mass equal to $0.3\times$ and $1\times$ initial water mass, respectively, to illustrate the impact of additional \ce{CO2} on the thermal and volatile evolution for a relatively coarse \ce{H2O} grid. For each evolution scenario, we further implement two albedo ($\alpha$) assumptions: $\alpha=0.75$ for better comparison with \citet{Barth2021} and $\alpha=0$, the clear-sky assumption.

We first present for TRAPPIST-1 g and the clear-sky scenario three example evolutions with 1, 5 and 100~TO initial water mass  (Sect.~\ref{sec: TRAPPIST-1g}) as in \citet{Barth2021}. Subsequently, a concise overview of solidification times and remaining water fraction in the solid mantle after the magma ocean evolution is given for TRAPPIST-1 e, f and g, for initial water mass between 1 and 100~TO and for the three \ce{CO2} ratios (Sect.~\ref{sec: Overview}). Finer evolution grids with respect to initial water and \ce{CO2} mass for TRAPPIST-1 e, f and g are found in  \href{https://doi.org/10.5281/zenodo.14442985l}{a detailed grid} for end states of \ce{H2O}, \ce{CO2} and \ce{O2} partial pressures as well as sequestered \ce{H2O} and \ce{O2} in the mantle. We further investigate the strong impact of \ce{CO2} on the desiccation and thus the magma ocean lifetime for water-poor composition ($\leq 10$~TO \ce{H2O}). 

We tie the explored magma ocean simulations to new constraints of the interior for all TRAPPIST-1 planets. Here, the inner planets (b,c,d) yield a constraint on the iron fraction of 27 wt-\% that allows us to place TRAPPIST-1e in the dry water regime ($\leq 10$~TO \ce{H2O}), for which a partly desiccated \ce{CO2}-atmosphere is expected. The amount of abiotically produced \ce{O2} in these scenarios depends strongly on the magma ocean lifetime. TRAPPIST-1 f and g are expected to be in the water-rich regime ($>>10$~TO \ce{H2O}), based on the interior models, for which the magma ocean ends in a ``wet'' \ce{CO2} atmosphere. 

Next, we explore the impact of an extended \ce{H2O} and \ce{CO2} atmosphere on the planetary radii and measured bulk density. We also present a compositional model for the refractory elements and volatiles present during planet formation adjusted for the metallicity of TRAPPIST-1 and compare to interior structures with respect to iron fraction and volatile ratio of TRAPPIST-1e,f, g that are compatible with the measured masses and radii of \citet{Agol2021} (Sect.~\ref{sec atm int model}). The results of the magma ocean, composition and interior models are presented in Sect.~\ref{sec: results} and  discussed in Sect.~\ref{sec: Discuss}. We conclude in Sect.~\ref{sec: Conclude} that comparative studies between TRAPPIST-1e and g and the inner planets are warranted to identify end states of volatile-poor formation (inner planets), desiccated evolution (TRAPPIST-1e) and volatile-rich formation with little modification by desiccation (TRAPPIST-1g). We outline future avenues to improve \magmoc{2.0} for a  better link to planet formation and also to tackle the full diversity of outgassed atmospheres during the crucial magma ocean stage (Sect.~\ref{sec: outlook}).


\section{Methods}
\label{sec: method}

We utilize and expand the open source \texttt{VPLANET} framework \citep{Barnes2020} that connects stellar and planetary processes in order to  simulate the evolution of stars and planets over time spans of Gyr. For the present paper we use the \texttt{stellar} module that tracks the bolometric luminosity of the star according to the \cite{Baraffe2015} stellar evolution model grid, and the XUV evolution according to the \cite{Ribas2005} model originally developed for solar twins, but which appears compatible with lower mass stars \citep[see e.g.][]{RicheyYowell2022}. The \texttt{VPLANET} \texttt{AtmEsc} module tracks water photolysis by the stellar radiation, hydrogen escape via energy- and diffusion-limited escape, and oxygen escape via hydrodynamic drag \citep{Watson1981,Hunten1987,Luger2015}. The \texttt{VPLANET} \texttt{EqTide} module simulates tidal effects, including frictional heating for the both the constant-phase-lag and constant-time-lag models \citep{FerrazMello2008,Leconte2010,Barnes2013}. The \texttt{RadHeat} module tracks the radiogenic heating of the unstable isotopes $^{40}$K, $^{232}$Th, $^{235}$U, and $^{238}$U. 

\subsection{\magmoc{}  approach}
\magmoc{} is a geophysical and geochemical model for the coupled mantle and atmosphere when the mantle is at least partially molten \citep{Barth2021}. We assume a bulk silicate Earth composition from \citet{O_neill_1998}, following \citet{Schaefer2016}. Table~\ref{Tab_geo} provides the most relevant geophysical parameters for this work.

Our magma ocean model assumes as outlined in \cite{Barth2021}
efficient cooling via convection of the magma ocean, which is only true until the melt fraction at the surface drops below 0.4. This condition is reached in our model typically with surface temperatures of 1650~K and a solidification radius $r_{\rm s}$ that already comprises about 99\% of the planetary radius. Below this temperature, we follow the argumentation of  \citet{Debaille2009} that a thick thermal boundary layer may be neglected towards the end of the magma ocean because crystallization of iron-rich minerals lead to overturning near the surface and a re-setting of the thermal boundary layer. The module thus switches to a solid-like viscosity \citep[][Eq.3]{Barth2021} and advances solidification and thermal evolution further until $r_{\rm s}$ is equal the planetary radius, which is typically for surface temperatures of 1400~K. At this point, our model does not further advance mantle and surface temperature evolution in contrast to other models \citep{Lebrun2013,Bower2019,Bower2022,Krissansen2022,Lichtenberg2021}.

We note that it is worthwhile to extend simulations beyond the mantle solidification time with \texttt{MagmOcV2.0} for cases with significant atmospheric erosion during the magma ocean stage to capture the final stages of complete desiccation via atmospheric escape that is an upper atmosphere process. The caveat is here that a) surface temperatures and heat fluxes are kept ``artificially'' high after solidification and b) no water condensation can occur, which would ``save'' the remaining water from being lost. Thus, water loss due to atmospheric erosion may be overestimated.

\begin{table}[h]
 \caption{Parameters of the geophysical model for \texttt{MagmOcV2.0}}
    \begin{tabular}{c|c}
    	\hline
    	Symbol & Parameter \\ 
    	\hline \hline
    	$T_\mathrm{p}$ [K]& Potential temperature of the mantle \\
    	$T_\mathrm{surf}$ [K] & Surface temperature \\
    	$r_\mathrm{s}$ [m]& Solidification radius \\
    	$\rho_\mathrm{m}$ [kg/m$^3$]& Mantle bulk density ($4000$) \\ 
    	$F_\mathrm{net}$ [W/m$^2$] & Net flux leaving atmosphere (OLR - ASR$^{a}$)\\
    	$\psi$ & Magma ocean averaged melt fraction \\
    	\hline
    \end{tabular}
    \\
    ${}^{a}$ OLR = Outgoing Longwave Radiation, ASR = Absorbed Stellar Radiation
    \label{Tab_geo}
\end{table}

\begin{figure*}[ht]
    \centering
    \includegraphics[width=0.95\textwidth]{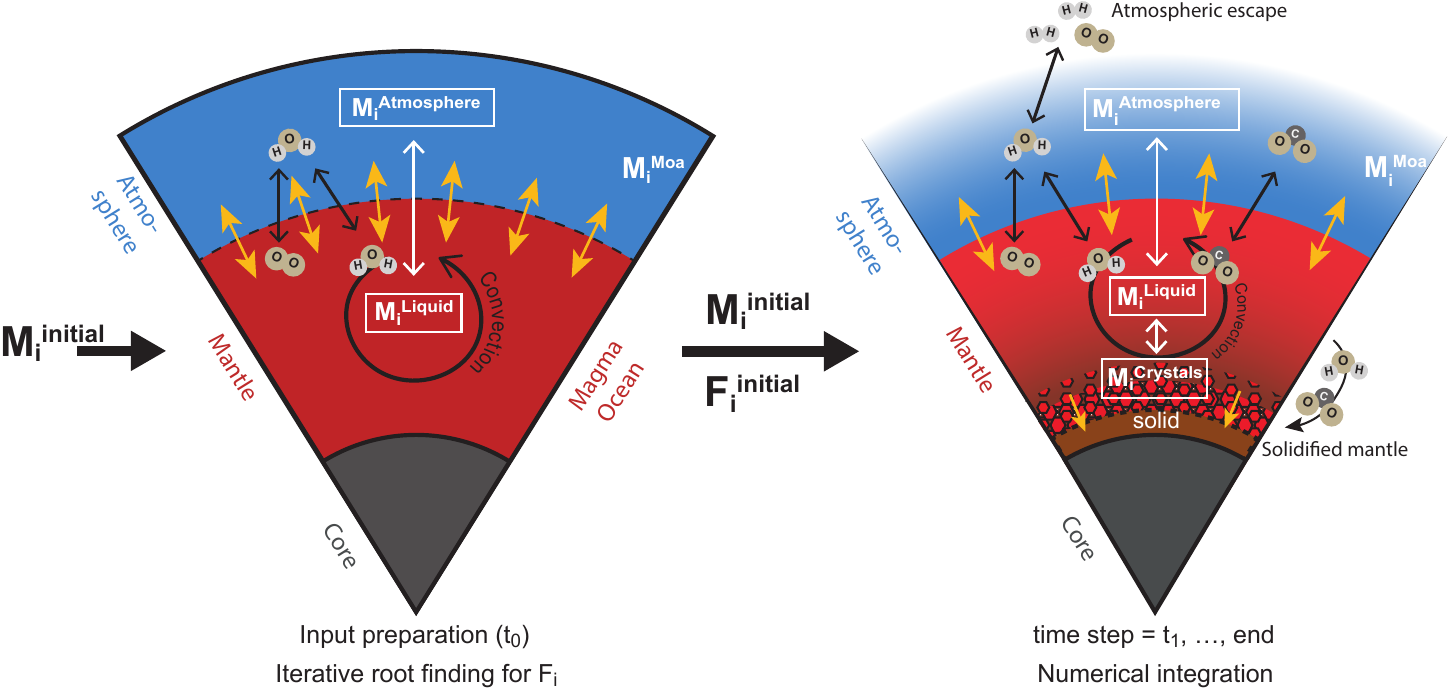}
   
    \caption{This schematic depicts the set-up of the volatile exchange during initialization (t=0) and run time. For initialization, a surface temperature of 4000~K and a completely molten magma ocean is assumed, where the dissolved volatiles are in balance with the outgassed volatile content, set by the volatile melt fraction $F_i$. As the magma ocean solidifies, part of the volatile budget is deposited in the solid mantle. Further, atmospheric escape can remove \ce{H2O}. These two sink terms thus reduce the amount of a volatile available in the fully coupled magma ocean-atmosphere system $M_i^{moa}$. The full overview of all included processes, including radiogenic heating, is shown in \citet[][Fig. 1]{Barth2021} }
    \label{fig: Volatile_Flowchart}
\end{figure*}

\subsection{H2O-CO2 Volatile budget}
\label{sec: outgas}
Here, we focus on advancing the volatile treatment of \texttt{MagmOcV1.0}, which is needed for coherent modelling of complete devolatilization on rocky exoplanets with a large variety of initial compositions and \ce{H2O}-\ce{CO2} ratios during the magma ocean stage. Table~\ref{Tab_Volat_Model} shows an overview of the relevant parameters.

\begin{table}[h]
 \caption{Parameters of the volatile model for \texttt{MagmOcV2.0}}
	\begin{tabular}{c|c}
		\hline
		Symbol & Parameter \\ 
		\hline \hline
    $M^{\mathrm{liq}}$ [kg]&  Total mass of liquid melt\\
         $ M^{\mathrm{crystal}}$ [kg] & Total mass of crystals in magma ocean\\
         \hline
		$F_i$ & Volatile $i$ mass fraction in liquid melt \\ 
		$M_i^{\mathrm{moa}}$ [kg] & Volatile $i$ mass in  magma ocean + atmosphere \\
		$M_i^{\mathrm{cyrstal}}$ [kg] & Volatile $i$ mass crystals in the magma ocean \\
		$M_i^{\mathrm{liq}}$ [kg] & Volatile $i$ mass in the liquid melt \\
		$M_i^{\mathrm{atm}}$ [kg] & Volatile  $i$ mass in the atmosphere \\
		$M_i^{\mathrm{sol}}$ [kg] & Volatile $i$ mass in the solidified mantle \\
		$k_{\ce{H2O}}$ & Water part. coeff. melt - solid (0.01) \\
        $k_{\ce{CO2}}$ & \ce{CO2} part. coeff. melt - solid (0.002) \\
		$\phi_1$ [kg/m${}^2$/s] & XUV-driven atm. mass-loss rate of H \\
		\hline
	\end{tabular}
	\label{Tab_Volat_Model}
\end{table}

The volatile budget is set up similarly to \citet{Barth2021}, where we assume at all times that a volatile $i$ (i=\ce{H2O},\ce{CO2}) is distributed in the fully coupled magma ocean-atmosphere system (MOA). This system consists of a liquid magma ocean $M^{liq}$, the atmosphere $M^{atm}$ and crystals forming within the liquid magma ocean  $M^{crystal}$, the latter is set to zero initially. 

We further assume initially that the volatiles $i$ in the liquid magma ocean and atmosphere are in balance with each other for a given initial volatile mass $M_i^{initial}$, which results in an initial volatile mass fraction $F^{initial}_i$, governing both the amount of volatile outgassed to build an atmosphere and the amount of volatile dissolved in the melt. As the mantle begins to solidify from bottom to top, that is, we assume that the crystals fall to the bottom of the magma ocean $M^{crystal}$, incorporating a small amount of the available volatile mass from the melt ($F_i M^{\mathrm{liq}}$). Sequestering of volatiles in the solidifying mantle $dM^{Sol}$ is regulated by the mantle-averaged constant partition coefficient $k_i$. The volatile mass in the solidified mantle ($M_i^{Sol}$) is inaccessible for the MOA and thus comprises a sink term. As the magma ocean depth, and thus the total mass of $M^{liq}$, decreases with increasing solidification radius $\frac{d r_s}{dt}$, the mass fraction $F_i$ of the volatile in the melt typically increases, leading to more outgassing. 

In summary, we assume mass balance in the MOA for a volatile $i$ of the form \citep[see also][]{Barth2021,Bower2019}:
\begin{align}
		M_i^{\mathrm{moa}} &= M_i^{\mathrm{crystal}} + M_i^{\mathrm{liq}} + M_i^{\mathrm{atm}} \nonumber\\
		&= k_i F_i M^{\mathrm{crystal}} + F_i M^{\mathrm{liq}} + \frac{4 \pi r_\mathrm{p}^2}{g} p_{i,\mathrm{mass}},
\label{eq: mass balance}
\end{align}
 where where $F_i$ is the volatile mass fraction of the liquid part of the magma ocean $M_i^{liq}$. In addition, we assume that the volatile $i$ is partly sequestered in the crystallized magma ocean $M_i^{cry}$. The mass of the volatile in the atmosphere is also set by $F_i$, where the partial pressure of volatile $i$ is determined by a Henrian fit to laboratory data (see Fig.~\ref{fig: OutgasLaws} and references in Sect.~\ref{sec: Derivation melt fraction}):
\begin{equation}
    p_{i,\rm{part}}=\left(\frac{F_i-c_i}{a_i} \right)^{b_i},\label{eq: Henrian fit}
\end{equation}
where $a_i,b_i$ and $c_i$ are the fit coefficients. We note that $c_i$ is a term that suppresses outgassing and is only used in the solubility law of \citet{Elkins-Tanton2008}. For the TRAPPIST-1 planet simulations, we use the solubility law of \citet{Niko2019}, for which $c_i$ is set to zero. Since we implemented and tested both solubility laws (Sect.~\ref{sec: Derivation melt fraction}), we show here the generalized form of the Henrian fit with the suppression term.

As pointed out by \citet{Bower2019}, the mass weighted pressure of volatile $i$ that is  $p_{i,\mathrm{mass}}$ enters into the mass balance. The relation between the partial and mass weighted pressure is:
\begin{equation}
 p_{i,\rm{mass}}= \frac{\mu_i}{\overline{\mu}_{\rm{atm}}} p_{i,\rm{part}},
\end{equation}
where $\mu_i$ is the molecular mass of volatile $i$ and $\overline{\mu}_{\rm{atm}}$ is the mean molecular mass of all atmosphere constituents.

In addition, atmospheric erosion can remove \ce{H2O} from the system and thus acts as another sink term to the volatile budget $M_i^{\mathrm{moa}}$ (Fig.~\ref{fig: Volatile_Flowchart}). See Section~\ref{sec: Atmesc} for a more detailed description of atmospheric escape. The two processes that remove volatiles from $M_i^{moa}$ are described by the following differential equations:
\begin{align}
\frac{d M_i^{sol}}{dt} & =  k_i F_i \underbrace{4\pi \rho_m r_s^2 \frac{d r_s}{dt}}_{d M^{\mathrm{crystal}}/ dt}\label{eq: sink_sol}\\
\frac{d M_{\ce{H2O}}^{esc}}{dt} & = 4 \pi r^2_p \phi_1 \frac{\mu_{\ce{H2O}}}{\mu_H}\label{eq: sink_H2O}\\
\frac{d M_{\ce{CO2}}^{esc}}{dt} & = 0\label{eq: sink-cO2}.
\end{align}
 We refer to Tables~\ref{Tab_geo} and \ref{Tab_Volat_Model}  for a concise description of all relevant parameters.

\subsubsection{Implementation of outgassing during run time}

We use the starting conditions for the magma melt fraction of volatile $i$, that is, $F_{i}(t=0)$ (Fig.~\ref{fig: Volatile_Flowchart} left) as input parameters for \magmoc{2.0}. We use a root-finding algorithm to solve the mass balance equations (Eq.~\ref{eq: mass balance}) with respect to $F_i$ for ($t=0$) for a given volatile initial mass $M^{initial}_i$ to find the starting conditions for a given initial magma ocean depth.

In \magmoc{1.0}, root finding is performed at all time steps to solve for $F_{i}(t)$ that satisfies mass balance, which is still relatively efficient for a single outgassed volatile.  
In \magmoc{2.0}, we opt instead to operate on the time derivatives of the mass balance equation $\frac{d M_i^{moa}}{dt}$, which is more numerically efficient for multiple volatiles. We also take into account outgassing feedback that occurs  due to changes of the mean molecular mass of the atmosphere, $\frac{d \overline{\mu}_{\rm atm}}{dt}$, when a magma ocean evolves from a \ce{CO2} towards an \ce{H2O}-dominated atmosphere \citep{Bower2019}.

During run-time, the derivatives of the volatile mass fractions are advanced using a set of coupled ordinary differential equations:
\begin{align}
\frac{d F_{\ce{H2O}}}{d t} &= \frac{C_{\ce{H2O}}A_{\ce{CO2}} -C_{\ce{CO2}} A_{\ce{H2O}}}{A_{\ce{H2O}}B_{\ce{CO2}}-C_{\ce{CO2}}B_{\ce{H2O}}}, \label{eq: dFH2Oup} \\
\frac{d F_{\ce{CO2}}}{d t} & = \frac{C_{\ce{CO2}}B_{\ce{H2O}} -C_{\ce{H2O}} B_{\ce{CO2}}}{A_{\ce{H2O}}B_{\ce{CO2}}-A_{\ce{CO2}}B_{\ce{H2O}}}  \label{eq: dFCO2up}, 
\end{align}
where $A_i, B_i$ and $C_i$ comprise different components of the complete derivative of the (volatile) mass balance equations. The full derivation is outlined in Sect.~\ref{sec: Derivation melt fraction}. The sink terms for the solidified mantle (Eq.~\ref{eq: sink_sol}) and atmospheric erosion (Eq.~\ref{eq: sink_H2O}) are substracted from $\frac{d M_i^{moa}}{dt}$.

We note that these coupled differential equations for outgassing of multiple volatiles differ from those of \citet{Bower2019}, because we choose the volatile melt fractions $F_i$ and not their partial pressures $p_{i,\rm{part}}(F_i)$ as primary variables of integration. We have further verified that \magmoc{2.0} with a pure \ce{H2O} atmosphere yields the same results as \magmoc{1.0} (not shown). Further, the benchmarking that we performed for different Earth scenarios with and without mixed atmospheres further confirms the validation of the code (Sect.~\ref{sec:Earth}).

\begin{figure*}[ht]
    \centering
    \includegraphics[width=0.45\textwidth]{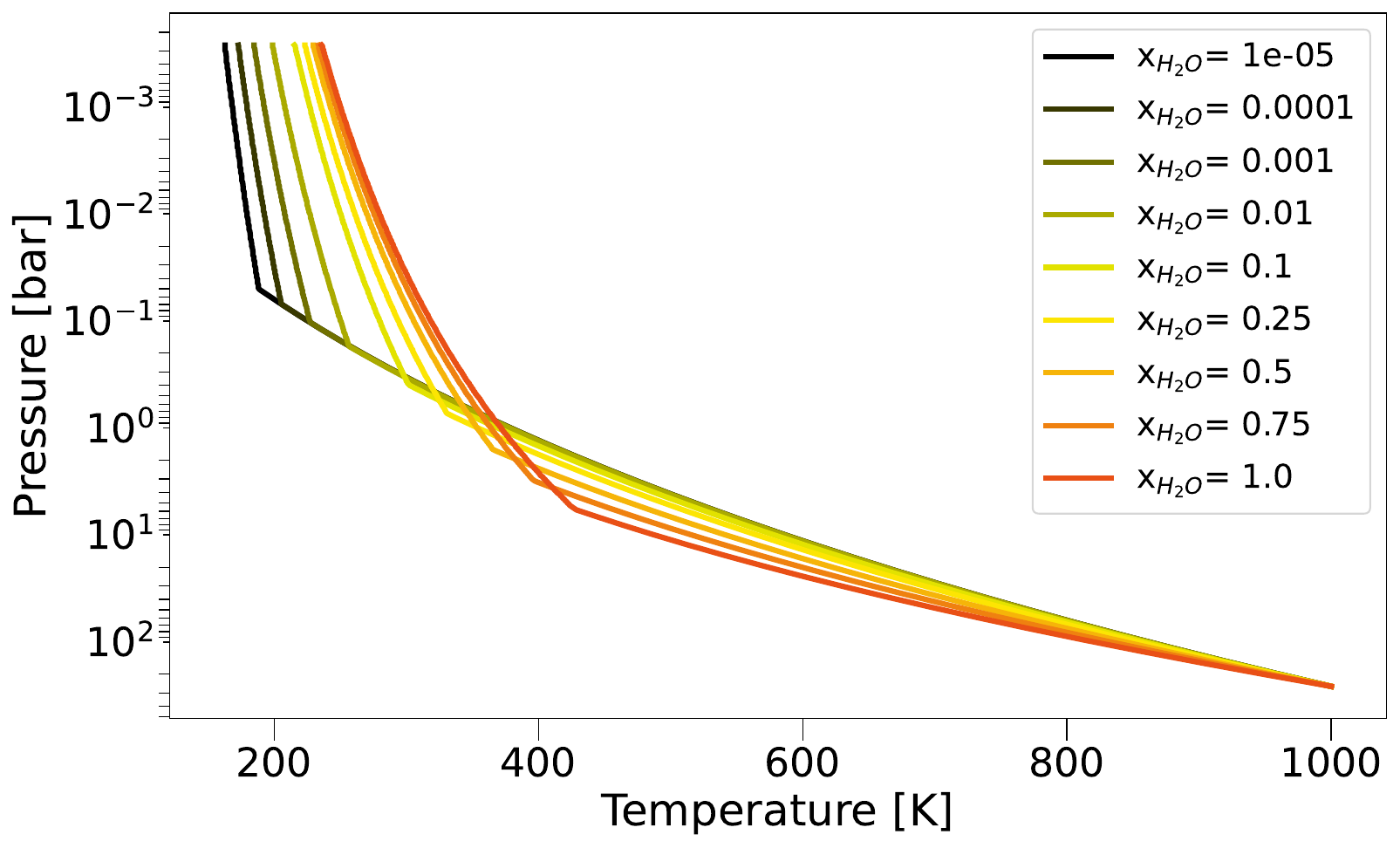}
    \includegraphics[width=0.45\textwidth]{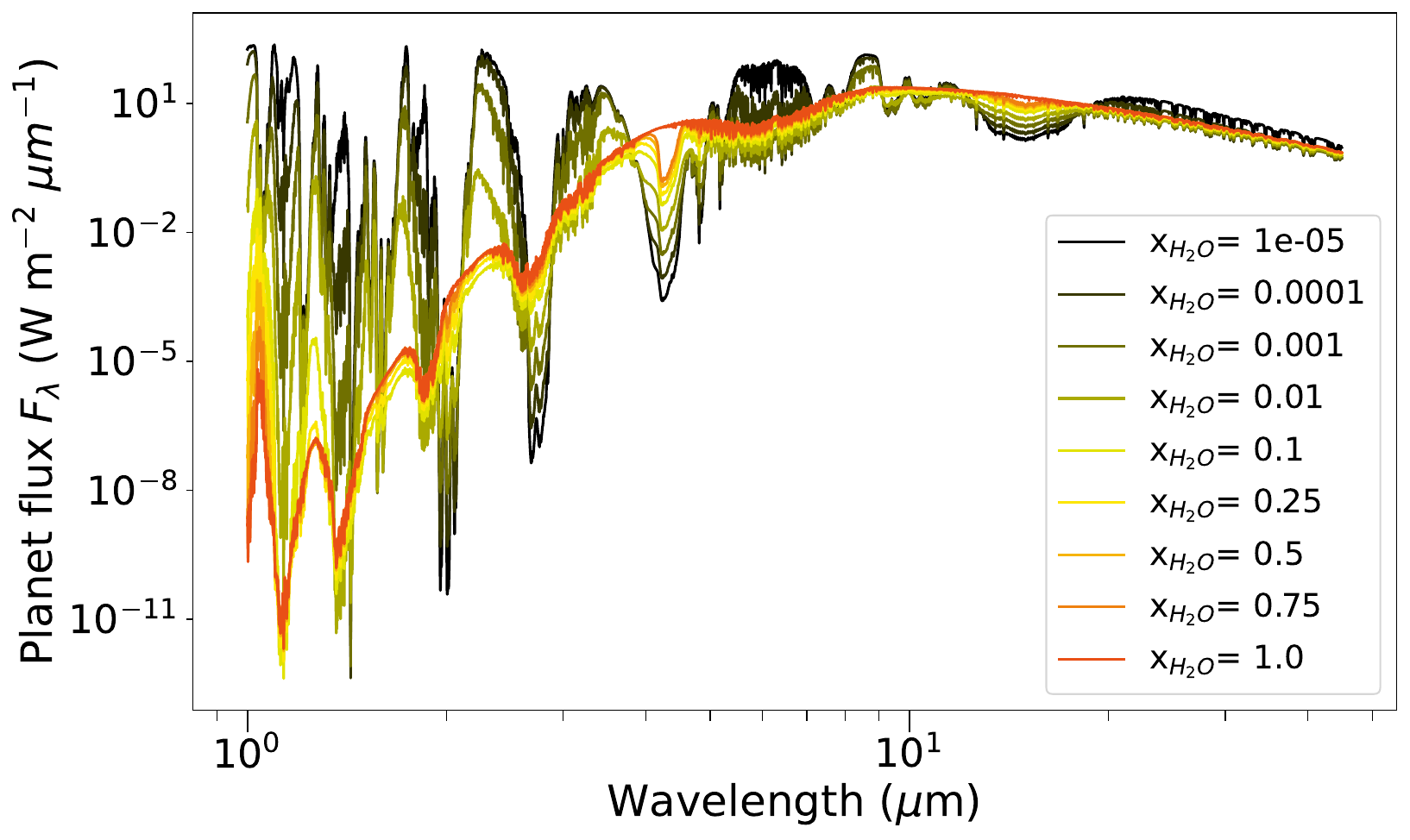}
    \caption{Vertically extended pressure-temperature ($p_{\rm gas},T_{\rm gas}$) profiles for $p_{surf}=260$~bar and $T_{surf}=1000$~K and different water mixing ratios $x_{\ce{H2O}}$ (left). The solid black and solid red line denote 100\% \ce{CO2} and 100\% \ce{H2O} atmosphere composition, respectively. The profiles converge in the upper atmosphere to the condensation curve, where we assume equilibrium between condensation and evaporation of \ce{H2O} and \ce{CO2} (supersaturation ratio $S=1$). Change in emission (right) from a pure \ce{H2O} atmosphere (red line) to \ce{CO2}-dominated (black line).}
    \label{fig: PT_profiles}
\end{figure*}

\begin{figure}[ht]
    \centering
    \includegraphics[width=0.45\textwidth]{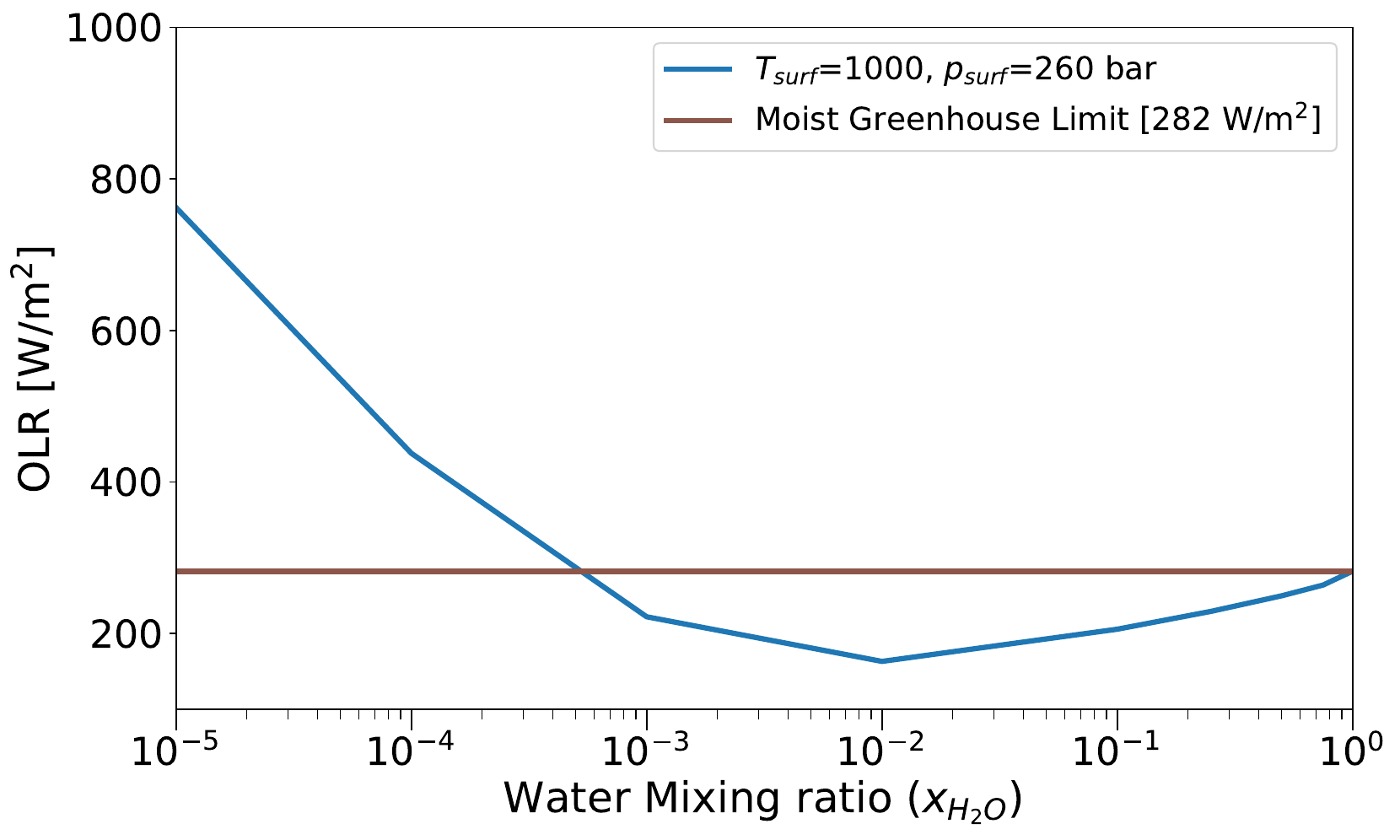}
    \caption{Integrated outgoing longwave radiation (OLR) for $p_{\rm surf}=260$~bar and $T_{\rm surf}=1000$~K and different \ce{H2O} to \ce{CO2} content. Mixing \ce{CO2} into the steam atmosphere mainly acts to cool the upper atmosphere layers towards the \ce{CO2} condensation curve. Only for $x_{\ce{H2O}}<10^{-3}$ (that is, $x_{\ce{CO2}}> 0.999$) will the overall thermal emission increase again as the steep adiabat of a \ce{CO2} dominated atmosphere extends into the upper thermally emitting part of the atmosphere.}
    \label{fig: GHlimit_CO2 mix}
\end{figure}

\subsection{Atmospheric escape}
\label{sec: Atmesc}
Atmospheric escape is calculated by the \texttt{VPLanet} module \texttt{AtmEsc}, which includes XUV photolysis of \ce{H2O} into H and O. We assume that the hydrogen produced from photolysis escapes into space based on the hydrodynamic escape mechanism described in \citet{Barnes2020}; see also \citet{Watson1981}, \cite{Zahnle1986}, and \citet{Luger2015}. Hydrogen may escape in one of two ways depending on the composition of the atmosphere. In water-dominated atmospheres, 
hydrogen atoms are liberated from water molecules where the optical depth is approximately unity, which is close to the exobase. Hence, these atoms can escape if the incident photons carry enough energy for the hydrogen atoms to achieve escape velocity, which is generally true for XUV photons. This escape mechanisms assumes as the limiting factor the amount of photon energy deposited in the atmosphere. In this energy-limited regime, hydrogen atoms can also carry away oxygen produced by photodissociation if the hydrogen escapes at sufficiently high velocity \citep{Hunten1987}. Our energy-limited escape model is identical to the one described in \citet{Barnes2020}, App. A.
 
If carbon dioxide and/or oxygen accumulates in the atmosphere, however, there may be fewer water molecules than photolyzing photons near the exobase. In this case, the limiting factor is the availability of water molecules, which must diffuse through the background gas(es) to reach the photolyzing layer of the atmosphere. In our model, we assume escape transitions from the energy-limited regime  to this diffusion-limited regime when the atmosphere consists of more than 60\% carbon dioxide or oxygen. We adopt the diffusion-limited escape model from \citet{Luger2015}, which is a hybrid of the models originally presented in \citet{Zahnle1986} and \citet{Hunten1987}. In our model, the flux is given by
\begin{align}
    F_\mathrm{diff} =  \frac{(m_\mathrm{bg} - m_\mathrm{H})(1 - X_\mathrm{bg})b_\mathrm{bg}gm_\mathrm{H}}{kT_\mathrm{flow}},
\end{align}
where $m_{bg}$ is the mass of the background gas (either oxygen or carbon dioxide), $m_H$ is the mass of a hydrogen atom, $X_{bg}$ is the molar mixing ratio of the background gas a the base of the flow, $b_{bg}$ is the binary diffusion coefficient for the dominant background gas, $k$ is the Boltzmann constant, and $T_{flow}$ is the temperature at the base of the flow. \citet{Mason1970} provides the values for the diffusion constants as 
\begin{equation}
b_\mathrm{O} = 4.8 \times 10^{19}T_\mathrm{flow}^{0.75}, 
\end{equation}
and
\begin{equation}
b_\mathrm{CO_2} = 8.4 \times 10^{17}T_\mathrm{flow}^{0.6}.
\end{equation}
We assume that $T_\mathrm{flow} = 400$~K for all cases \citep{Luger2015}. Note that the only difference between this diffusion-limited escape model from the one described in App.~A of \citet{Barnes2020} is that here we include \ce{CO2}.

\subsection{Vertically extended mixed atmospheres of magma oceans}
\label{sec: PT}

Another improvement of \magmoc{2.0} is its treatment of the atmosphere as vertically extended rather than as a single layer. Following the methodology established in previous studies \citep{Goldblatt2013, Lichtenberg2021,Schaefer2016}, we construct pressure-temperature profiles for a given surface pressure  $p_{surf}$, surface temperature $T_{surf}$, and atmosphere composition.

For atmospheres consisting of a single volatile, we assume a dry adiabat for the lower atmospheric layers. For the upper atmospheric layers,  we incorporate latent heat release via condensation, described by:
\begin{align}
\frac{d \ln T}{d \ln p}  & = R/c_{p,i} (T)\quad \textrm{dry} \\ 
                         & = RT/L_i\quad \quad       \textrm{condensation},
\end{align}
where $p$ is the gas pressure, $T$ is the gas temperature, $R$ is the ideal gas constant, $c_{p,i}(T)$ is the specific heat capacity of the volatile $i$ (either \ce{H2O} or \ce{CO2}) using the Shomate equation with values obtained from the NIST data base \citep{Cox,Chase}. Values for latent heat $L_i$ are taken from \citet{Lichtenberg2021} and \citet{Pierrehumbert}. 

A condensate becomes thermally stable when the p-T profile intersects the relevant condensation curves. For the \ce{H2O} condensation curve, we adopt the August-Roche-Magnus formulation as described in \citet{Alduchov1996}. For pure \ce{CO2}, we use the Clausius-Clayperon relation. We validated this method by comparing pressure-temperature profiles for \ce{H2O} and \ce{CO2} and  $T_{\rm surf}=500 - 2000$~K with previous work \citep{Goldblatt2013,Lichtenberg2021}.

For mixed \ce{H2O} and \ce{CO2} atmospheres, we adopt the multi-species adiabat formulation with condensation of \citet{Graham2021}, where we assume for simplicity again that \ce{CO2} and \ce{H2O} are not removed from an atmospheric layer when condensation occurs (that is, with a supersaturation ratio $S = 1$). Thus, we assume in a specific atmosphere layer equilibrium between condensation and instant re-vaporization of volatiles. This approach is valid for a hot steam atmospheres with vigorous mixing. We note that we use like \citet{Lichtenberg2021} for the calculation of the atmosphere's pressure-temperature profile the ideal gas law for the volatiles, which needs to be revisited for very volatile-rich compositions\footnote{More specifically, \ce{H2O} is supercritical for $T>646$~K and $p>221$~bar and \ce{CO2} is supercritical for $T>304$~K and $p>74$~bar. }. However, we establish here first a basic framework to expand the boundaries of Solar-System magma ocean simulations by starting with simplified assumptions and offering it as an open-source project to the community. We further note that for the majority of the magma ocean simulations, the radiative properties are set by the water vapor opacities. Water, however, makes the atmosphere in the infrared, that is, for the outgoing long wave irradiation optically thick already for low atmospheric pressures $p>0.1$~bar. Further, we find that the magma ocean solidification occurs, when the planet is in the runaway greenhouse limit. Thus, the cooling of the magma ocean is set by the radiative and thermodynamics properties of upper atmosphere layers, the temperature of which is mostly determined by the latent heat release of water vapor that is not impacted by supercritical conditions at the surface. Also \citet{Marcq2017} have verified that deviations from the ideal gas law are of minor importance for calculating the outgoing long wave irradiation of their magma ocean atmospheres. Supercritical surface layers may impact, however, the dry adiabat and thus the very hot initial conditions of the magma ocean stage. Differences in the initial conditions of the magma ocean, however, were found already for Earth to be of minor importance for the magma ocean lifetime and volatile evolution, as the magma ocean spends the majority of the evolution in the runaway greenhouse regime (Sect.~\ref{sec:Earth}). 

In all cases, we utilize a fourth-order Runge-Kutta integrator to compute the pressure-temperature ($p_{\rm gas},T_{\rm gas}$) profile by integrating upwards from surface temperature $T_{\rm surf}= 500$ - 4000~K and $p_{\rm surf}=0.26$ - 26,000~bar\footnote{Note that the assumption of a dry lapse rate is no longer valid beyond the triple point of \ce{H2O} and \ce{CO2}. However, for consistency, we maintain the dry adiabatic assumption even for extreme temperatures and pressures. We acknowledge the crossing of the triple points as a limitation and area for potential improvement in Sect.~\ref{sec: outlook}.}.

Figure~\ref{fig: PT_profiles} (left) shows example profiles for $p_{\rm surf}=p_{\ce{H_2O}}+p_{\ce{CO_2}}=260$~bar for different water volume mixing ratios $ x_{\ce{H2O}}$. We define the vertically uniform $ x_{\ce{H2O}}$ as the ratio of \ce{H2O} partial pressure over the total surface pressure:
\begin{equation}
 x_{\ce{H2O}}=p_{\ce{H2O}}/p_{\rm surf}   
\end{equation}
assuming a well-mixed two component \ce{H2O}-\ce{CO2} atmosphere:
\begin{equation}
p_{\ce{H2O}}+p_{\ce{CO2}}=p_{\rm surf}.
\end{equation}
We stress again that we operate under the assumption of a highly mixed, hot atmosphere, with instant re-evaporation of condensates. Thus, we assume to first order that almost all water is in the gasphase.

The mixed \ce{H2O}-\ce{CO2} pressure-temperature profiles indicate that water vapor releases significantly more latent heat per volume than \ce{CO2}, causing the profiles to closely resemble the pure \ce{H2O} profile even when 90\% of the atmosphere is comprised of \ce{CO2} (Fig.~\ref{fig: PT_profiles} left). The resulting pressure-temperature profiles for a given \ce{H2O}-\ce{CO2} content are used as input for radiative transfer calculations with the open source, atmosphere model \texttt{petitRADTRANS} \citep{Molliere2019} described in the next subsection.
 
\subsection{Radiative transfer in a vertically extended atmosphere}
\label{sec: RT}

\begin{table*}
 	\caption{Opacity sources used in this work.}
	\centering
	\begin{tabular}{lllll} 
		\hline
		\rule{0pt}{3ex}Data type & Data source &  Reference(s) \\
		\hline
	\rule{0pt}{3ex}H$_2$O broadened by H$_2$O & \multirow{ 2}{*}{HITRAN2020}  & \multirow{ 2}{*}{\cite{22GoRoHa}} \\
	 H$_2$O broadened by CO$_2$$^a$ &  &  &  \\
			\rule{0pt}{3ex}CO$_2$ broadened by CO$_2$ & \multirow{ 2}{*}{HITRAN2020}& \multirow{ 2}{*}{\cite{22GoRoHa}} \\
	CO$_2$ broadened by H$_2$O &  &  &  \\
\rule{0pt}{3ex} \multirow{ 4}{*}{H$_2$O continuum} & CAVIAR$^b$  & \cite{11PtMcSh,16ShCaMo,09PaPtSh} \\
 & MT\_CKD & \cite{12MlPaMo} \\
& Baranov 2008 &  \cite{08BaLaMa} \\
& Odintsova 2020 & \cite{20OdTrSi} \\
\rule{0pt}{3ex}CO$_2$ continuum & MT\_CKD & \cite{12MlPaMo} \\
\hline
	\end{tabular}
	\flushleft{$^a$ We used air broadening coefficients as a proxy for CO$_2$ broadening here, as CO$_2$ broadening of H$_2$O parameters were not available in the HITRAN2020 database at the time of computing these opacities.}\\
 	\vspace{-2ex}
	\flushleft{$^b$ More details on the H$_2$O continuum data used from the CAVIAR laboratory experiment can be found in \cite{22AnChEl}. }\\
  	\label{t:opacities}
\end{table*}

\begin{table*}
\caption{Definition of the thermal emission grid}
\centering
\begin{tabular}{ |c | c| c | m{25em}| } 
\hline
   Quantity & Min & Max & Stepsize\\
  \hline
  $T_{\rm{surf}}$& 500~K & 4000~K & 100~K \\ 
  \hline
  $p_{\rm{surf}}$ & 0.26 bar & 26000 bar & 5 per log scale\\ 
  \hline
  $x_{\ce{H2O}}$ & 0 & 1 & 0, $10^{-5}$, $10^{-4}$,  $10^{-3}$, $10^{-2}$, 0.1, 0.2, 0.4, 0.6, 0.8, 1.0\\ 
  \hline
\end{tabular}

\label{tab: grid}
\end{table*}

The thermal evolution of the magma ocean planet is determined by the net outgoing emission $F_{\rm net}$, which is defined as the difference between outgoing long wave radiation (OLR) and incoming absorbed stellar radiation (ASR) at the top of the atmosphere:
\begin{equation}
 F_{\rm net}=  F_{\rm OLR} -  F_{\rm ASR}\label{eq: FnetOr}.
\end{equation}

For the absorbed stellar radiation flux, we use
\begin{equation}
F_{\rm ASR} =    \sigma T_{\rm eff}^4,  
\end{equation}
where the planet's effective or black body temperature $T_{\rm eff}$ is calculated as 
\begin{equation}
 T_{\rm eff}= \left(\frac{1-\alpha}{4 \sigma} \frac{L(t_*)}{4\pi a^2}\right)^{\frac{1}{4}} \label{eq: Teff}
\end{equation}
with the bolometric luminosity of the star $L(t_*)$ at stellar age $t_*$, where $t_*=t+t_{\rm ini}$, that is,  simulation $t$ plus initial stellar age of 5 Myrs, the semi major $a$ and the planetary albedo $\alpha$ of the planet. In this work, we we assumed two albedos 0.75 and 0. The albedo of 0.75 is mostly used to facilitate comparison between \magmoc{1.0} and  \magmoc{2.0}. However, our assumption of a well mixed atmosphere with immediate re-evaporation of condensates is more consistent with a cloud-free atmosphere with low albedo. Moreover, for planets orbiting ultra cool M dwarfs like TRAPPIST-1, very low scattering on top of the atmosphere (albedo= 0-0.1) is predicted \citep{2013Kopparapu}. Thus, we adopt a clear sky albedo ($\alpha$) of 0 as the new default.

We calculate the OLR with \texttt{petitRADTRANS} \citep{Molliere2019} using opacities sampled with the correlated k-method for a wavelength resolution of R=1000. We further assume constant \ce{H2O} and \ce{CO2} volume mixing ratios throughout the atmosphere together with the pressure-temperature profiles as outlined in the previous section. The OLR is equal to the integrated emission on top of the atmosphere between 0.2 and 35 microns, similar to the wavelength coverage in \citet{Goldblatt2013}.

Similar to previous work \citep{Goldblatt2013, Lichtenberg2021,Boukrouche2021}, we emphasize the importance of using the continuum opacities for the dominant greenhouse gases, here \ce{H2O} and \ce{CO2} are vital ingredients to correctly model the atmospheres of rocky (exo-)planets in the habitable zone. The opacity sources and their references are provided in Table~\ref{t:opacities}. We compute the H$_2$O and CO$_2$ k-tables using the HITRAN2020 line list data and broadening coefficients~\citep{22GoRoHa}, with a constant line-wing cutoff of 25cm$^{-1}$. These opacities were converted to a format for input into \texttt{petitRADTRANS} as outlined in \cite{20ChRoAl.exo}. We have verified that for a pure \ce{H2O} atmosphere, the OLR yields the canonical runaway greenhouse limit of 282~W/m$^{2}$ for surface temperatures between 500 - 1800~K \citep{Goldblatt2013}. We further note that for \ce{CO2} we use an additional continuum opacity compared to \citet{Lichtenberg2021}, which leads to an overall reduction in emission of 20\% in a pure \ce{CO2} atmosphere with $p_{\rm surf}=260$~bar compared to \citet{Lichtenberg2021}. Already, \citet{Marcq2017} preformed non-gray radiative transfer calculations for mixed \ce{H2O}-\ce{CO2} atmospheres on magma oceans. They also retrieve the blanketing effect of the runaway greenhouse limit of 280~W/m$^{}2$ that is indeed the major factor shaping the magma ocean evolution also for mixed \ce{H2O}-\ce{CO2} atmospheres in the simulations presented here. In fact, \citet[][]{Marcq2017} also point out that the runaway greenhouse limit can be lowered by tens of W/m${}^2$ when \ce{CO2} starts to dominate over \ce{H2O} (see also Sections \ref{sec: grey} and \ref{sec: Grey_exo}).

The emission curves for $p_{\rm surf}=260$~bar and $T_{\rm surf}$ for different water volume mixing ratios $x_{\ce{H2O}}$ (Fig.~\ref{fig: PT_profiles} right) are very similar to the emission curve of a pure \ce{H2O} atmosphere for $x_{\ce{H2O}}\geq 10^{-2}$. There are, however, two notable differences: A \ce{CO2} absorption band at 4.3 $\mu$m appears as soon as \ce{CO2} is added and its amplitude increases with higher \ce{CO2} abundances. In addition, the overall thermal emission decreases with lower $x_{\ce{H2O}}$ because less water per volume is available that can condense out and thus heat the upper atmospheric layers.  Consequently, the upper atmosphere layers cool down and emit less flux (see Fig.~\ref{fig: PT_profiles} left). 

For $x_{\ce{H2O}}\leq 10^{-3}$, \ce{CO2} emission begins to dominate, causing the overall emission to increase as $x_{\ce{H2O}}$ decreases, eventually exceeding the runaway greenhouse limit.  This effect occurs because the hot dry adiabat of a \ce{CO2}-dominated atmosphere extends up to $p \leq 50$~mbar into the emitting atmosphere layers. 

A closer inspection reveals that the runaway greenhouse limit for a mixed \ce{H2O}-\ce{CO2} atmosphere with $x_{\ce{H2O}}<1$ is restricted to a lower surface temperature regime compared to an atmosphere composed of 100\% \ce{H2O} \citep[see e.g.][]{Goldblatt2013,Lichtenberg2021} and further becomes smaller than the canonical value of 282~W/m${}^2$. Further details on this phenomenon are provided in Sections~\ref{sec: grey} and \ref{sec: Grey_exo}.

During the runtime of \magmoc{2.0}, we do not compute full radiative transfer; instead, we construct for a given planet a thermal emission grid (See Table~\ref{tab: grid} for the set-up). As the magma ocean evolves, the model interpolates surface temperature linearly, while surface pressure and water volume mixing ratio are interpolated logarithmically within the range $10^{-6}\geq x_{\ce{H2O}} \leq 1$. For $x_{\ce{H2O}} <10^{-6}$, thermal emission for $x_{\ce{H2O}} = 0$ is used, representing a pure \ce{CO2} atmosphere. This is done because we interpolate in log-space and thus cannot interpolate to log(0). We also found that for $x_{\ce{H2O}} =10^{-6}$, the integrated thermal emission deviates by less than 10\% from the that of a  pure \ce{CO2} atmosphere.

To speed-up computation time and as a `sanity check´ for the thermal evolution based on full radiative transfer calculations, we have also developed an analytic approximation of the thermal emission of a mixed \ce{H2O}-\ce{CO2} steam atmosphere, including modifications to the runaway greenhouse atmosphere limit with increasing \ce{CO2} content. The equations describing the approximation are outlined in the Appendix~\ref{sec: grey}.

\FloatBarrier
\section{Magma ocean evolution of TRAPPIST-1 e, f, and g} 
\label{sec: Simulations}

 \begin{table*}[ht]
\begin{center}
    \caption{Physical and run parameters for TRAPPIST-1~e, f, and g used by \magmoc{2.0}.}
	\begin{tabular}{cccc}
		\noalign{\smallskip}
		\hline
		\noalign{\smallskip}
		Parameter & TRAPPIST-1 e  & TRAPPIST-1 f & TRAPPIST-1 g \\ 
		\noalign{\smallskip}
		\hline \hline
		\noalign{\smallskip}
		$r_\mathrm{p} \: [R_\Earth]$ ${}^{a}$ &0.920 & 1.045 & 1.129  \\
		$r_\mathrm{c} \: [R_\Earth]$ ${}^{b}$ &0.490 & 0.557 & 0.602  \\
		$M_\mathrm{p} \: [M_\Earth]$ ${}^{a}$  & 0.692 & 1.039 & 1.321 \\
		$a \: [\si{\astronomicalunit}]$ ${}^{a}$ & 0.0293 & 0.0385 & 0.0468 \\
		$e$ ${}^{c}$ & 0.005 & 0.01 & 0.002 \\
		Initial Radiogenic Power [TW] ${}^{e}$ & 57 & 69 & 85 \\
		\noalign{\smallskip}
		\hline
       HZ entry [Myrs] & 253 & 130 & 76 \\
         \hline
		Albedo $\alpha$ &  \multicolumn{3}{c}{0.75 (\texttt{MagOcV1.0}) \& 0 (clear-sky)} \\
		$M_{\ce{H2O}}^\mathrm{ini}$ &  \multicolumn{3}{c}{$1-\SI{100}{TO}$} \\
        $M_{\ce{CO2}}^\mathrm{ini}$ scaled with $M_{\ce{H2O}}^\mathrm{ini}$ &  \multicolumn{3}{c}{$0, 0.3, 1$ (main) \& 0, 0.1, 0.2, ... , 1 (See~\href{https://doi.org/10.5281/zenodo.14442985l}{detailed grid}) }  \\
		$T_\mathrm{surf}^\mathrm{ini} = T_\mathrm{p}^\mathrm{ini}$  & \multicolumn{3}{c}{\SI{4000}{\kelvin}} \\
		$\epsilon_\mathrm{XUV}$ &  \multicolumn{3}{c}{0.3}  \\ 
		Stellar age at $t=0$ & \multicolumn{3}{c}{5 Myrs}  \\ 
		Atmospheric model &  \multicolumn{3}{c}{RT grid, corr. gray} \\
		\vplanet{} modules & \multicolumn{3}{c}{\magmoc{2.0}, \texttt{AtmEsc}, \texttt{RadHeat}, \texttt{EqTide}, \texttt{STELLAR}} \\
		\noalign{\smallskip}
		\hline
	\end{tabular}
	\\
	${}^{a}$ \citet{Agol2020}, ${}^{b} r_\mathrm{c} = r_\mathrm{p} \times r_{\mathrm{c,\Earth}}/R_\Earth$, ${}^{c}$ \citet{Grimm2018}, ${}^{d}$ based on \citet{Grimm2018} data, ${}^{e}$ For the abundances of radioactive isotopes, we use Earth abundances, scaled by the mass of the planet. 
	\label{Tab_Input_TRAPPIST-1}
\end{center}
\end{table*}
We tested \magmoc{2.0} for different Earth scenarios (Sect.~\ref{sec:Earth}), confirming previous research findings on the substantial \ce{H2O}-\ce{CO2} outgassing feedback \citep{Bower2019}. Further, we found that the addition of \ce{CO2} has a minor impact on the solidification timescale compared to a pure \ce{H2O} atmosphere when atmospheric escape is negligible. This similarity between mixed \ce{H2O}-\ce{CO2} and pure \ce{H2O} simulations is primarily due to the dominance of the runway greenhouse radiation limit for the majority of the magma ocean lifetime. In contrast to the Earth's magma ocean stage \citep[e.g.][]{Hamano2013,Barth2021}, atmospheric erosion is not negligible for the potentially habitable TRAPPIST-1 planets e, f, and g.
 
In this work, we systematically revisit evolution trajectories investigated by \citet{Barth2021} with pure \ce{H2O} atmospheres for 1 - 100 TO initial water mass. We focus on simulations with no \ce{CO2}, an initial \ce{CO2} mass equal to $0.3\times$ the initial \ce{H2O} mass, and an extreme scenario, where the initial \ce{CO2} mass is assumed to be equal to the initial \ce{H2O} mass. See Table~\ref{Tab_Input_TRAPPIST-1} for the relevant parameters. We further note that we neglect tidal interactions in accordance to results by \citet{Barth2021}.
 
 The wide range in initial volatile composition encompasses the large uncertainties in volatile content and the \ce{H2O}-\ce{CO2} ratio acquired during planet formation \citep[e.g.][]{Bitsch2019}. These diverse scenarios allow us to assess to what extent \ce{CO2} impacts the magma ocean lifetime and volatile distribution on magma oceans with oxidized outgassing in close proximity to an M dwarf star. 

We adopt the \ce{H2O} and \ce{CO2}  outgassing laws of \citet{Niko2019}, which better represent the current understanding that \ce{CO2} is already outgassed during the initial magma ocean stage (see Sect.~\ref{sec: outgas}), whereas \ce{H2O} only builds up when the mantle begins to fully solidify. Unless otherwise specified, we adopt the RT atmosphere grid model.

\subsection{TRAPPIST-1 g evolution}
\label{sec: TRAPPIST-1g}
\begin{figure}[htb]
    \centering
    \includegraphics[width=0.49\textwidth]{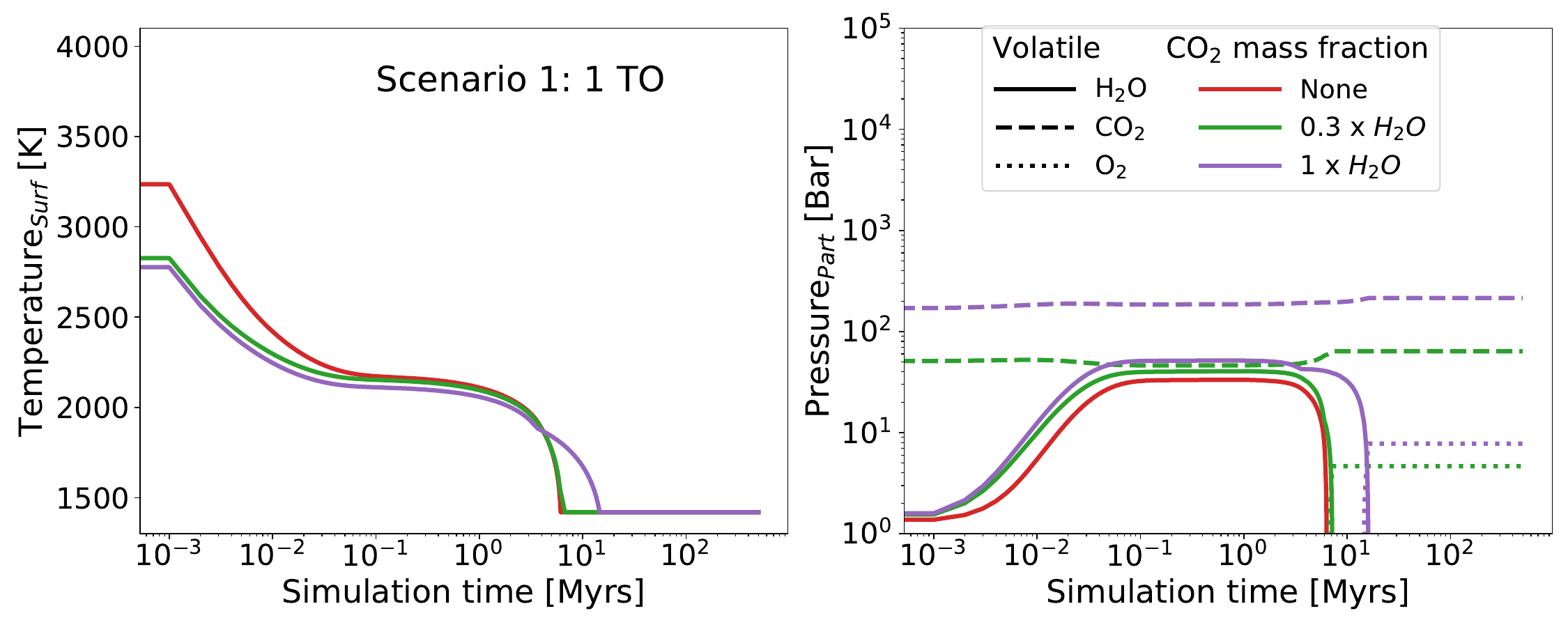}
     \includegraphics[width=0.49\textwidth]{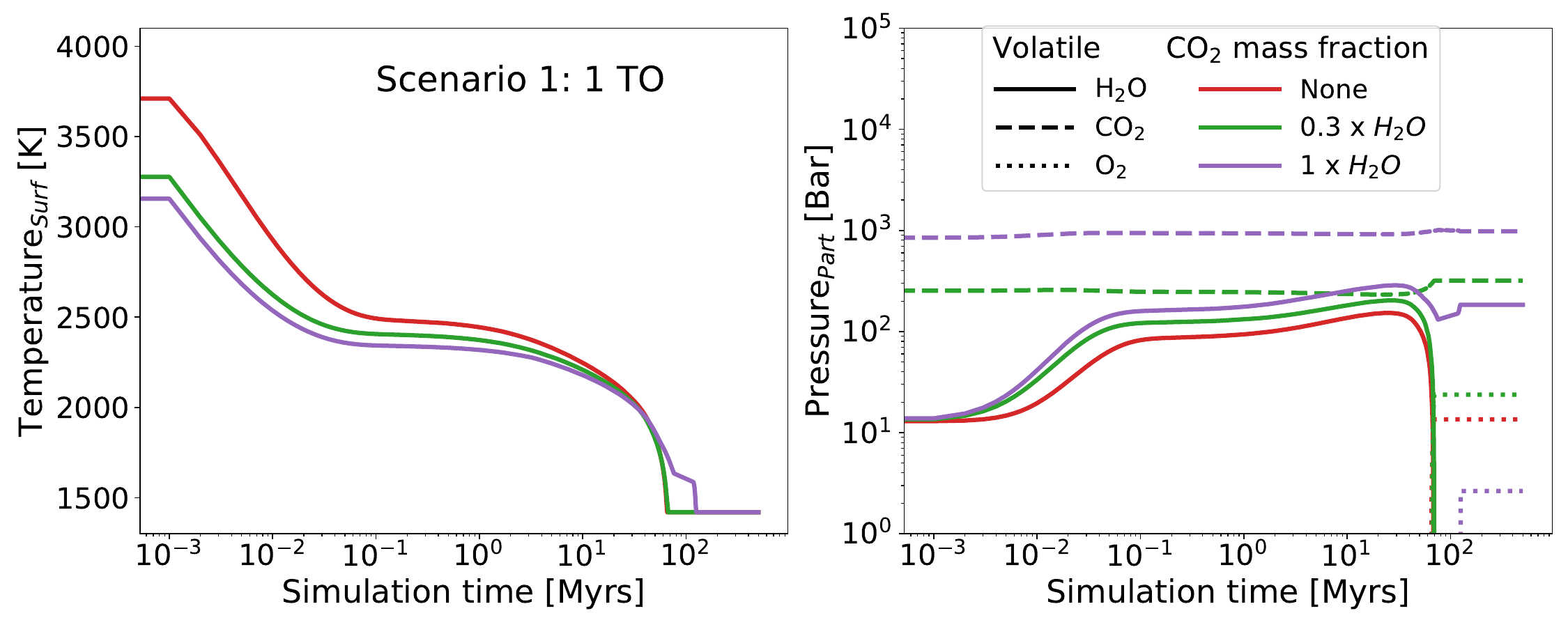}
   \includegraphics[width=0.49\textwidth]{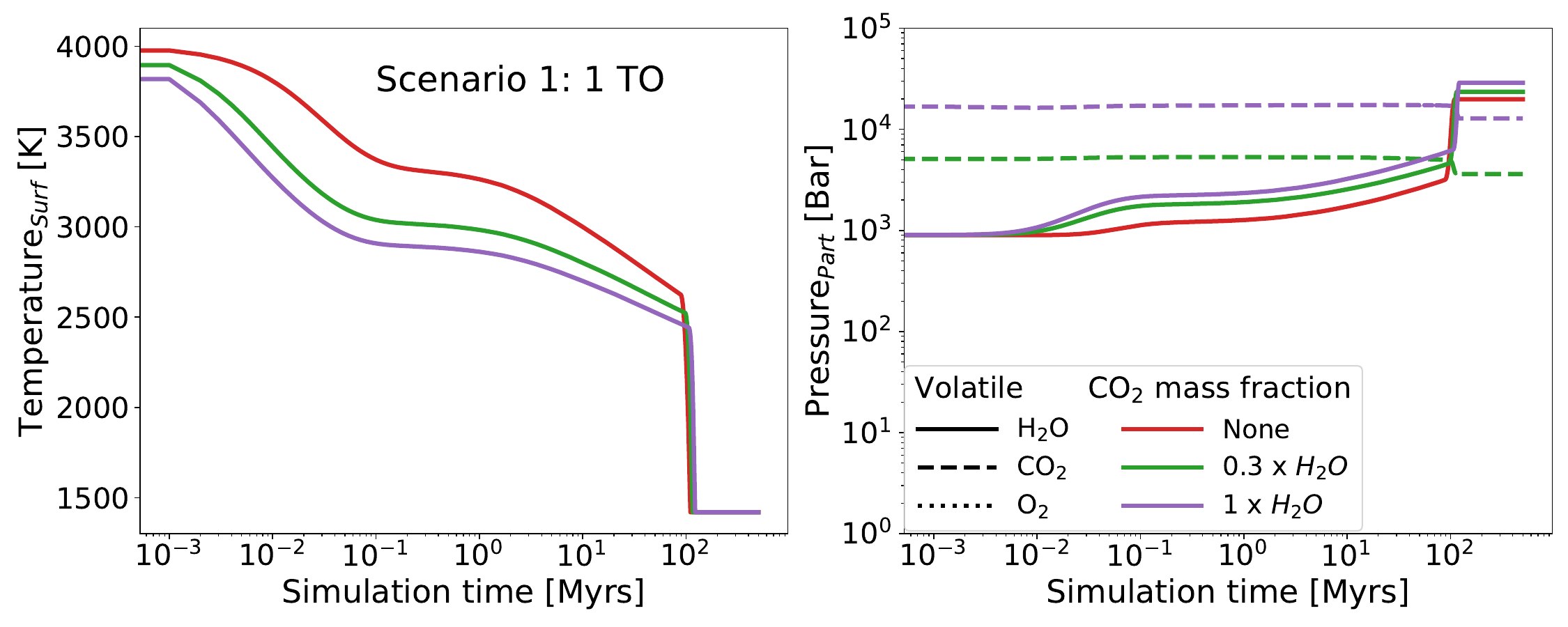}
    \caption{Magma ocean evolution for initial \ce{H2O} of 1 TO, 5 TO, and 100 TO , respectively. All scenarios are for albedo=0. Initial \ce{CO2} mass content is scaled relative to the \ce{H2O} content by a factor of 0, 0.3 and 1, denoted by red, green and purple lines, respectively. Surface temperature (left) and volatile content evolution (right) are shown.\ce{H2O}, \ce{CO2}, \ce{O2} are denoted by solid, dashed and dotted lines, respectively. }
    \label{fig: TR1g-Evolution_A0}
\end{figure}

\begin{figure*}[ht]
     \includegraphics[width=0.95\textwidth]{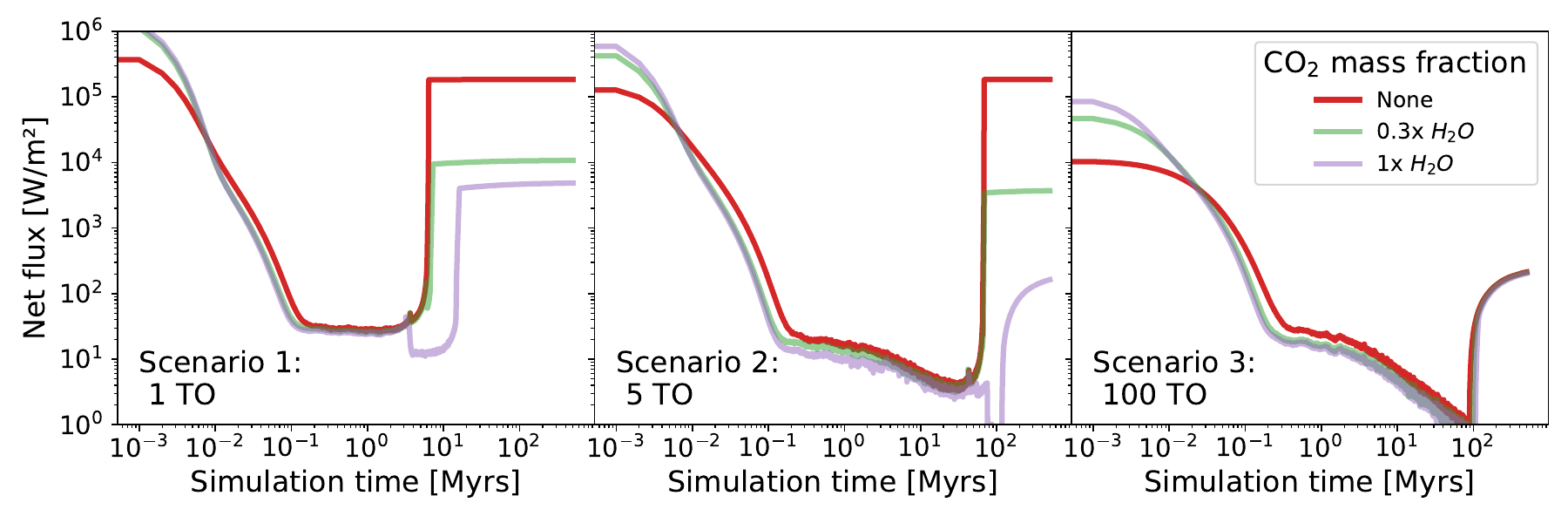}
    \caption{TRAPPIST-1 g: Net flux (OLR-ASR) evolution for the magma ocean scenarios shown in Fig.~\ref{fig: TR1g-Evolution_A0}, that is, for 1 TO, 5 TO, and 100 TO initial \ce{H2O} and various initial \ce{CO2} mass fractions for an albedo 0. We note simulations with an assumed albedo of 0.75 (not shown) are qualitatively similar but have a shorter regime with net flux limited by the runaway greenhouse radiation limit. }
    \label{fig: TR1g-Emission}
\end{figure*}
Following \citet{Barth2021}, we select TRAPPIST-1 g simulations for initial water masses of 1, 5, and 100~TO, respectively, to discuss critical stages in the magma ocean evolution for planets in the habitable zone of TRAPPIST-1. \citet{Barth2021} identified roughly three scenarios for the magma ocean evolution of TRAPPIST-1~g that represent significant stages for understanding the impact of atmospheric erosion on an oxidized magma ocean. We thus explore here the same initial water scenarios for better comparison. To assess the impact of \ce{CO2} on the three outlined scenarios (Scenario 1, 2 and 3), we compare the surface temperature and evolution tracks without \ce{CO2} to those with \ce{CO2} (Fig.~\ref{fig: TR1g-Evolution_A0}).  We note, however, that \citet{Barth2021} assumed an albedo = 0.75.  Here, we focus mostly on simulations with a clear-sky albedo of 0, which is more consistent with our clear-sky radiative transfer as well as with the low scattering efficiency calculated for planets around M dwarfs \citep{2013Kopparapu}. 

We first report the general trends for all scenarios: There is a significant impact on the general volatile distribution (Sect.~\ref{sec: feedback}) in simulations with additional \ce{CO2}: The magma ocean starts with a \ce{CO2}-dominated atmosphere, then \ce{H2O} tends to become the most dominant species as the mantle solidifies. The planet can end its magma ocean stage again with a \ce{CO2} dominated atmosphere, when the planet desiccates. The feedback on the volatile distribution between \ce{H2O} and \ce{CO2} during outgassing is mostly evident in changes in \ce{CO2} partial pressures.
When \ce{H2O} starts to become dominant during late stage mantle outgassing, \ce{CO2} partial pressures drop. If the planet desiccates, \ce{CO2} partial pressures rise. These changes generally align with our results for the Earth magma ocean simulations (Sect.~\ref{sec: Earth -Mixed CO2 scenarios}), as well as with those derived by \citet{Bower2019}.

\citet{Bower2019} also reported that the presence of a thick \ce{CO2} atmosphere delays the outgassing of \ce{H2O} during the magma ocean evolution for Earth. We can't reproduce this claim in our simulations. Instead, we find delayed \ce{H2O} outgassing only in the volatile poor `Scenario 1' with albedo 0.75, specifically in the extreme case of adding 1~TO \ce{CO2} (not shown). Closer inspection reveals that the delay in \ce{H2O} outgassing arises solely from the slower cooling of the magma ocean and is not attributable to the feedback effect.

While we don't find significant changes in time of \ce{H2O} outgassing, the thermal evolution and volatile distribution on TRAPPIST-1g can be impacted with \ce{CO2}, depending on initial water content.

In the `dry' scenario with 1~TO \ce{H2O}, the atmosphere becomes completely desiccated, which leads to rapid mantle solidification within 7~Myrs with no \ce{CO2} and moderate \ce{CO2} content. If an extreme amount of \ce{CO2} is added, then diffusion limited escape delays desiccation such that complete solidification occurs after 17 Myrs. Further, we find that even in the case of moderate \ce{CO2} content, even a slight delay of solidification by 1~Myrs due to diffusion limited escape facilitates moderate abiotic \ce{O2} build-up of a few bar. Without any \ce{CO2}, all abiotically created \ce{O2} is deposited in the mantle before complete solidification.

In the `intermediate' scenario with 5~TO, complete desiccation occurs at a much later date compared to the `dry' scenario, that is, after 70~Myrs simulation time. Moderate \ce{CO2} content leads again to a slight delay in desiccation such that two times the amount of abiotically created \ce{O2} can accumulate in the atmosphere compared to the \ce{CO2}-free simulation. Leaving a few 10s of bar. With extreme \ce{CO2} content, complete desiccation is delayed even further by \ce{CO2} diffusion limited escape to 120~Myrs. In this simulation, the planet enters the habitable zone of its host star at 76~Myrs with about 1~TO of \ce{H2O} in the atmosphere (200~bar). Beyond 200 Myrs simulation time, the model is no longer suitable to follow the further evolution, as evidenced by the plateauing of the surface temperature  evolution (See Sect.~\ref{sec: method}).

For the `wet' scenario with 100~TO, all simulations enter the habitable zone at 76~Myrs with more than 2000~bars of \ce{H2O} in the atmosphere and the mantle solidifies very late after 150~Myrs.  Additional \ce{CO2} extends the solidification time by only few \% compared to the \ce{H2O}-free simulation time. Generally, additional \ce{CO2} has little effect on the final magma ocean state. The final state is characterized by an accumulation of more than 10000~bars of water in the atmosphere as the mantle solidifies. All abiotically created \ce{O2} enters the mantle already during the magma ocean evolution and thus no \ce{O2} build-up occurs even after several 100~Myrs. The lack of final \ce{O2} build-up beyond solidification may be, however, due to the limitation of the model that is strictly valid ``only'' until the magma ocean solidification state at 150~Myrs is reached.

A closer look on the evolution of the net fluxes for the three scenarios (Fig.~\ref{fig: TR1g-Emission}) reveals the key processes determining the magma ocean lifetime. For the `dry scenario 1', the final thermal evolution is primarily determined by the fate of the water, even in the presence of 200~bar \ce{CO2}. The mixed \ce{H2O}-\ce{CO2} atmosphere enters the runaway greenhouse radiation limit already after 100 000 years, which  reduces the outgoing thermal radiation to less than 300~W/m${}^2$ as long as water is present. Once all water has been eroded, the remaining 60 to 200~bar thick \ce{CO2} atmosphere with a 1400~K surface temperature yields net outgoing flux of more than 1000~W/m${}^2$ that quickly leads to mantle solidification (Sect.~\ref{sec: RT}, \citet{Lichtenberg2021}). All initial differences with thermal evolution due to differences in the \ce{CO2} content are removed, once the simulations enter the runaway greenhouse limit.

For the `intermediate' scenario 2, both, the evolution of the runaway greenhouse radiation limit and atmospheric erosion determines the mantle solidification time. Here, again the runaway greenhouse radiation limit with less than 300~W/m${}^2$ is reached after 100 000 years. Because there is more water in the system compared to the `dry' scenario, the radiation limit is maintained for at least 70 Myrs. The net flux continuously drops during that time mainly because the stellar luminosity continuously decreases as the host star evolves to the main sequence. We further note that simulations with additional \ce{CO2} consistently show a lower runaway greenhouse limit (in particular around 1~Myrs evolution time) by 10s of W/m${}^2$ as long as the atmosphere remains \ce{CO2} dominated (see Sect.~\ref{sec: Grey_exo}). When the majority of \ce{H2O} is outgassed during the final stages of mantle solidification (at 70~Myrs), the thermal evolution curves converge for the moderate \ce{CO2} and \ce{CO2}-free simulation until complete desiccation occurs that triggers quasi-instant solidification. For extreme \ce{CO2} content, however, diffusion limited escape can prevent desiccation and thus delays solidification to 120~Myrs. 

The ``wet scenario 3''  with 100~TO \ce{H2O} is entirely determined by the runaway greenhouse limit. Additional \ce{CO2} leads only initially to differences in evolution. Once, the simulations enter the runaway greenhouse limit (after a few 100 000~Myrs), these differences seize to matter. We note here again that the simulations with additional \ce{CO2} have a lower runaway greenhouse radiation limit, as long as \ce{CO2} remains the dominant atmosphere constituent. All simulations converge, however, once the majority of the water is outgassed after 100~Myrs. Even with extreme \ce{CO2} content, solidification occurs only slightly later compared to the \ce{CO2}-free simulation. 

In \citet{Barth2021}, the radiation limit was assumed to apply only for surface temperatures cooler than 1800~K, regardless of surface pressure. Within that framework, thick \ce{H2O}-dominated atmospheres entered the runaway greenhouse radiation limit very late leading to an extension of the magma ocean lifetime. In this work, however, we find that magma oceans with thick \ce{H2O} dominated atmospheres enter the greenhouse limit for hotter surface temperatures (Sect.~\ref{sec: grey}). Thus, we don't reproduce the magma ocean lifetime extension proposed in \citet{Barth2021} for very high volatile content.

In summary, the fate of water effectively dominates magma ocean solidification lifetimes for all cases. \ce{CO2} generally ``only'' influences the time scale of water erosion. Because the atmosphere keeps more water vapor, in particular for extreme \ce{CO2} content, the magma ocean lifetime of TRAPPIST-1g can be significantly delayed for the `dry' and `intermediate' scenarios with 1~TO and 5~TO \ce{H2O} respectively. 

We caution, however, that we cannot predict how long the 1400~K surface temperatures can be maintained, because the geophysical model terminates the evolution of the mantle at this point. It thus remains to be confirmed if the extreme jump in outgoing flux as the planet desiccates that triggers immediate solidification can be re-captured with other models that follow the mantle evolution beyond the magma ocean solidification state.

\subsection{Overview of magma ocean evolution in TRAPPIST-1 e, f, and g}
\label{sec: Overview}
\begin{figure*}
    \centering
        Albedo 0.75 \hspace{7.5cm} Albedo 0 \par
         \includegraphics[width=0.49\textwidth]{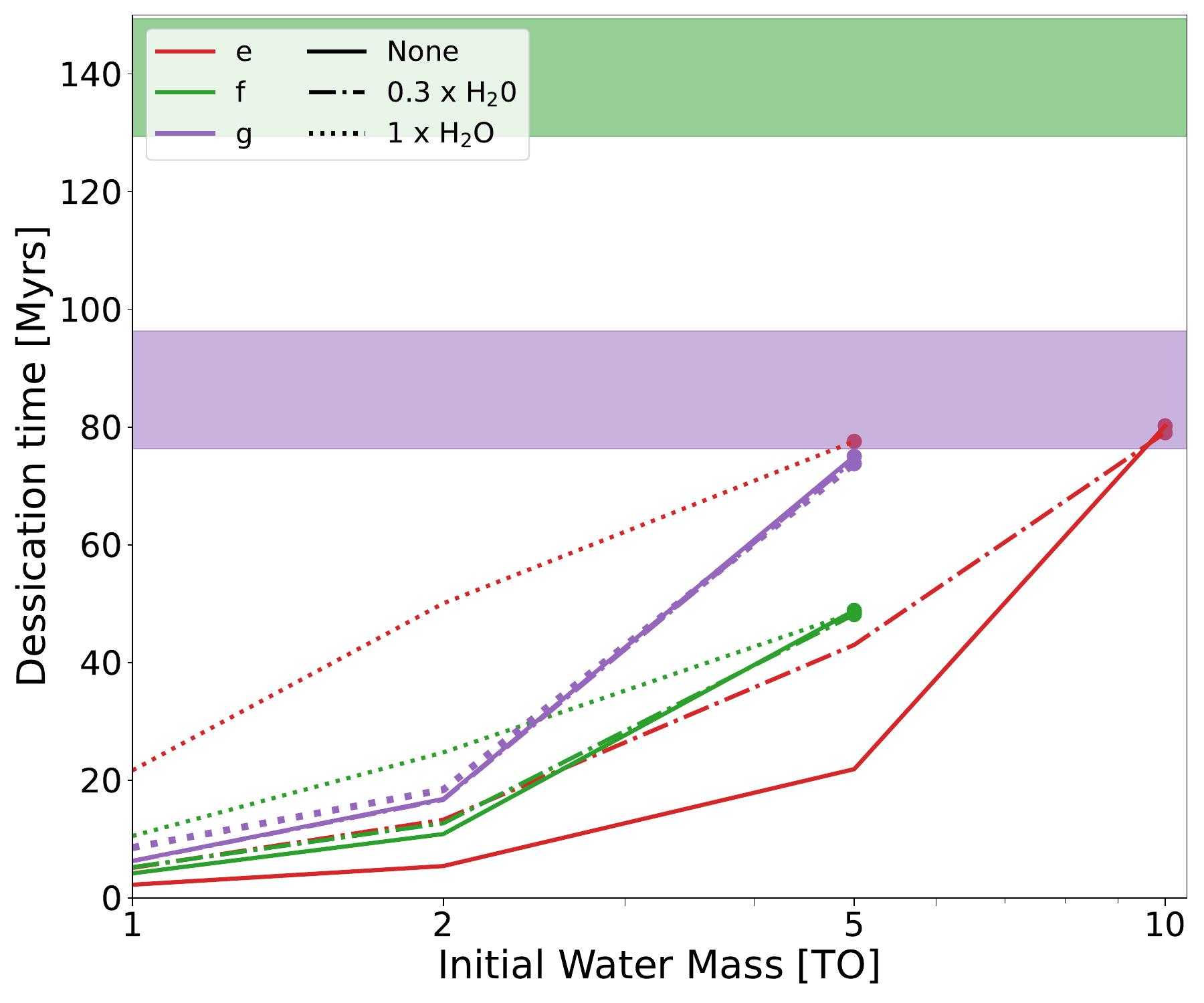}
         \includegraphics[width=0.49\textwidth]{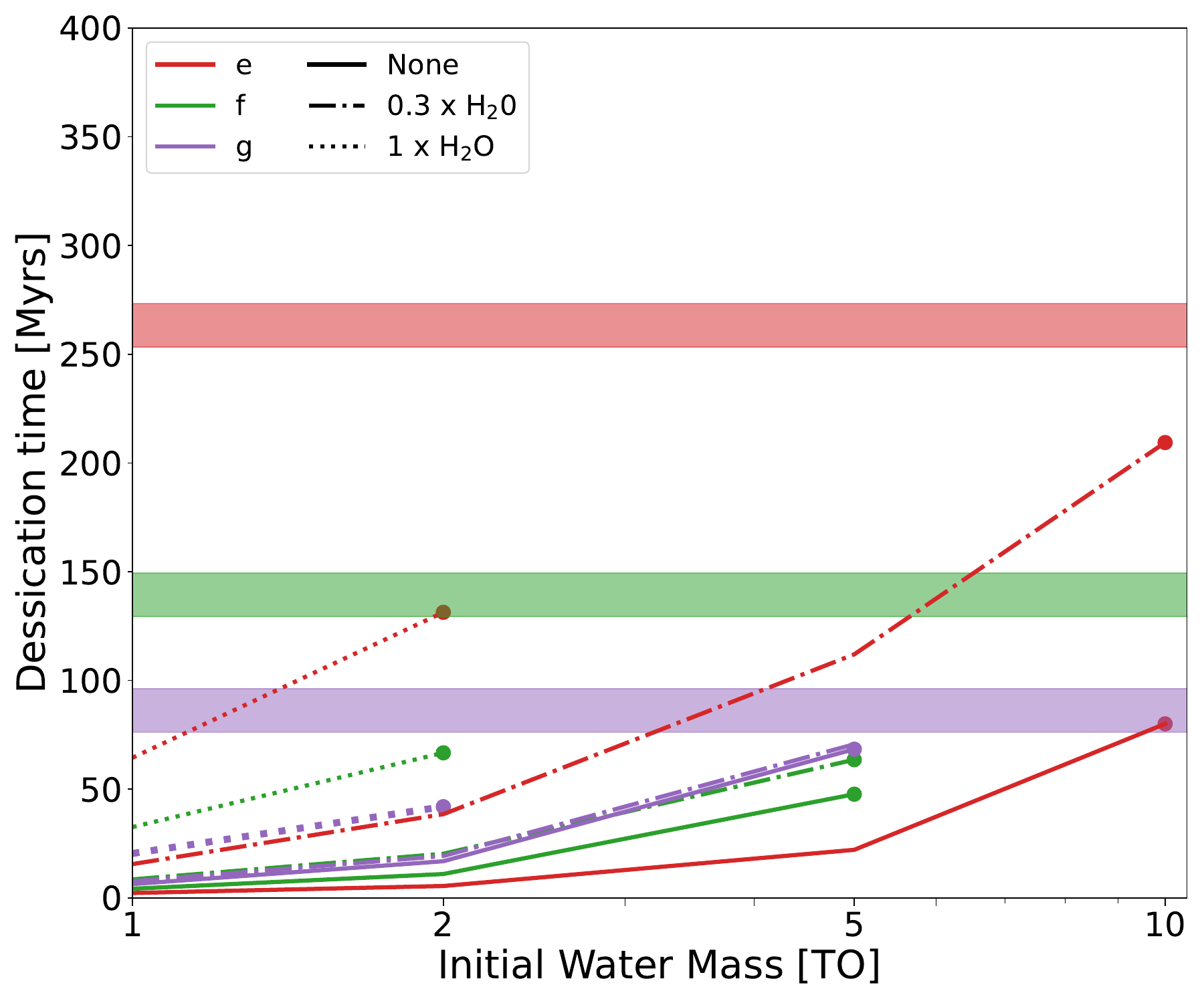}
   \caption{Overview of total atmospheric desiccation times of TRAPPIST-1 e (red), f (green) and g (purple) for a dry composition with 1-10 TO of initial \ce{H2O}, assuming an albedo of 0.75 (left) and 0 (right). Colored shaded regions denote the time, when the respective planet enters the habitable zone.  Solid lines represent scenario with pure \ce{H2O} atmospheres, dashed-dotted lines denote scenarios with added \ce{CO2} scaled by $0.3 \times$ compared
to the initial \ce{H2O} mass, and dotted lines show scenarios with added \ce{CO2} equal to the initial water mass. The solid circles denote for a given \ce{H2O}-\ce{CO2} mass fraction and planet the maximum initial water mass, for which total atmosphere desiccation occurs before the respective planet enters the habitable zone. Please note the change in y-axis scale between the plots. }
    \label{fig: TR1e 10 TO}
\end{figure*}

We find that additional \ce{CO2} has a particular strong impact on the timescale of atmospheric water loss for simulations with dry to intermediate initial water content ($\leq 10$~TO \ce{H2O}). Because desiccation is in our model followed by complete mantle solidification,  the impact of \ce{CO2} on atmospheric erosion also impacts the magma ocean solidification time scale. In addition, the magma ocean duration changes with different scattering (albedo) assumptions, which may also impact atmospheric water loss. Thus, we first inspect the change in atmospheric desiccation time for TRAPPIST-1 e, f and g  and for our nominal clear-sky (albedo=0) and the high albedo simulations (albedo =0.75) (Fig.~\ref{fig: TR1e 10 TO}).

In our simulations that assume clear-sky albedo, $F_{\rm net}$ is low and thus the magma ocean cools and evolves over longer timescales compared to the high albedo simulations. This difference in cooling timescale by itself does not appear, however, to strongly impact atmospheric mass loss rate as is evident by comparing pure \ce{H2O} simulations for albedo 0, and 0.75, respectively (Fig.~\ref{fig: TR1e 10 TO}, solid lines). The addition of \ce{CO2}, however, leads to significant changes in atmospheric desiccation.

If a clear sky is assumed, scenarios with extreme \ce{CO2} content only lead to complete atmospheric desiccation with very dry initial water content ($\leq 2$ TO \ce{H2O}). For the high albedo simulations, the shorter magma ocean duration compared to the clear-sky simulations appears to limit the impact of \ce{CO2}. Desiccation is only significantly delayed for TRAPPIST-1e, even with extreme \ce{CO2} content.

\begin{figure}
    \centering  
    {\small Albedo=0.75}\par\medskip
    \includegraphics [width=0.49\textwidth]{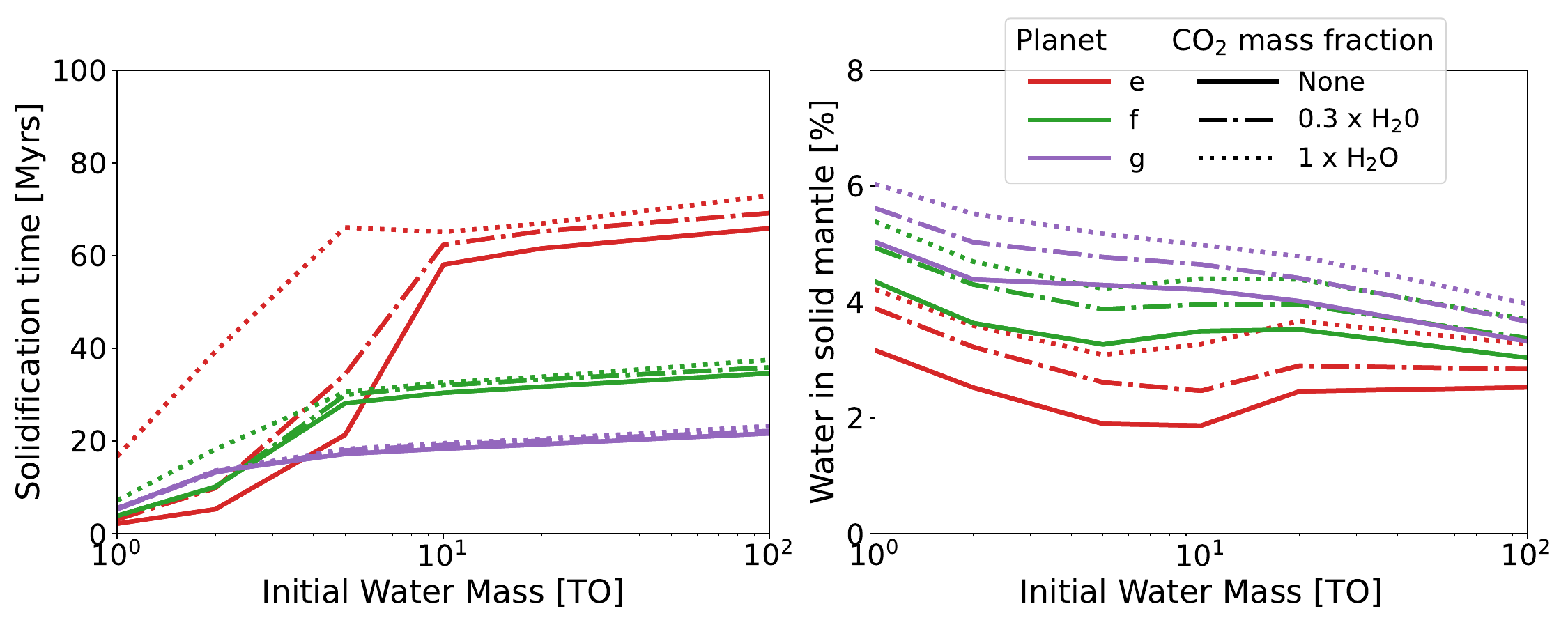}
        {\small Albedo=0}\par\medskip
    \includegraphics [width=0.49\textwidth]{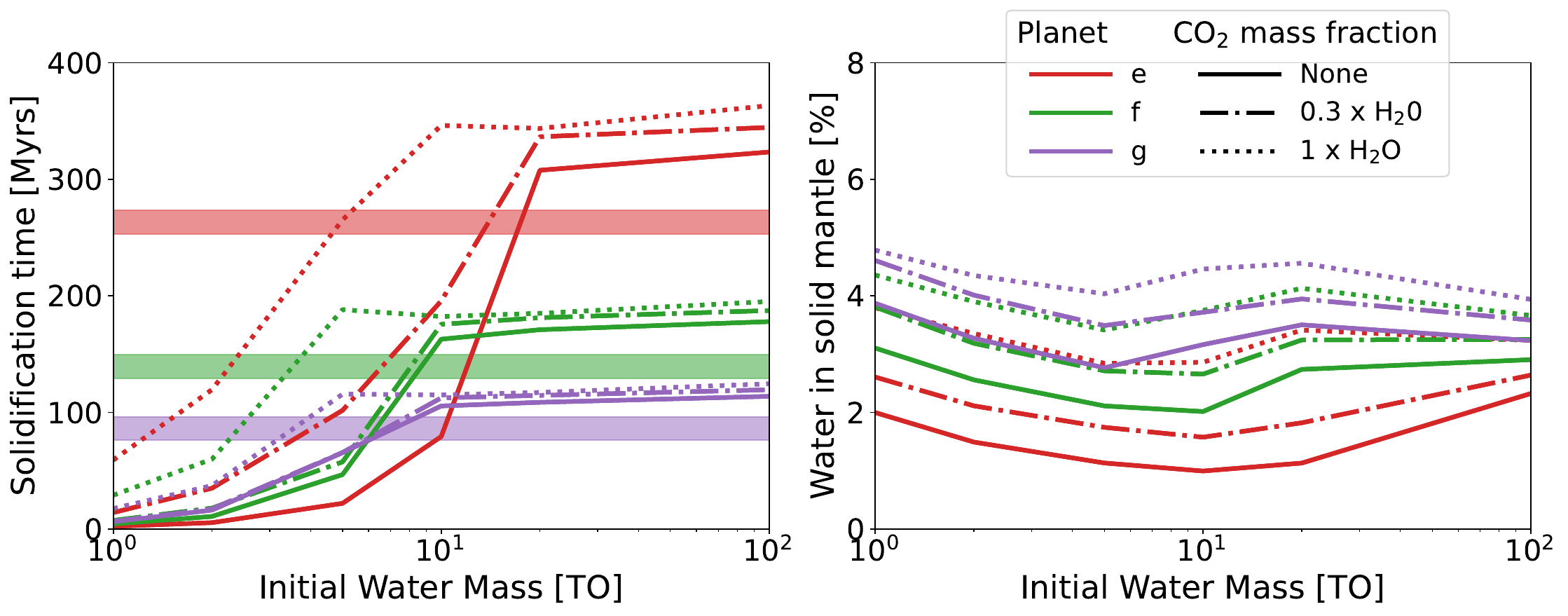}
   \caption{Overview of magma ocean solidification time (left) and the remaining water in the solid mantle compared to initial water mass in percent (right) for TRAPPIST-1 e (red), f (green), and g (purple), considering various \ce{CO2} contents for albedo 0.75 (top panels) and albedo 0 (bottom panels). Solid lines represent scenarios with pure \ce{H2O} atmospheres, dashed-dotted lines denote scenarios with added \ce{CO2} scaled by $0.3x$ compared to the initial \ce{H2O} mass, and dotted lines show scenarios with added \ce{CO2} equal to the initial water mass. Colored shaded regions denote the time, when the respective planet enters the habitable zone.}
    \label{fig: Overview TR1efg}
\end{figure}

The combination of delayed atmospheric water mass loss and low magma ocean cooling rate in the clear-sky case also leads to a significant extension of the magma ocean solidification timescale with intermediate initial water content (5-10 TO \ce{H2O})\footnote{A more \href{https://doi.org/10.5281/zenodo.14442985l}{detailed grid} displaying critical properties at the end of the magma ocean can be found on Zenodo.} (Fig.~\ref{fig: Overview TR1efg}). This is true for TRAPPIST-1 e, f and g with extreme \ce{CO2} content (Fig.~\ref{fig: Overview TR1efg} dotted lines, bottom left panel) compared to the pure \ce{H2O} and high albedo simulations (Fig.~\ref{fig: Overview TR1efg} solid lines, upper left panel). The magma ocean solidification times are so long that the planets can enter their habitable zone before atmospheric desiccation is complete.

For the volatile-rich scenarios ($\geq 10$ TO \ce{H2O}) and with clear skies (albedo = 0), TRAPPIST-1 e, f and g  enter the habitable zone with partly molten surfaces and a mixed \ce{H2O}-\ce{CO2} dominated atmosphere, because atmospheric escape is of lesser importance. But even there,  more \ce{CO2} leads to an extension of mantle solidification time in particular for TRAPPIST-1e. In these scenarios, the addition of \ce{CO2} reduces the runaway greenhouse radiation limit by a few 10s of W/m${}^2$\citep[see also][]{Marcq2017}. For simulations with albedo=0.75, the extension of the magma ocean lifetime for the volatile-rich scenarios is less evident and is at most  10\% even for TRAPPIST-1e.

The duration of the magma ocean also shapes the amount of remaining water in the solidified mantle due to atmospheric erosion (Fig.~\ref{fig: Overview TR1efg}, right panels). The impact of atmospheric erosion is already evident by comparing the remaining water fraction for the individual planets. TRAPPIST-1 g can retain the most water because it is farther away from its host star and thus experiences less erosion compared to TRAPPIST-1f. TRAPPIST-1f experiences less atmospheric erosion than TRAPPIST-1 e and thus retains more water than the latter.

The impact of water mass loss over different time scales explains the differences for remaining water in the solidified mantle when comparing high albedo (short magma ocean phase $<100$~Myrs) with clear sky simulations (long magma ocean phase, up to 350~Myrs). As already pointed out by \citet{Barth2021}, the volatiles in a magma ocean planet are embedded in a strongly coupled system. If this system is subject to continuous water erosion for a longer time, then also more water can be removed from the planet, because more water is outgassed from the melt to compensate for the mass loss from the atmosphere. Consequently, for the same planet and the same initial water input, there is more remaining water in the melt in the high albedo simulations compared to clear sky simulations with longer magma ocean lifetimes.

However, only 2--6\%\footnote{However, it's important to consider that we are exploring a large range of initial water masses and that the small percentages of remaining water in the mantle has to be scaled with initial water mass to diagnose how much total water mass is in the mantle.} of the initial water can be retained in the melt in the pure \ce{H2O} outgassing scenario for albedo 0.75 and 1-5\% for albedo 0. Apparently, the impact of extended water erosion over timescales of 100~Myrs and longer does not lead to a drastic reduction in remaining water. This relatively minor impact on the water mantle content with extended magma ocean lifetime indicates that most water is sequestered in the solidified part of the mantle during the very first  million years of simulation time, before atmospheric erosion has had time to remove significant amounts of water from the atmosphere and thus from the remaining molten part of the mantle. It takes typically more than a few million years to significantly erode the atmosphere, except for the very 'dry' scenarios with 1 TO \ce{H2O}.

Adding \ce{CO2}, tends to increase the percentage of remaining water fraction by up to 2\%. For TRAPPIST-1~e, the total amount of water stored in the  solid mantle can thus be doubled with extreme initial \ce{CO2} content, assuming a  clear-sky albedo. The impact of \ce{CO2} on the remaining water is so strong, because more \ce{CO2} in the system leads to higher water melt fractions for the whole duration of the magma ocean evolution. Consequently, more \ce{H2O} can be sequestered in the solid part of the mantle throughout the evolution. 

In any case, our work shows that it is important to capture the feedback of \ce{CO2} on the atmospheric water loss and the magma solidification timescale. Large amounts of \ce{CO2} ($>1000$~bar) can prevent or at least delay complete desiccation for the TRAPPIST-1 e,f and g planets. We further find that among all investigated planets, TRAPPIST-1~e, orbiting closest to its host star, is more strongly impacted by various processes that shape the magma ocean stage: These are the assumed albedo, which represents here cases with low scattering (albedo=0) versus high scattering of incoming stellar light (albedo =0.75), the reduction in the runaway greenhouse radiation limit with high \ce{CO2} abundances,  as well as the reduction of atmospheric water loss because atmospheric escape of \ce{H2O} is diffusion limited with high amounts of \ce{CO2} in the atmosphere.

\subsection{Atmosphere extent and interior modelling constraints}\label{sec atm int model}

To be able to understand the impact of an extended atmosphere on the observed radii of the TRAPPIST-1 planets, we first make use of an interior structure model \citep{Noack2017} employing look-up tables created with Perple\_x \citep{Connolly2009} for thermodynamic properties of the silicate mantles. For a first-order estimate on the potential chemical composition of the TRAPPIST-1 planets, we use an adapted version of the condensation model presented in \citet{Bitsch2020}, and employed the measured metallicity of TRAPPIST-1 to derive the likely chemical composition of the star and planets (depending on their distance of the star), see Section \ref{sec chem comp} for more information on the numerical models. For the silicate mantle of the TRAPPIST-1 planets, and under the assumption of an Earth-like mantle iron number of 0.1 (i.e. a magnesium number of 0.9), our model predicts the following molar composition: 5.35\% FeO, 48.15\% MgO, 39.09\% SiO$_2$, 2.85\% CaO, 1.83\% Al$_2$O$_3$, and 1.0\% Na$_2$O. Residual iron as well as any condensed FeS contributes to the metal core, with a predicted mass fraction of 27\%. However, in our interior structure model, we vary the core-mass fraction further to obtain realistic interior structure profiles for the TRAPPIST-1 planets that can match their observed radii. For simplicity, in our model we consider only pure iron, see Section \ref{sec chem comp}. For the innermost planets b, c and d, a core-mass fraction of 27\% leads to a model radius of 1.114  Earth radii (compared to an observed radius of 1.116$^{+0.012}_{-0.014}$, \citealp{Agol2021}), 
1.098~Earth radii (compared to 1.097$^{+0.012}_{-0.014}$) and 0.774~Earth radii (compared to 0.788$^{+0.01}_{-0.011}$), and therefore our model radii match the measured radii within the observational error. This suggests that our compositional model is able to correctly predict the planetary composition of the TRAPPIST-1 planets. For the outer planets of the system, the appearance of volatiles adds a degeneracy to our interior structure.

\begin{figure}
    \centering
    \includegraphics [width=0.5\textwidth]{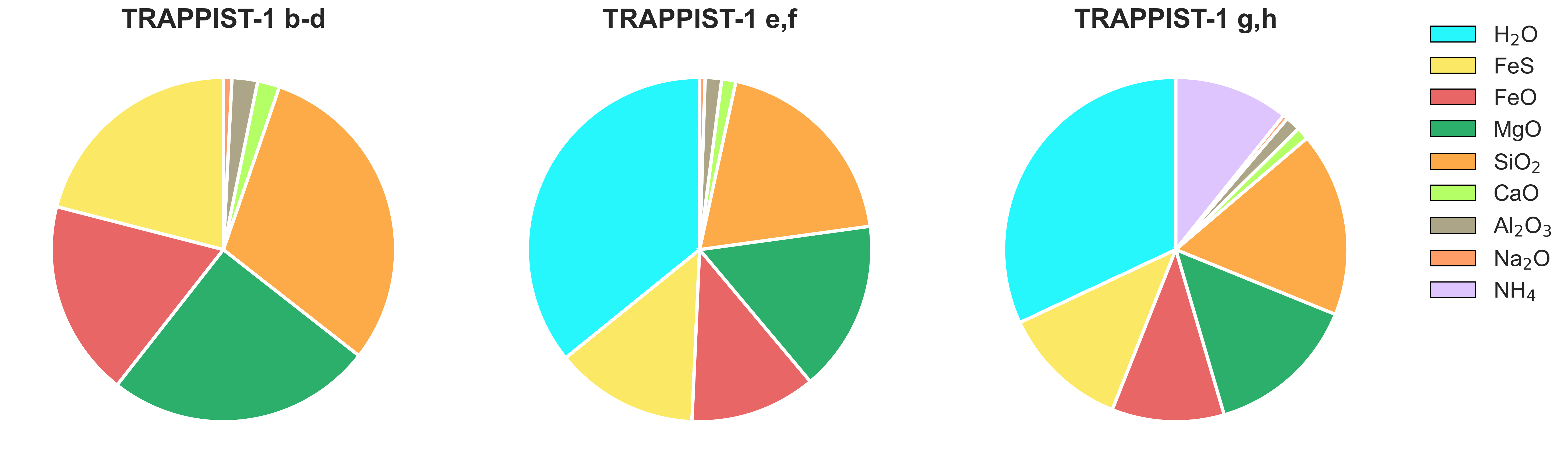}
   \caption{Predicted composition of the building blocks of the different TRAPPIST-1 planets based on the stellar metallicity and the assumption that the planets did not strongly migrate during accretion. Due to secondary processes including collisions and stripping of material, as well as melting and evaporation processes, especially for the outer planets, the final planetary composition is expected to be considerably less volatile-rich than predicted here for the planetary building blocks. 
   }
    \label{fig: composition}
\end{figure}

For the building blocks of the outer planets TRAPPIST-1e, f, and g (and h with almost exactly the same composition as g), our model predicts that the fraction of silicates and metals in the planetary building blocks would decrease to 64 wt-\% (e,f) and 57 wt-\% (g,h), see Fig. \ref{fig: composition}. For e and f, the remainder of the mass is water in the form of ice layers (including high-pressure ice) as well as liquid water. For g and h, volatiles are separated into 32 wt-\% H$_2$O as well as 11 wt-\% NH$_4$. However, during planet accretion, due to high-energy impacts and melting processes in the interior, we expect the final planetary compositions to be drier. In addition, planetary orbits may change during the planet formation process, leading to an additional uncertainty in accreted planetary building material and hence fraction of volatile materials.

\begin{table}[]
\caption{Combinations of core mass fractions and water mass fractions for TRAPPIST-1e, f and g that match the observed radii from \citet{Agol2021}.}
    \centering
    \begin{tabular}{c||c|c|c}
         & \multicolumn{3}{c}{H$_2$O [wt-\%]}  \\
         Fe [wt-\%]  & e & f & g  \\
         \hline
         15 & - & - & 0 \\
         20 & - &	0 &     1.1 \\
         25 & 0 &	1.5 &	3.1 \\
         27 & 0.6${}^a$ &  2.3 &   3.8 \\
         30 & 1.7 &	3.4 &	5.0 \\
         35 & 3.5 &	5.5 &	7.9 \\
         40 & 5.5 &	8.5 &	10.9 \\
         45 & 8.3 &	11.4 &	14.1 \\
         50 & 11.3 &	14.3 &	17.0
    \end{tabular}
    \newline a: Desiccated (0~wt\%) and Earth-like water composition of 0.03-0.27~wt\% are also possible within error bars. 
    \label{tab:planet comp}
\end{table}

Table~\ref{tab:planet comp} lists the range of iron fractions and water/ice fractions that would match the observed masses and radii of TRAPPIST-1e, f and g while taking into account the star-derived mantle chemical setup. Figure~\ref{fig:H2O_content_1e,f,g} shows the possible \ce{H2O} ranges for each planet within error bars. While for all three planets, the measured radii could be explained by a dry composition (with 25.2, 20.4 and 15.9 wt-\%, respectively), increasing iron fractions are possible when adding a water/ice layer of increasing extend with increasing distance to the star. It should be noted that here we only considered water ice and did not take into account any contribution of NH$_4$, which may be abundant in addition to H$_2$O in TRAPPIST-1g following our compositional model. Given the apparent lower iron fraction in the system compared to the solar system (as suggested also in \citealp{Unterborn2018}), we only investigate core-mass fractions of up to 50 wt-\%. Within this range, the maximum water fraction is 17.1 wt-\% and therefore below 20 wt-\%, in accordance with \citet{Dorn2018}, \citet{Unterborn2018}, \citet{Agol2021}, \citet{Barth2021} and \citet{Acuna2021}, even though here we apply our TRAPPIST-1 adapted compositional model instead of an Earth-like mineralogy. 

\begin{figure}
    \centering
    \includegraphics [width=0.5\textwidth]{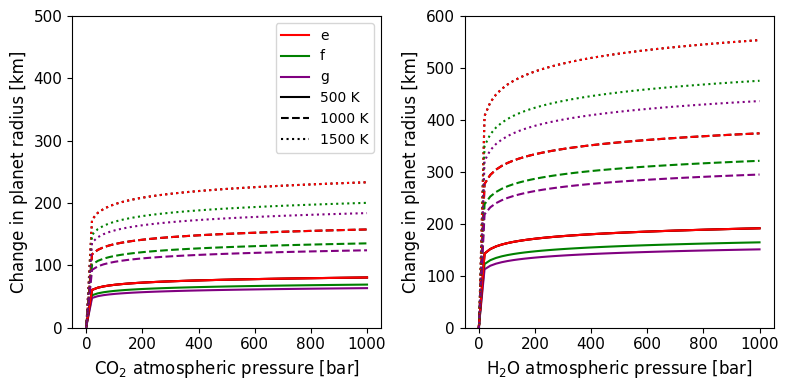}
    \includegraphics [width=0.5\textwidth]{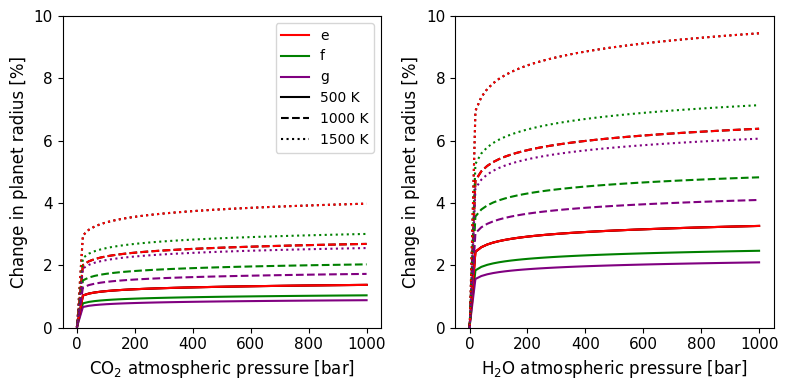}
    \includegraphics [width=0.5\textwidth]{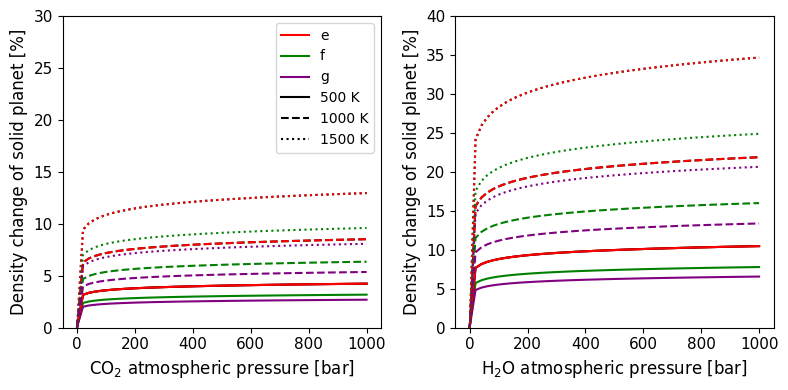}
   \caption{
   Predicted increase in planet radius (in km and \%) as well as resulting increase in density in \% calculated for the sub-atmospheric planet layers (i.e. core, mantle and water/ice layers) when considering CO$_2$ or H$_2$O atmospheres of variable average atmospheric temperature and atmospheric pressures up to 1000 bar.}
    \label{fig: atm Rp}
\end{figure}

In Fig. \ref{fig: atm Rp} we display how the planetary radius and density of the planet underneath the atmosphere would change when considering extended atmospheres of different compositions and pressures, calculated as in \citet{Ortenzi2020}. We first calculate the predicted atmospheric thickness for the measured mass and radii of each planet and calculate the related radius change. To give examples on how the density of the planets would be affected, we subtract the predicted radius change from the planet radius and calculated the increased density of the sub-atmospheric planet. We note that for today's measured planet masses and radii, extended steam atmospheres as depicted here are not realistic and the density variations should therefore only be seen as exemplary values.

The atmospheric pressure has a logarithmic effect on the thickness of the atmosphere, leading to increasing atmospheric pressures showing only a weak effect for pressures above a few tens of bars. Water steam atmospheres have approximately twice as strong an influence on the atmosphere thickness as the heavier CO$_2$ atmospheres. In addition, the atmospheric temperature strongly impacts the atmospheric thickness, with an almost linear effect (a doubled absolute temperature leads to an approximately doubled atmosphere thickness).

\section{Results Summary}
\label{sec: results}

In this study, we introduce \magmoc{2.0}, a magma ocean model with versatile multi-volatile outgassing and thermal cooling informed by radiative transfer calculations in a vertically extended atmosphere. Comparison of different atmosphere models for the Earth with mixed \ce{H2O}-\ce{CO2} content shows that using radiative transfer calculation in a vertically extended atmosphere generally yields longer solidification times compared to a gray atmosphere model. The solidification times still remain within one order of magnitude when atmospheric escape is neglected (Sects.~\ref{sec: Earth - pure H2O} and~\ref{sec: Earth -Mixed CO2 scenarios}). 

For a pure \ce{CO2} atmosphere, however, the magma ocean reaches the solidification surface temperature of 1400~K significantly faster (by more than one order of magnitude) compared to a gray atmosphere model (Sect.~\ref{sec: Earth - pure CO2}). We find that a vertically extended \ce{CO2} dominated atmosphere with surface temperatures larger than 2000~K emits more thermal flux on top of the atmosphere than is captured with the original gray model. Consequently, adding \ce{CO2} to a water steam atmosphere leads to faster cooling of the magma ocean during the initial magma ocean stage. 

As soon as the runaway greenhouse limit is reached ($T_{surf}\leq 1800 K$), however, a second thermal evolution stage begins, lasting much longer than the initial stage. Here, the addition of \ce{CO2} leads to slower cooling because its presence cools the upper atmosphere and this reduces overall thermal emission compared to a pure \ce{H2O} atmosphere.

 Application to the potentially habitable TRAPPIST-1 planets e, f and g shows that the albedo is of major importance in determining magma ocean lifetimes. If an albedo of $0.75$ is assumed, as for the Earth-like simulations (Sect.~\ref{sec:Earth}), than the magma ocean lifetime is always shorter than 100~Myrs for TRAPPIST-1 e,f, and g. An albedo of zero, which is probably more appropriate for hot \ce{H2O}-\ce{CO2} dominated atmospheres around M dwarfs \citep{2013Kopparapu} extends the magma ocean lifetime for all investigated TRAPPIST-1 planets such that they can even enter their respective habitable zone with intermediate to water-rich compositions ($T > 10$~TO initial water mass). The presence of \ce{CO2}, on the other hand, results for water-rich scenarios overall only in a small (1\% -10\%) net extension of the magma ocean lifetime. For dry compositions ($\leq 10$~TO initial \ce{H2O}) and large amounts of \ce{CO2}, \ce{CO2} diffusion limited escape can also significantly extend the magma ocean lifetime for TRAPPIST-1e ( albedo 0.75) and TRAPPIST-1 e, f, and g (albedo 0).

The presence of \ce{CO2}, however, modifies the volatile distribution in all Earth and TRAPPIST-1 e, f and g simulations. The magma ocean evolves from an initially \ce{CO2}-dominated atmosphere into a \ce{H2O}-dominated atmosphere and for water poor scenarios ( $\leq 10$~TO \ce{H2O}) the end state is again a \ce{CO2}-dominated atmosphere. This shift in atmosphere composition is accompanied by a feedback in outgassing as outlined by \citet{Bower2019}. Generally, the addition of \ce{CO2} increases the water melt fraction. Larger melt fractions lead further to larger partial pressures of \ce{H2O} and may also shift the on-set of water outgassing. The impact of \ce{O2} is less clear, because in our simulations \ce{O2} only accumulates as the magma ocean solidifies. Here, the magma ocean lifetime that can be extended with low albedos and/or diffusion limited escape is important. We thus find that an extension of the magma ocean lifetime with a clear-sky albedo reduces the amount of abiotically created \ce{O2} strongly, as already outlined by \citet{Barth2021}.

The increase of the water melt fraction due to the \ce{H2O}-\ce{CO2}  feedback is particularly evident for TRAPPIST-1e  (Fig.~\ref{fig: Overview TR1efg}, right panels). Larger water melt fractions also increase the amount of water sequestered in the solid mantle. Here, the amount of remaining water can be increased by 100\% and more for the clear-sky albedo with equal amounts of \ce{H2O} and \ce{CO2} compared to a pure \ce{H2O} scenario.  However, the fraction of initial water retained in the solid mantle still remains low: only between 1\% and 6\% of initial water mass can be retained in the mantle.


New interior composition constraints for all TRAPPIST-1 planets reveals that at least the inner planets (b,c, d) are apparently dry and exhibit an iron fraction of 25 wt-\%. If the sane iron fraction is assumed, than TRAPPIST-1e has currently a low water content, consistent with a dry evolution scenario ($\leq 10$ initial \ce{H2O}.  We find that for such a dry composition, the TRAPPIST-1e magma ocean lifetime can be significantly extended due to \ce{CO2} diffusion limited escape. The planetary atmosphere may not completely desiccate even with initial water masses as low as 5~TO \ce{H2O} within the first 280~Myrs of its evolution, that is until it enters its habitable zone. TRAPPIST-1 e is thus the ideal planet to test formation theories and the impact of atmospheric erosion on the water budget during the early evolution of rocky planets orbiting M dwarfs. 

In this work, we further developed an analytical corrected gray atmosphere model. This model incorporates information from a thermal emission grid obtained via full radiative transfer calculations. Compared to the traditional gray atmosphere model, this corrected gray atmosphere represents a significant improvement. It yields sufficiently similar  results for all investigated scenarios in this work compared to the RT atmosphere model, including the special scenario for TRAPPIST-1 e. Both methods are computationally efficient, however, the corrected gray model is 10 to 26 times faster compared to the RT atmosphere model (Sect.~\ref{sec: Stability}). 

\section{Discussion}
\label{sec: Discuss}

The magma ocean model \magmoc{2.0} that is introduced in this work is designed to facilitate efficient testing of various scenarios and assumptions. This capability is particularly valuable in the era of detailed characterization of rocky planets beyond our Solar System. \magmoc{2.0} provides a fast approach to study the evolution of rocky planets in a large parameter space. On the other hand, \magmoc{2.0} can benefit from more complex magma ocean and atmosphere models, leading to continuous improvement of this model for better overall understanding of planetary processes. 

In this work, specifically, \magmoc{2.0} contributes to a better understanding of the potential final atmospheric state and its impact on inferred interior structure for the potentially habitable TRAPPIST-1 e, f, and g planets. This research is crucial for unraveling the complex interplay between planet formation, planetary structure, atmospheric composition and ultimately their potential habitability. TRAPPIST-1~e may be very volatile-poor compared to TRAPPIST-1 f and g if we assume an iron mass fraction of 27 wt\% as for the inner planets.  TRAPPIST-1 f and g, on the other hand, must have formed volatile-rich to explain large water fractions that we estimate for the same iron mass fraction, resulting potentially in thick \ce{CO2} atmospheres that may prevent the existence of surface liquid water.

If TRAPPIST-1 e is indeed (partly) desiccated, then it may have formed close to the water-ice line, during which the planet migrated rapidly inwards, while f and g formed exterior to the ice line and migrated later. Alternatively, the planets did not undergo significant migration, instead the water ice line shifted during the disk evolution outward starting from a location near TRAPPIST-1 e, leading to the same outcome \citep{Bitsch2019}.

\subsection{Mutual benefit between versatile \magmoc{2.0} and other magma ocean models}

The investigation of mixed outgassing on magma oceans in the habitable zone of TRAPPIST-1 underscores the potential synergies between magma oceans models with a more comprehensive treatment of geophsical processes, more complex models that capture the geochemistry more completely and simpler frameworks such as \vplanet{}/\magmoc{2.0} that can explore a broad range of initial volatile content and the impact of various treatments of atmospheric and outgassing processes. For instance, the incorporation of feedback between \ce{H2O} and \ce{CO2} outgassing was motivated by the magma ocean model results of \citet{Bower2019}. Despite employing different differential equation and a simpler geophysical model, we generally reproduce the findings of \citet{Bower2019} for Earth (Sect.~\ref{sec: feedback}). 

Comparison between results from the Earth magma ocean model of \citet{Elkins-Tanton2008} and \magmoc{2.0} further yields agreement in volatile distribution when the same outgassing laws are used that suppress \ce{CO2} outgassing (Table~\ref{Tab_Results_Earth}). However, when modern outgassing laws are applied, there is less agreement in volatile distribution, in particular for \ce{CO2}. Suppressed \ce{CO2} outgassing leads to 10--100 times higher \ce{CO2} melt fractions at the end of the magma ocean stage compared to a magma ocean evolution, where the majority of \ce{CO2} is outgassed already at the beginning (Table~\ref{Tab_Results_Earth_Niko}). Thus, \magmoc{2.0} can shed light on how initial outgassing conditions can alter the volatile distribution in diverse magma ocean models.

Application of \magmoc{2.0} to TRAPPIST-1 e, f and g reveals that additional \ce{CO2} also significantly influences these planets' volatile distributions, resulting in increased abiotic \ce{O2} build-up and a higher remaining water fraction in the solid mantle at the end of the magma ocean stage. Furthermore, we find that the inclusion of \ce{CO2} does not consistently lead to delayed \ce{H2O} outgassing, as inferred by \citet{Bower2019} for an Earth-like scenario. Instead, the onset of \ce{H2O} outgassing may be delayed, start even earlier, or remain unaffected compared to simulations without \ce{CO2}. In our model, it is the radiative properties of the mixed \ce{H2O}-\ce{CO2} atmosphere at the time of mantle solidification that primarily determine when \ce{H2O} outgassing occurs. This insight is facilitated by exploring more volatile scenarios with different atmosphere models (Sections~\ref{sec: Earth -Mixed CO2 scenarios-Outgas} and\ref{sec: Earth -Mixed CO2 scenarios- CO2}).

We also demonstrate that \magmoc{2.0} can identify unexpected feedbacks between atmospheric erosion and mantle solidification due to diffusion limited escape in  mixed \ce{H2O}-\ce{CO2} outgassing. For TRAPPIST-1 e with 10~TO \ce{H2O} initially, the reduction of the runaway greenhouse radiation limit by 10s of W/m${}^2$ as water erosion from the atmosphere substantially prolongs the magma ocean stage and also diminishes water sequestration in the mantle.

We note that this special case was not identified in a similar magma ocean study with mixed \ce{H2O}-\ce{CO2} outgassing for the TRAPPIST-1 planets conducted by \citet{Krissansen2022}. Their model did not consider the reduction of the runaway greenhouse limit with \ce{CO2}, which is a prerequisite for this scenario to occur. Additionally, it does not account for changes in the volatile distribution due to the evolution in atmospheric composition. Apart from these differences, the model of \citet{Krissansen2022} yields similar magma ocean solidification lifetimes for TRAPPIST-1 e, f and g compared to our study.

\subsection{The concept of solidification time}
\label{sec: Solid time}

When comparing different magma oceans simulations, it is important to discuss the concept of `solidification time' with care. In \magmoc{}, the magma ocean effectively ends with the convectively dominated viscosity regime. This point is reached when the critical the surface melt fraction $ \psi= 0.4$, following \citet{Lebrun2013} with surface temperatures of 1650~K. At this point 99\% of the planetary radius is solidified. In principle, the remaining melt is no longer well mixed and in full equilibrium with the atmosphere. For the last \% of radius solidification, thus, \magmoc{} switches to a high viscosity treatment of the mantle with no further outgassing but with further surface cooling and ongoing atmosphere erosion. Complete solidification then occurs when the solidification radius is equal to the planetary radius, which is in our model typically for surface temperatures of 1400~K (see Sect.~\ref{sec: method}). Thus, solidification temperatures in our model effectively comprises between 1650 - 1400~K.

Since our magma ocean effectively ends at $ \psi= 0.4$ and complete solidification occurs shortly thereafter, \magmoc{2.0} stops earlier and at higher surface temperatures of 1400~K compared to the model of \citet[e.g.][]{Lichtenberg2021}. In this model solidification is defined when the rheological front reaches the surface with $ \psi= 0.22$. Since these authors present the full thermal evolution for pure \ce{H2O} and \ce{CO2} atmospheres, respectively, we can still compare \magmoc{2.0} to their model state at $ \psi= 0.4$ which is  reached in their models with surface temperatures between 1700 K and 1500~K, depending on atmosphere composition, with a similar strong temperature decrease in this regime as also observed in our models  \citep[][Fig. 5]{Lichtenberg2021}. Thus, we identify general agreement for the thermal evolution of our Earth magma ocean model and that of \citet{Lichtenberg2021}  (Sections~\ref{sec: Earth - pure CO2}, \ref{sec: Earth - pure CO2}). 

Our effective solidification temperature of 1400~K is similar to that of the model of \citet{Krissansen2022} and only 100 to 150~K cooler compared to the solidification temperatures in the models of \citet{Niko2019} and \citet{Hamano2013}, respectively. In the magma ocean model of \citet{Elkins-Tanton2008} solidification is assumed when 98\%  of the planet's radius is solid, which aligns well with our choice of  $\psi= 0.4$ with 99\% mantle solidification, allowing comparisons between \magmoc{2.0} and their Earth simulations (Section ~\ref{sec:Earth}).

\subsection{The impact of clouds}
\label{sec: Discuss_alb}

Efficient cooling of a magma ocean requires a large net flux $F_{Net}$ leaving the top of the atmosphere, which is the difference between outgoing long wave radiation (OLR) and absorbed stellar atmosphere (ASR). Our simulation results show that mainly two effects  determine the magma ocean lifetime: The presence of water vapor and scattering on top of the atmosphere via Rayleigh scattering and clouds, encapsulated in our work via the bond albedo $\alpha$. The presence of water 'sets' the outgoing longwave radiation (OLR), the albedo $\alpha$ sets the absorbed stellar radiation.

When comparing simulations with $\alpha=0.75$ (Venus-like scattering) to $\alpha=0$ (cloud-free scattering), we find that the cloud-free assumption yields a much larger magma ocean lifetime for all planets even without adding \ce{CO2}. The addition of \ce{CO2} is in this respect of minor importance, it may prolong, however, the timescale of water loss as outlined in the next section. We further note that \citet{Marcq2017} propose in their model that the inclusion of clouds lowers the runaway greenhouse radiation limit to 197~W/m$^{2}$, whereas for the majority of our simulations the limit of 280~W/m${}²$ holds unless the atmosphere is highly water-depleted. \citet{2013Kopparapu}, on the other hand, argue that water clouds can be neglected for the calculation of the OLR.

In the clear-sky simulations, the magma ocean lifetime is so prolonged that all planets may enter the habitable zone\footnote{According to \citet{2013Kopparapu}.} before complete mantle solidification for water-rich composition ($>10$~TO \ce{H2O}). This result is of outermost importance because many geophysical models assume by default that water can condense out if the incoming stellar flux is low enough. While this is a safe assumption for Earth, this is apparently no longer true for planets that orbit an M dwarf star.

Instead, volatile-rich planets can be 'nominally' in the habitable zone, but condensation on the surface is not yet possible. In addition, atmospheric water loss continues for several hundred years after the habitable zone entrance for TRAPPIST-1e. In principle, this result is similar to that of \citet{Turbet2021} who showed that even in the Solar system, Venus probably never had a chance to form a surface ocean because thick water clouds prevented efficient cooling of the surface. On the other hand, \citet{Yang2014} showed that for tidally locked planets around M dwarf stars, the high albedo from water clouds forming predominantly over the dayside of an Earth-like planet, may allow to shift the inner edge of the habitable zone closer to its host star.

In this work, we simplified the problem by adopting two extreme values to outline the scope of the impact. In our model-setup, however, the cloud-free assumption is the most physically consistent, because we assume efficient re-evaporation after condensation (that is, with a supersaturation ratio S=1). Further, \citet{2013Kopparapu} point out that water clouds in a water dominated atmosphere have only a minor impact because of the high intrinsic optical thickness of water vapor in the infrared. Conversely, \citet{Marcq2017} find that the runaway greenhouse radiation limit is lowered to 200~W/m$^{2}$, when water clouds are included. \ce{CO2} clouds, on the other hand, may allow for strong scattering. This effect may be relevant in our framework for desiccated \ce{CO2} atmospheres. However, even for relatively low water abundances of $x_{H2O}=10^{-3}$, the upper atmosphere is warmed by the latent heat release of water condensation such that \ce{CO2} clouds are not expected to form (see Fig.~\ref{fig: H2O_PT}). In any case, clearly more work is warranted to assess the impact of clouds in particular to investigate if and when the TRAPPIST-1 e, f and g planets can form surface oceans after the magma ocean stage. This will require further work on the details of precipitation and evaporation of water on rocky planets around M dwarfs. Already for 'Earth-like' conditions with temperate temperatures, different cloud models can yield very different results \citep{Sergeev2022}. Further, \citet{Sergeev2024} point out that resolving moist convection on the dayside of potentially habitable planets requires a higher resolution than is adopted so far. In any case, clearly more work is needed to explore the role of clouds on the outgoing long wave irradiaton and albedo for the magma ocean stage on rocky planets around red dwarfs.

\subsection{Volatile evolution}
\label{sec: Discuss_Vol}

A magma ocean evolution with a mixed \ce{H2O}-\ce{CO2} atmosphere can exhibit feedback between \ce{H2O} and \ce{CO2} outgassing that shapes the overall volatile distribution, that is, the distribution of \ce{H2O}, \ce{CO2} and \ce{O2}. In addition, in a \ce{CO2}-dominated atmosphere that is prone to emerge at the beginning and the end of the magma ocean stage, \ce{CO2}-diffusion limited escape can limit water loss, thus preventing complete desiccation on the investigated planets for a longer time. In the following subsections, we will discuss the feedback impact for each volatile species in more detail.

\subsubsection{H$_2$O}

The fate of water during the magma ocean evolution on rocky planets around M dwarf stars is of utmost importance for constraining their habitability. If there is too little water ($<10$~TO \ce{H2O}), the planets completely desiccate during the magma ocean stage with only a few percent initial water mass remaining in the mantle. 

We find  that the mantle is driest for our TRAPPIST-1 e simulations compared to simulations of TRAPPIST-1 f and g because planet e is more strongly affected by UV-photolysis of \ce{H2O}. Additionally, the mantle tends to become even drier when the magma ocean stage is prolonged, which is the case for the clear-sky (albedo=0) simulations. The desiccation of the mantle  of TRAPPIST-1e can be, however, compensated by large amounts of \ce{CO2} as is evident by comparing the remaining mantle water inventory for the albedo $0$ and $0.75$ evolution scenario  with assumed initial \ce{CO2} mass being equal to the initial \ce{H2O} content (Fig.~\ref{fig: Overview TR1efg}). Despite much longer desiccation times, the remaining water content remains of similar order. 

A desiccated evolution scenario is further of particular interest for TRAPPIST-1 e because its composition is apparently consistent with a dry Earth-like \ce{H2O} content ($<<1$~wt-\%) (see Sect.~\ref{sec atm int model}). Thus, here our evolution scenarios with 1-10~TO initial water mass, as estimated for Earth (see Sections~\ref{sec:Earth}) are particularly relevant. They show that the atmosphere of TRAPPIST-1 e may retain 5~TO water until the star's luminosity dropped such that it enters the habitable zone (Fig.~\ref{fig: TR1e 10 TO}). Similarly, even if the atmosphere is desiccated, up to 4\% of the initial water mas could be stored in the mantle, allowing in principle to build-up a secondary surface water reservoir\citep{Godolt2019}. Since such a relatively wet mantle would require large amounts of \ce{CO2}, the surface may be Venus-like and thus too warm for surface liquid water. On the other hand, there may be several other ways to retain more water in the melt like delayed \ce{H2O} outgassing \citep{Ikoma2018,Bottinga1990,Lensky2006} and a basal magma ocean that doesn't require huge amounts of \ce{CO2} (see also Sect.~\ref{sec: Discuss_Vol}). Conversely, the outer planets TRAPPIST-1 f and g may have the opposite problem of having too much water for a habitable surface ocean. A very thick water ocean not only dilutes nutrients, the further enrichment from the crust may also be suppressed due to the high pressures at the bottom of the ocean \citep{Noack2016}.

The result that an extension of the magma ocean stage generally leads to a drier planet is in marked contrasts with the findings from the 0D magma ocean model by \citet{Moore2023}. Fig.~\ref{fig: Volatile_Flowchart} illustrates the reason for this discrepancy: The amount of dissolved water in the magma ocean, our primary variable of integration variable $F_{\ce{H2O}}$, is in balance with the outgassed atmosphere. As previously pointed out by \citet{Barth2021}, as long as such a balance exists, atmospheric escape will drive additional outgassing from the magma ocean to establish a new equilibrium. It is thus mainly the reduction of water loss in the presence of a thick \ce{CO2} atmosphere that can prevent atmosphere desiccation, in particular on TRAPPIST-1e. It is then the retention of significant amount of water vapor in the atmosphere ($>100$~bar) that extends the magma ocean lifetime, and not the other way around.


We note that there are alternative mechanisms to 'save' water from atmospheric escape during the magma ocean stage apart from \ce{CO2} diffusion limited escape. In \magmoc{2.0}, water is sequestered by partitioning 1\% of available water mass in the solid mantle during solidification ($k_{\ce{H2O}}$), consistent with values for relevant minerals such as lherzolite and  peridotite \citep{Elkins-Tanton2008,Johnson1992}. This water is saved from atmospheric escape because it is no longer part of the fully coupled volatile system. Nevertheless, the assumption of a constant $k_{\ce{H2O}}$ at the bottom of a highly convective magma ocean may be overly simplistic \citep{Ikoma2018}. Additionally, whether crystals settle at the bottom of the magma ocean (fractional solidification), as is assumed in this model, or remain suspended in the melt (batch solidification) may further impact how much water is incorporated into the solid mantle. Moreover, the role of minerals capable of incorporating higher water mass fractions like phyllosilicates has not been accounted for yet \citep{Herbort2020}. Lastly, the efficiency of outgassing in a highly convective magma ocean may be reduced compared to 'static' laboratory experiments \citep{Salvador2023}. 

In this study, we identify an \ce{H2O}-\ce{CO2} feedback as one mechanism that increases the amount of water in the mantle. Adding \ce{CO2}-outgassing leads to changes in the mean molecular weight of the atmosphere during the magma evolution  \citep{Niko2019,Bower2019,Lichtenberg2021}: Initially, the magma ocean begins its evolution with a \ce{CO2}-dominated rather than a \ce{H2O}-dominated one because \ce{CO2} is less soluble in the melt compared to \ce{H2O}. One major consequence is that the water melt fraction $F_{\ce{H2O}}$ increases in a magma ocean with a \ce{CO2}-dominated atmosphere, leading to increased \ce{H2O} outgassing with higher \ce{H2O} partial pressures of \ce{H2O} at later times. Simultaneously, because there is more water in the melt, a greater amount of water can be partitioned and sequestered into the solidifying mantle. Generally it holds that the more \ce{CO2} is in the magma ocean-atmosphere system initially, the more \ce{H2O} remains in the mantle. Consequently, the addition of \ce{CO2} results in 15\% to 50\% more remaining water in the mantle at the end of the magma ocean stage for all investigated cases. 

Other possibilities to retain more water in the melt include the delay or suppression of \ce{H2O} outgassing \citep{Ikoma2018} through mechanisms such a bubble nucleation \citep{Bottinga1990,Lensky2006}. Further, \citet{Bower2022,Maurice2017} that use more complex models pointed out that the mode of crystallization can significantly increase water retention. In addition, \citet{Maurice2017} suggest that Rayleigh-Taylor instabilities may replenish the mantle with volatiles from the deeper interior. Further, \citet{Hier2017} suggest that volatile-rich may be trapped at the solidification front. Another mechanism to retain water in the melt may also be the early stratification of the magma ocean \citep{Orourke2020,Samuel2023} such that part of the magma ocean solidifies downward, resulting in a basal magma ocean. If stratification occurs before the majority of \ce{H2O} is outgassed, then a basal magma ocean could, in principle, retain large amounts of water over the course of billion of years \citep{Moore2023}. Therefore, it is very likely that we underestimate the amount of water that can be retained in the planet. Still, we emphasize that increasing the amount of water with the addition of \ce{CO2} will in all these cases also increase the amount of volatiles that can be stored in the planet. It is thus highly important to diagnose in the near-future the \ce{CO2} atmosphere budget on the potentially habitable TRAPPIST-1 planets in the future.

Here, our magma ocean simulations can help as they also incorporate atmospheric escape. Most importantly we find that the inclusion of \ce{CO2} can significantly reduce atmospheric escape compared to a pure \ce{H2O} atmosphere, as already pointed out by \citet{Moore2023}. On the other hand, Joule heating, which is not considered in this work, may instead lead to increased atmospheric erosion compared to our model, at least for TRAPPIST-1e \citep{Cohen2024}.

\subsubsection{CO$_2$}
 
The addition of \ce{CO2} has a profound impact on the overall volatile budget, as it is much less soluble in magma and less prone to be partitioned into the solid mantle than \ce{H2O} (Table~\ref{Tab_geo}). Further, \ce{CO2} can create a 'diffusion barrier' for atmospheric water loss as outlined above for water-poor compositions ($<10$~TO \ce{H2O}) and in particular for TRAPPIST-1e at the inner edge of the habitable zone of its host star. Still, even in this case, generally large amounts of \ce{CO2} are required for a significant impact of the magma ocean, that is, \ce{CO2} mass $\geq 0.5 \times$ initial water mass (See \href{https://doi.org/10.5281/zenodo.14442985l}{detailed grid}). Such extreme scenarios can lead to very thick ($\geq 10^3$~bar) \ce{CO2}-dominated atmospheres. In very volatile-rich cases ($>>10$~TO \ce{H2O} and \ce{CO2}), the magma ocean may not even completely solidify, retaining surface temperatures above 1400~K (Fig.~\ref{fig: GHlimit_CO2 mix}). 

In contrast, \citet{Krissansen2022} assume less extreme values of additional \ce{CO2}, with partial pressures of up to $10^3$~bars, allowing at least the crust to fully solidify.  Another difference is that the runaway greenhouse radiation limit depends in our model on the water volume mixing ratio and can be lowered by 10s of W/m$^{2}$, when \ce{CO2} is added (Sections~\ref{sec: grey} and \ref{sec: Grey_exo}). Instead,  \citet{Krissansen2022} assume that all water condenses out of the atmosphere onto the solidified crust of TRAPPIST-1 e, f and g, once the ASR drops below the constant runaway greenhouse radiation limit of 282 W/m$^{2}$. Due to the presence of liquid surface water in their model, surface weathering and the carbon cycle remove hundreds of bars of atmospheric \ce{CO2}. In our model, the presence of liquid surface water is questionable for the thick remaining \ce{CO2} atmospheres, which appears to be a reasonable outcome for the outer planets TRAPPIST-1 f and g. Further, in contrast to the model of \citet{Krissansen2022}, the magma ocean stage in our cloud-free (albedo=0) model can be extended well into the habitable zone. Thus, we caution to automatically assume rainout of water from the atmosphere, only due to low stellar luminosity.

We further note again that the role of clouds \citep{Turbet2021,Yang2014} is of outermost importance (see Sect.~\ref{sec: Discuss_alb}). If water cannot condense out on the surface after the magma ocean stage, then the remaining \ce{H2O}-\ce{CO2}-\ce{O2} atmospheres would only be modified by further atmospheric loss. Constraints of \ce{CO2}-\ce{H2O} dominated atmospheres on TRAPPIST-1 e, f and g may thus provide insights into the abundances of volatiles that were present during the magma ocean stage and on conditions to form surface liquid water, as also pointed out by \citet{Krissansen2023}.

\subsubsection{O$_2$}

In our work, atmospheric \ce{O2} begins to accumulate as soon as more than 5~TO \ce{H2O} initial water are in the system for albedo=0.75, reaching several 100 bars partial pressures for 100~TO initial \ce{H2O} as already outlined by \citet{Luger2015} and \citet{Barth2021}. \ce{O2} can only be stored in the FeO-buffer in the mantle, which remains accessible only as long as the magma ocean is not solidified. For the high albedo simulations, the magma ocean solidifies within 100~Myrs and thus tends to yield large partial pressures of \ce{O2}. Conversely, for magma ocean lifetimes larger than 100~Myrs, less than 50~bars of \ce{O2} build-up abiotically. Instead, more \ce{O2} is stored in the mantle (see \href{https://doi.org/10.5281/zenodo.14442985l}{detailed grid}). This grid also shows the tendency of more \ce{O2} being stored in the mantle for extended magma ocean lifetimes even for simulations with a high albedo of 0.75  and initial water masses between 5 - 10 TO.

\citet{Krissansen2022} find in their model that even if there is significant abiotically build-up of oxygen, up to 100~bar \ce{O2} can be removed due to reduced outgassing later in the evolution and crustal oxidation. However, the authors also find that in some scenarios 100~bar of atmospheric \ce{O2} may remain on TRAPPIST-1 e, f and g.

Both models neglect, however, possible \ce{O2} sequestration in the iron-core \citep{Wordsworth2018}. Moreover, abiotic \ce{O2} build-up occurs when the majority of \ce{H2O} is outgassed, that is, when the mantle is mostly solidified, already limiting the exchange of volatiles between the atmosphere and the mantle. Therefore, it is questionable if removal of atmospheric \ce{O2} via the even deeper iron-core can play a significant role this late in the magma ocean evolution. 

\subsection{The need for tighter link to planet formation models}

The prospects of obtaining reliable atmospheric constraints for rocky exoplanets in the near future holds great promise. Establishing a coherent link  between planet formation, rocky planet evolution (including the crucial magma ocean stage), interior models and atmospheric composition constraints has the potential to unveil the complete history of rocky exoplanets. Studies of the TRAPPIST-1 system are particularly promising because it is postulated to be only weakly modified by late accretion \citep{Raymond2022}.

The water content of the Trappist planets needs to have been delivered during the protoplanetary disc phase. Planet formation during this initial stage can be driven either by pebble accretion \citep[e.g.][]{Ormel2017,Schoonenberg2019,Bitsch2019} or planetesimal accretion \citep[e.g.][]{Coleman2019}, where both scenarios are capable of roughly reproducing the observed structure of the system. Both scenarios require planet migration as a key ingredient due to the observed resonance configuration of the system \citep{Pichierri2024}. This begs the question of whether the planets formed interior or exterior to the water and \ce{CO2} ice lines in the disc, and how their water content is influenced by their formation location and disc cooling. Consequently, planet formation theories can give a range of initial water contents for the TRAPPIST-1 planets, depending on the exact stellar composition of Trappist-1 and on how efficient atmospheric recycling of incoming water rich pebbles is \citep{Muller2024}.

In the \citet{Schoonenberg2019} scenario, the planets are formed originally at the water ice line and the migrate inwards, which sets their water content. The migration direction and speed in comparison with the growth rate then sets the final water content of planets forming exterior to the water ice line \citep[e.g.][]{Bitsch2019}. Planets that start their formation exterior to the water ice line, but then migrate inwards very quickly will have a low water content despite their water-rich formation location. Planets that finish their accretion in the outer regions of the disc and that then only migrate across the water ice line once their accretion is complete will have the maximal water content that the system allows, which should be around 35-50\% for solar composition, depending on the exact chemical model and stellar composition \citep[][]{Bond2010,Bitsch2020,Cabral2023}. The exact volatile ratio could also be affected by atmospheric recycling of incoming water ice rich pebbles, where water fractions above 15\% are hard to achieve. A high volatile mass fraction appears to be the currently favored scenario for TRAPPIST-1 g inferred from interior models \citep{Unterborn2018, Barth2021,Raymond2022}.

To assess the initial \ce{CO2} content of the TRAPPIST-1 planets, the \ce{CO2} evaporation line needs to be considered, which is located further away from the host star due to the lower evaporation temperature of \ce{CO2} compared to water. This difference inherently implies that a planet forming in the outer regions of the disc that accretes \ce{CO2} ice, will also accrete water ice. The ratio of the water to \ce{CO2} content is then set by the initial \ce{CO2} to water ratio of the system and also by the growth and migration behavior of the planet exterior to the \ce{CO2} line, similar as described above for the water content. 

In addition, the effect of pebble evaporation and condensation can dramatically change the water and \ce{CO2} content of growing planets. As pebbles drift inwards and cross evaporation fronts, they release their volatile content into the gas phase \citep[e.g.][]{Piso2015,Aguichine2020,Schneider2021}, enriching the disc with vapor. This vapor can then diffuse outwards and re-condense, resulting in very large fractions of the corresponding molecule in the solids around the evaporation front \citep{Ros2013}. This effect can potentially explain the low water to CO ratio of the Comet C/2016 R2 \citep{Mousis2021}. This effect occurs for all volatiles, indicating that planets that form close to the \ce{CO2} evaporation front could have very large \ce{CO2} to water ratios. 

Here, we build upon previous work \citep{Bitsch2020} to constrain the composition of the building blocks of the TRAPPIST-1 planets thus to derive constraints on the iron fractions 25 wt\% and thus on the volatile composition of the inner planets. Based on our results, TRAPPIST-1 b,c,d already formed volatile-poor and are thus not expected to currently exhibit any atmosphere.

In contrast to that TRAPPIST-1e, f and g started with volatile-rich building blocks, of which they may have lost a significant fraction already before the magma ocean phase. However, of these planets TRAPPIST-1e is particularly prone to complete desiccation during the magma ocean evolution - in particular, when the already long magma ocean stage of 50~Myrs is extended even further. Interior models confirm that currently measured radii and masses are compatible with a dry interior and (moderately) thick \ce{CO2} atmosphere of 100~bar. 
TRAPPIST-1g, on the other hand, may exhibit up to 17 wt\% of water, which indicates an evolution that allows to retain significant fractions of volatiles until today.

This work thus shows that a strong link between \ce{CO2} and \ce{H2O} content during formation that is modified during the rocky planet evolution justifies the need for a coherent chain between long term planet evolution models, current atmosphere characterization efforts of rocky exoplanets, and planet formation theories. 

The TRAPPIST-1 planetary system may be key here to compare the outcome of three different scenarios in volatile accretion during formation and their subsequent loss or retainment via JWST observations and future missions like LIFE, ARIEL. PLATO may be highly useful to constrain extended steam atmospheres (up to 10\%) in young rocky exoplanets that are at the end of their magma ocean stage \citep{Turbet2019,Schlecker2024}.

\section{Conclusion}
\label{sec: Conclude}
\magmoc{V2.0} can provide novel insights into  feedback between \ce{H2O} and \ce{CO2} for outgassing in magma oceans on Earth-mass rocky planets in the habitable zone around TRAPPIST-1. Together with modeling of the refractory and volatile abundances during formation and of the current interior composition, we derive the following key insights for the TRAPPIST-1 system:

\begin{itemize}
 \item A compositional model adjusted by the measured metallicity of TRAPPIST-1 yields a dry composition with an iron fraction of 27 wt\% for the inner TRAPPIST-1 planets (b,c,d), which is compatible with the measured radii and masses from \citet{Agol2021}. This result is also compatible with recent JWST measurements that may indicate an absence of a substantial atmosphere on TRAPPIST-1b and c \citep{Zieba2023,Greene2023}.
 \item Assuming an iron fraction of 27~wt\%, interior structure modeling for TRAPPIST-1f and g, yields a substantial water fraction of 2.3~wt\%, and 3.8~wt\%, respectively. This result aligns well with a volatile-rich formation environment that is inferred for these planets in this work and also aligns with formation modeling  \citep[See also][]{Barth2021,Raymond2022,Miguel2020,Schoonenberg2019}. 
 \item Without additional sink terms, thick ($\geq 1000$~bar), mixed \ce{CO2}-\ce{H2O} atmospheres remain at the end of the magma ocean stage for volatile-rich scenarios ( $>10$ TO initial water mass) that are favored for TRAPPIST-1 f and g.
 \item More than 1000~bar of \ce{CO2} retained after the magma ocean stage may contribute up to 2\% to the total planetary radius if water vapor is removed after the magma ocean stage. An \ce{H2O}-dominated atmosphere would comprise even up to 10\% of the planetary radius, as already pointed out by \citet{Turbet2019,Schlecker2024}. While such extended steam atmospheres are incompatible with expectations for the evolved TRAPPIST-1 system, they may be identifiable for younger systems with future space missions like PLATO.
 \item In all cases, we find that adding \ce{CO2}, increases the \ce{H2O} mass fraction in the melt. Thus, between 15\% to 100\% more water can be retained in the solidified crust, which results in sequestration of between 2\% and 6\% of initial water mass in the solidified mantle for TRAPPIST-1 f, g and also e.
 \item TRAPPIST-1~e may have formed like TRAPPIST-1 f, and g in a volatile-rich formation environment.  If, however, an iron fraction of 27~wt\% is assumed as derived for the inner planets, then its current radius and mass is consistent with a more desiccated evolution scenario (water content of $0.6^{+2.1}_{-0.6}$~wt\%). The values also include Earth-like water content ($\leq 10$~TO \ce{H2O}).
 \item TRAPPIST-1~e magma ocean simulations with Earth-like water content are very sensitive to assumptions about scattering (albedo) as well as to changes in atmospheric water loss brought about via diffusion limited escape in a \ce{CO2} background atmosphere. Here, \ce{CO2} can prevent, or at least delay, complete atmospheric water loss even for a long magma ocean stage of 350~Myrs.  The atmospheric end result for TRAPPIST-1e after the magma ocean stage could thus be a either a dry \ce{CO2} or a mixed \ce{H2O}-\ce{CO2} atmosphere with several 100 to 1000~bar surface pressure. The mixed atmosphere could, in principle, further evolve into a more Earth-like state if water condenses out of the atmosphere and if up to 1000~bar of atmospheric \ce{CO2} can be sequestered in the mantle as proposed e.g. by \citet{Krissansen2022}.
  \item TRAPPIST-1~e thus emerges as a `Rosetta stone' for deciphering formation as well as evolution processes for rocky planets in the habitable zone of active M dwarfs.
\end{itemize}

\section{Outlook}
\label{sec: outlook}
In this work, we have illustrated how combined \ce{H2O}-\ce{CO2} outgassing can alter the magma ocean evolution and volatile distribution on rocky planets in the habitable zone of their M dwarf host stars, using TRAPPIST-1 e, f and g as examples. However, future investigation is necessary to test the impact of other factors that we either neglected or assumed to be constant. For example, we kept the albedo fixed at $\alpha=0.75$ and $\alpha=0$, respectively), neglecting albedo variations as the atmosphere evolves \citep{Pluriel2019}. We also assumed in the vertically extended atmosphere immediate re-evaporation after condensation, which is a strong simplification, in particular towards the end of the magma ocean stage. Additionally, exploring the evolution of the mantle's redox state \citep{Katyal2020} will be important because it may result in outgassing of reduced volatiles, such as \ce{H2} and \ce{CO} \citep{Ortenzi2020,Deng2020}. 

\ce{H2} is proposed to be dissolved into the mantle and core during the planetary accretion state, establishing a more direct connection between magma ocean and formation models \citep[][e.g.]{Johansen2023,Young2023}. A significant atmospheric \ce{H2} content during the magma ocean phase may lead to even more substantial outgassing feedback effects as tackled in this study, because it would significantly reduce the mean molecular weight of the atmosphere. Moreover, \ce{H2} is a potent greenhouse gas that can significantly prolong the magma ocean stage \citep{Lichtenberg2021} even more. At the same time, it can allow for surface liquid water on rocky planets at the outer edge of the habitable zone such as Mars \citep{Pahlevan2022}. Our versatile outgassing formalism coupled with the RT atmosphere model, allows adoption of other outgassing scenarios.

While our current setup lacks a thermal boundary layer (as in \citet{Barth2021}),  future investigations are planned that incorporate now more advanced geophysical \vplanet~modules, like \texttt{ThermInt}, for the exploration of an Earth-like planet with stagnant lid configuration \citep{Driscoll2014}. Additionally, plans include integration the 'mush stage' described by \citet{Lebrun2013} after $ \psi= 0.4$ is reached.
 
Similarly, we recognize that necessity for a modification of the atmospheric equation of state that is in this work based on the ideal gas law at the surface for particularly volatile-rich evolution scenarios, e.g. those with initial volatile mass much larger than 10 TO that is $p_{surf}>1000$~bar.

Still, our work strongly indicates that TRAPPIST-1e is a particularly interesting  planet, the atmospheric composition of which may shed light on key processes of formation and early rocky planet evolution. Further, it will be worthwhile to compare TRAPPIST-1e with TRAPPIST-1g and f  that are expected to be more volatile-rich than e. Here, the \ce{CO2} atmospheric content will be key to investigate if it represents the primordial volatile content or if it is modified after the magma ocean phase, e.g., by a carbonate-silicate cycle.

In summary, \magmoc{2.0} presents a valuable tool to gain insights into the early stages for rocky planet evolution and to understand the impact of various processes on the outgassed secondary atmosphere. These insights may facilitate rocky planet atmosphere characterization with the James Webb Space Telescope and future missions like the Habitable World Observatory and LIFE to understand the formation and evolution history of rocky planets \citep{Bonati2019,Turbet2019,Way2022,Krissansen2022,Schlecker2024}.

\section{Data availability}
https://doi.org/10.5281/zenodo.14442985

\begin{acknowledgements}
     We thank the anonymous referee for a highly constructive discussion that improved this work substantially.
     We thank Tim Lichtenberg and his group for excellent discussions about the opacities and atmospheres on rocky planets. We also thank Paul Mollière and Thomas Henning for support during the master thesis of P. Barth that was the starting point for the whole \magmoc{} development. L.C. acknowledges support by the DFG priority programme  SP1833 "Building a habitable Earth" Grant CA 1795/3 and the Royal Astronomical Society University Fellowship URF R1 211718 hosted by the University of St Andrews. R.B. acknowledges support from NASA grants numbered 80NSSC20K0229, 80NSSC18K0829 and 80NSSC18K0261. P.B. acknowledges financial support from the Austrian Academy of Science.  Ch.H. and L.C. acknowledge funding from the European Union H2020-MSCA-ITN-2019 under Grant Agreement no. 860470 (CHAMELEON). L. N. acknowledges funding from the European Union (ERC, DIVERSE, 101087755).
\end{acknowledgements}


%
%

 \bibliographystyle{aa} 
   \bibliography{Outgasing}

\begin{thebibliography}{137}
\expandafter\ifx\csname natexlab\endcsname\relax\def\natexlab#1{#1}\fi

\bibitem[{{Acu{\~n}a} {et~al.}(2023){Acu{\~n}a}, {Deleuil}, \& {Mousis}}]{Acuna2023}
{Acu{\~n}a}, L., {Deleuil}, M., \& {Mousis}, O. 2023, \aap, 677, A14

\bibitem[{Acu{\~n}a {et~al.}(2021)Acu{\~n}a, Deleuil, Mousis, Marcq, Levesque, \& Aguichine}]{Acuna2021}
Acu{\~n}a, L., Deleuil, M., Mousis, O., {et~al.} 2021, Astronomy \& Astrophysics, 647, A53

\bibitem[{Agol {et~al.}(2021)Agol, Dorn, Grimm, Turbet, Ducrot, Delrez, Gillon, Demory, Burdanov, Barkaoui, Benkhaldoun, Bolmont, Burgasser, Carey, de~Wit, Fabrycky, Foreman-Mackey, Haldemann, Hernandez, Ingalls, Jehin, Langford, Leconte, Lederer, Luger, Malhotra, Meadows, Morris, Pozuelos, Queloz, Raymond, Selsis, Sestovic, Triaud, \& Grootel}]{Agol2020}
Agol, E., Dorn, C., Grimm, S.~L., {et~al.} 2021, The Planetary Science Journal, 2, 1

\bibitem[{{Agol} {et~al.}(2021){Agol}, {Dorn}, {Grimm}, {Turbet}, {Ducrot}, {Delrez}, {Gillon}, {Demory}, {Burdanov}, {Barkaoui}, {Benkhaldoun}, {Bolmont}, {Burgasser}, {Carey}, {de Wit}, {Fabrycky}, {Foreman-Mackey}, {Haldemann}, {Hernandez}, {Ingalls}, {Jehin}, {Langford}, {Leconte}, {Lederer}, {Luger}, {Malhotra}, {Meadows}, {Morris}, {Pozuelos}, {Queloz}, {Raymond}, {Selsis}, {Sestovic}, {Triaud}, \& {Van Grootel}}]{Agol2021}
{Agol}, E., {Dorn}, C., {Grimm}, S.~L., {et~al.} 2021, The Planetary Science Journal, 2, 1

\bibitem[{{Aguichine} {et~al.}(2020){Aguichine}, {Mousis}, {Devouard}, \& {Ronnet}}]{Aguichine2020}
{Aguichine}, A., {Mousis}, O., {Devouard}, B., \& {Ronnet}, T. 2020, \apj, 901, 97

\bibitem[{{Alduchov} \& {Eskridge}(1996)}]{Alduchov1996}
{Alduchov}, O.~A. \& {Eskridge}, R.~E. 1996, Journal of Applied Meteorology, 35, 601

\bibitem[{Anisman {et~al.}(2022)Anisman, Chubb, Elsey, Al-Refaie, Changeat, Yurchenko, Tennyson, \& Tinetti}]{22AnChEl}
Anisman, L.~O., Chubb, K.~L., Elsey, J., {et~al.} 2022, J. Quant. Spectrosc. Radiat. Transfer, 278, 108013

\bibitem[{{Baraffe} {et~al.}(2015){Baraffe}, {Homeier}, {Allard}, \& {Chabrier}}]{Baraffe2015}
{Baraffe}, I., {Homeier}, D., {Allard}, F., \& {Chabrier}, G. 2015, \aap, 577, A42

\bibitem[{Baranov {et~al.}(2008)Baranov, Lafferty, Ma, \& Tipping}]{08BaLaMa}
Baranov, Y., Lafferty, W., Ma, Q., \& Tipping, R. 2008, J. Quant. Spectrosc. Radiat. Transfer, 109, 2291

\bibitem[{{Barnes} {et~al.}(2020){Barnes}, {Luger}, {Deitrick}, {Driscoll}, {Quinn}, {Fleming}, {Smotherman}, {McDonald}, {Wilhelm}, {Garcia}, {Barth}, {Guyer}, {Meadows}, {Bitz}, {Gupta}, {Domagal-Goldman}, \& {Armstrong}}]{Barnes2020}
{Barnes}, R., {Luger}, R., {Deitrick}, R., {et~al.} 2020, \pasp, 132, 024502

\bibitem[{{Barnes} {et~al.}(2013){Barnes}, {Mullins}, {Goldblatt}, {Meadows}, {Kasting}, \& {Heller}}]{Barnes2013}
{Barnes}, R., {Mullins}, K., {Goldblatt}, C., {et~al.} 2013, Astrobiology, 13, 225

\bibitem[{{Barth} {et~al.}(2021){Barth}, {Carone}, {Barnes}, {Noack}, {Molli{\`e}re}, \& {Henning}}]{Barth2021}
{Barth}, P., {Carone}, L., {Barnes}, R., {et~al.} 2021, Astrobiology, 21, 1325

\bibitem[{{Bitsch} \& {Battistini}(2020)}]{Bitsch2020}
{Bitsch}, B. \& {Battistini}, C. 2020, \aap, 633, A10

\bibitem[{{Bitsch} {et~al.}(2019){Bitsch}, {Raymond}, \& {Izidoro}}]{Bitsch2019}
{Bitsch}, B., {Raymond}, S.~N., \& {Izidoro}, A. 2019, \aap, 624, A109

\bibitem[{{Blank} {et~al.}(1993){Blank}, {Stloper}, \& {Carroll}}]{Blank1993}
{Blank}, J.~G., {Stloper}, E.~M., \& {Carroll}, M.~R. 1993, Earth and Planetary Science Letters, 119, 27

\bibitem[{{Bonati} {et~al.}(2019){Bonati}, {Lichtenberg}, {Bower}, {Timpe}, \& {Quanz}}]{Bonati2019}
{Bonati}, I., {Lichtenberg}, T., {Bower}, D.~J., {Timpe}, M.~L., \& {Quanz}, S.~P. 2019, \aap, 621, A125

\bibitem[{{Bond} {et~al.}(2010){Bond}, {O'Brien}, \& {Lauretta}}]{Bond2010}
{Bond}, J.~C., {O'Brien}, D.~P., \& {Lauretta}, D.~S. 2010, \apj, 715, 1050

\bibitem[{{Bottinga} \& {Javoy}(1990)}]{Bottinga1990}
{Bottinga}, Y. \& {Javoy}, M. 1990, Chemical Geology, 81, 255

\bibitem[{{Boukrouche} {et~al.}(2021){Boukrouche}, {Lichtenberg}, \& {Pierrehumbert}}]{Boukrouche2021}
{Boukrouche}, R., {Lichtenberg}, T., \& {Pierrehumbert}, R.~T. 2021, \apj, 919, 130

\bibitem[{{Bower} {et~al.}(2022){Bower}, {Hakim}, {Sossi}, \& {Sanan}}]{Bower2022}
{Bower}, D.~J., {Hakim}, K., {Sossi}, P.~A., \& {Sanan}, P. 2022, The Planetary Science Journal, 3, 93

\bibitem[{{Bower} {et~al.}(2019){Bower}, {Kitzmann}, {Wolf}, {Sanan}, {Dorn}, \& {Oza}}]{Bower2019}
{Bower}, D.~J., {Kitzmann}, D., {Wolf}, A.~S., {et~al.} 2019, \aap, 631, A103

\bibitem[{Buder {et~al.}(2021)Buder, Sharma, Kos, Amarsi, Nordlander, Lind, Martell, Asplund, Bland-Hawthorn, Casey, {et~al.}}]{Buder2021}
Buder, S., Sharma, S., Kos, J., {et~al.} 2021, Monthly Notices of the Royal Astronomical Society, 506, 150

\bibitem[{{Cabral} {et~al.}(2023){Cabral}, {Guilbert-Lepoutre}, {Bitsch}, {Lagarde}, \& {Diakite}}]{Cabral2023}
{Cabral}, N., {Guilbert-Lepoutre}, A., {Bitsch}, B., {Lagarde}, N., \& {Diakite}, S. 2023, \aap, 673, A117

\bibitem[{{Carone} {et~al.}(2014){Carone}, {Keppens}, \& {Decin}}]{Carone2014}
{Carone}, L., {Keppens}, R., \& {Decin}, L. 2014, \mnras, 445, 930

\bibitem[{Carroll \& Holloway(1994)}]{Caroll1994}
Carroll, M.~R. \& Holloway, J.~R., eds. 1994, Volatiles in Magmas (Berlin, Boston: De Gruyter)

\bibitem[{{Catling} \& {Kasting}(2017)}]{CatlingBook}
{Catling}, D.~C. \& {Kasting}, J.~F. 2017, {Atmospheric Evolution on Inhabited and Lifeless Worlds}

\bibitem[{{Chao} {et~al.}(2021){Chao}, {deGraffenried}, {Lach}, {Nelson}, {Truax}, \& {Gaidos}}]{Chaso2021}
{Chao}, K.-H., {deGraffenried}, R., {Lach}, M., {et~al.} 2021, Chemie der Erde / Geochemistry, 81, 125735

\bibitem[{{Chase}(1998)}]{Chase}
{Chase}, M. 1998, {NIST-JANAF Themochemical Tables, Fourth Edition} ({J. Phys. Chem. Ref. Data, Monograph 9})

\bibitem[{{Chen} \& {Kipping}(2017)}]{ChenKipping2017}
{Chen}, J. \& {Kipping}, D. 2017, \apj, 834, 17

\bibitem[{Chubb {et~al.}(2021)Chubb, Rocchetto, Al-Refaie, Waldmann, Min, Barstow, Molli{\'e}re, Phillips, Tennyson, \& Yurchenko}]{20ChRoAl.exo}
Chubb, K.~L., Rocchetto, M., Al-Refaie, A.~F., {et~al.} 2021, A\&A, 646

\bibitem[{{Cohen} {et~al.}(2024){Cohen}, {Glocer}, {Garraffo}, {Alvarado-G{\'o}mez}, {Drake}, {Monsch}, \& {Fauth Puigdomenech}}]{Cohen2024}
{Cohen}, O., {Glocer}, A., {Garraffo}, C., {et~al.} 2024, \apj, 962, 157

\bibitem[{{Coleman} {et~al.}(2019){Coleman}, {Leleu}, {Alibert}, \& {Benz}}]{Coleman2019}
{Coleman}, G.~A.~L., {Leleu}, A., {Alibert}, Y., \& {Benz}, W. 2019, \aap, 631, A7

\bibitem[{Connolly(2009)}]{Connolly2009}
Connolly, J. 2009, Geochemistry, geophysics, geosystems, 10

\bibitem[{{Cox} {et~al.}(1984){Cox}, {Wagman}, \& {Medvedev}}]{Cox}
{Cox}, J., {Wagman}, D., \& {Medvedev}, V. 1984, {CODATA Key Values for Thermodynamics} ({Hemisphere Publishing Corp., New York})

\bibitem[{{Debaille} {et~al.}(2009){Debaille}, {Brandon}, {O'Neill}, {Yin}, \& {Jacobsen}}]{Debaille2009}
{Debaille}, V., {Brandon}, A.~D., {O'Neill}, C., {Yin}, Q.~Z., \& {Jacobsen}, B. 2009, Nature Geoscience, 2, 548

\bibitem[{{Deng} {et~al.}(2020){Deng}, {Du}, {Karki}, {Ghosh}, \& {Lee}}]{Deng2020}
{Deng}, J., {Du}, Z., {Karki}, B.~B., {Ghosh}, D.~B., \& {Lee}, K. K.~M. 2020, Nature Communications, 11, 2007

\bibitem[{Dixon {et~al.}(1995)Dixon, Stolper, \& Holloway}]{Dixon1995}
Dixon, J.~E., Stolper, E.~M., \& Holloway, J.~R. 1995, Journal of Petrology, 36, 1607

\bibitem[{{Dorn} {et~al.}(2018){Dorn}, {Mosegaard}, {Grimm}, \& {Alibert}}]{Dorn2018}
{Dorn}, C., {Mosegaard}, K., {Grimm}, S.~L., \& {Alibert}, Y. 2018, \apj, 865, 20

\bibitem[{{Driscoll} \& {Bercovici}(2014)}]{Driscoll2014}
{Driscoll}, P. \& {Bercovici}, D. 2014, Physics of the Earth and Planetary Interiors, 236, 36

\bibitem[{Dullemond \& Monnier(2010)}]{Dullemond2010}
Dullemond, C. \& Monnier, J. 2010, Annual Review of Astronomy and Astrophysics, 48, 205

\bibitem[{{Elkins-Tanton}(2008)}]{Elkins-Tanton2008}
{Elkins-Tanton}, L.~T. 2008, Earth and Planetary Science Letters, 271, 181

\bibitem[{Ferraz-Mello {et~al.}(2008)Ferraz-Mello, Rodr{\'\i}guez, \& Hussmann}]{FerrazMello2008}
Ferraz-Mello, S., Rodr{\'\i}guez, A., \& Hussmann, H. 2008, Celestial Mechanics and Dynamical Astronomy, 101, 171

\bibitem[{{Gardner} {et~al.}(1999){Gardner}, {Hilton}, \& {Carroll}}]{Gardner1999}
{Gardner}, J.~E., {Hilton}, M., \& {Carroll}, M.~R. 1999, Earth and Planetary Science Letters, 168, 201

\bibitem[{Gillon {et~al.}(2017)Gillon, Triaud, Demory, Jehin, Agol, Deck, Lederer, De~Wit, Burdanov, Ingalls, {et~al.}}]{Gillon2017}
Gillon, M., Triaud, A.~H., Demory, B.-O., {et~al.} 2017, Nature, 542, 456

\bibitem[{{Godolt} {et~al.}(2019){Godolt}, {Tosi}, {Stracke}, {Grenfell}, {Ruedas}, {Spohn}, \& {Rauer}}]{Godolt2019}
{Godolt}, M., {Tosi}, N., {Stracke}, B., {et~al.} 2019, \aap, 625, A12

\bibitem[{{Goldblatt} {et~al.}(2013){Goldblatt}, {Robinson}, {Zahnle}, \& {Crisp}}]{Goldblatt2013}
{Goldblatt}, C., {Robinson}, T.~D., {Zahnle}, K.~J., \& {Crisp}, D. 2013, Nature Geoscience, 6, 661

\bibitem[{{Gordon} {et~al.}(2022){Gordon}, {Rothman}, {Hargreaves}, {Hashemi}, {Karlovets}, {Skinner}, {Conway}, {Hill}, {Kochanov}, {Tan}, {Wcis{\l}o}, {Finenko}, {Nelson}, {Bernath}, {Birk}, {Boudon}, {Campargue}, {Chance}, {Coustenis}, {Drouin}, {Flaud}, {Gamache}, {Hodges}, {Jacquemart}, {Mlawer}, {Nikitin}, {Perevalov}, {Rotger}, {Tennyson}, {Toon}, {Tran}, {Tyuterev}, {Adkins}, {Baker}, {Barbe}, {Can{\`e}}, {Cs{\'a}sz{\'a}r}, {Dudaryonok}, {Egorov}, {Fleisher}, {Fleurbaey}, {Foltynowicz}, {Furtenbacher}, {Harrison}, {Hartmann}, {Horneman}, {Huang}, {Karman}, {Karns}, {Kassi}, {Kleiner}, {Kofman}, {Kwabia-Tchana}, {Lavrentieva}, {Lee}, {Long}, {Lukashevskaya}, {Lyulin}, {Makhnev}, {Matt}, {Massie}, {Melosso}, {Mikhailenko}, {Mondelain}, {M{\"u}ller}, {Naumenko}, {Perrin}, {Polyansky}, {Raddaoui}, {Raston}, {Reed}, {Rey}, {Richard}, {T{\'o}bi{\'a}s}, {Sadiek}, {Schwenke}, {Starikova}, {Sung}, {Tamassia}, {Tashkun}, {Vander Auwera}, {Vasilenko}, {Vigasin}, {Villanueva}, {Vispoel}, {Wagner}, {Yachmenev}, \&
  {Yurchenko}}]{22GoRoHa}
{Gordon}, I.~E., {Rothman}, L.~S., {Hargreaves}, R.~J., {et~al.} 2022, {J. Quant. Spectrosc. Radiat. Transfer}, 277, 107949

\bibitem[{{Graham} {et~al.}(2021){Graham}, {Lichtenberg}, {Boukrouche}, \& {Pierrehumbert}}]{Graham2021}
{Graham}, R.~J., {Lichtenberg}, T., {Boukrouche}, R., \& {Pierrehumbert}, R.~T. 2021, The Planetary Science Journal, 2, 207

\bibitem[{{Greene} {et~al.}(2023){Greene}, {Bell}, {Ducrot}, {Dyrek}, {Lagage}, \& {Fortney}}]{Greene2023}
{Greene}, T.~P., {Bell}, T.~J., {Ducrot}, E., {et~al.} 2023, \nat, 618, 39

\bibitem[{Grimm {et~al.}(2018)Grimm, Demory, Gillon, Dorn, Agol, Burdanov, Delrez, Sestovic, Triaud, Turbet, Bolmont, Caldas, de~Wit, Jehin, Leconte, Raymond, Grootel, Burgasser, Carey, Fabrycky, Heng, Hernandez, Ingalls, Lederer, Selsis, \& Queloz}]{Grimm2018}
Grimm, S.~L., Demory, B.-O., Gillon, M., {et~al.} 2018, Astronomy {\&} Astrophysics, 613, A68

\bibitem[{{Hamano} {et~al.}(2013){Hamano}, {Abe}, \& {Genda}}]{Hamano2013}
{Hamano}, K., {Abe}, Y., \& {Genda}, H. 2013, \nat, 497, 607

\bibitem[{{Herbort} {et~al.}(2020){Herbort}, {Woitke}, {Helling}, \& {Zerkle}}]{Herbort2020}
{Herbort}, O., {Woitke}, P., {Helling}, C., \& {Zerkle}, A. 2020, \aap, 636, A71

\bibitem[{{Hier-Majumder} \& {Hirschmann}(2017)}]{Hier2017}
{Hier-Majumder}, S. \& {Hirschmann}, M.~M. 2017, Geochemistry, Geophysics, Geosystems, 18, 3078

\bibitem[{{Holtz} {et~al.}(1995){Holtz}, {Behrens}, {Dingwell}, \& {Johannes}}]{Holtz1995}
{Holtz}, F., {Behrens}, H., {Dingwell}, D.~B., \& {Johannes}, W. 1995, American Mineralogist, 80, 94

\bibitem[{{Hunten} {et~al.}(1987){Hunten}, {Pepin}, \& {Walker}}]{Hunten1987}
{Hunten}, D.~M., {Pepin}, R.~O., \& {Walker}, J.~C.~G. 1987, Icarus, 69, 532

\bibitem[{{Ikoma} {et~al.}(2018){Ikoma}, {Elkins-Tanton}, {Hamano}, \& {Suckale}}]{Ikoma2018}
{Ikoma}, M., {Elkins-Tanton}, L., {Hamano}, K., \& {Suckale}, J. 2018, \ssr, 214, 76

\bibitem[{{Johansen} {et~al.}(2023){Johansen}, {Ronnet}, {Schiller}, {Deng}, \& {Bizzarro}}]{Johansen2023}
{Johansen}, A., {Ronnet}, T., {Schiller}, M., {Deng}, Z., \& {Bizzarro}, M. 2023, \aap, 671, A76

\bibitem[{{Johnson} \& {Dick}(1992)}]{Johnson1992}
{Johnson}, K. T.~M. \& {Dick}, H. J.~B. 1992, \jgr, 97, 9219

\bibitem[{{Johnstone} {et~al.}(2021){Johnstone}, {Bartel}, \& {G{\"u}del}}]{Johnstone2021}
{Johnstone}, C.~P., {Bartel}, M., \& {G{\"u}del}, M. 2021, \aap, 649, A96

\bibitem[{Jorge {et~al.}(2022)Jorge, Kamp, Waters, Woitke, \& Spaargaren}]{Jorge2022}
Jorge, D.~M., Kamp, I., Waters, L., Woitke, P., \& Spaargaren, R.~J. 2022, Astronomy \& Astrophysics, 660, A85

\bibitem[{{Katyal} {et~al.}(2019){Katyal}, {Nikolaou}, {Godolt}, {Grenfell}, {Tosi}, {Schreier}, \& {Rauer}}]{Katyal2019}
{Katyal}, N., {Nikolaou}, A., {Godolt}, M., {et~al.} 2019, \apj, 875, 31

\bibitem[{{Katyal} {et~al.}(2020){Katyal}, {Ortenzi}, {Lee Grenfell}, {Noack}, {Sohl}, {Godolt}, {Garc{\'\i}a Mu{\~n}oz}, {Schreier}, {Wunderlich}, \& {Rauer}}]{Katyal2020}
{Katyal}, N., {Ortenzi}, G., {Lee Grenfell}, J., {et~al.} 2020, \aap, 643, A81

\bibitem[{{Kopparapu} {et~al.}(2013){Kopparapu}, {Ramirez}, {Kasting}, {Eymet}, {Robinson}, {Mahadevan}, {Terrien}, {Domagal-Goldman}, {Meadows}, \& {Deshpande}}]{2013Kopparapu}
{Kopparapu}, R.~K., {Ramirez}, R., {Kasting}, J.~F., {et~al.} 2013, \apj, 765, 131

\bibitem[{{Krissansen-Totton}(2023)}]{Krissansen2023}
{Krissansen-Totton}, J. 2023, \apjl, 951, L39

\bibitem[{{Krissansen-Totton} \& {Fortney}(2022)}]{Krissansen2022}
{Krissansen-Totton}, J. \& {Fortney}, J.~J. 2022, \apj, 933, 115

\bibitem[{{Lammer} {et~al.}(2018){Lammer}, {Zerkle}, {Gebauer}, {Tosi}, {Noack}, {Scherf}, {Pilat-Lohinger}, {G{\"u}del}, {Grenfell}, {Godolt}, \& {Nikolaou}}]{Lammer2018}
{Lammer}, H., {Zerkle}, A.~L., {Gebauer}, S., {et~al.} 2018, \aapr, 26, 2

\bibitem[{{Lebrun} {et~al.}(2013){Lebrun}, {Massol}, {Chassefi{\`e}Re}, {Davaille}, {Marcq}, {Sarda}, {Leblanc}, \& {Brandeis}}]{Lebrun2013}
{Lebrun}, T., {Massol}, H., {Chassefi{\`e}Re}, E., {et~al.} 2013, Journal of Geophysical Research (Planets), 118, 1155

\bibitem[{{Leconte} {et~al.}(2010){Leconte}, {Chabrier}, {Baraffe}, \& {Levrard}}]{Leconte2010}
{Leconte}, J., {Chabrier}, G., {Baraffe}, I., \& {Levrard}, B. 2010, Astro.~\& Astrophys., 516, A64+

\bibitem[{{Lensky} {et~al.}(2006){Lensky}, {Niebo}, {Holloway}, {Lyakhovsky}, \& {Navon}}]{Lensky2006}
{Lensky}, N.~G., {Niebo}, R.~W., {Holloway}, J.~R., {Lyakhovsky}, V., \& {Navon}, O. 2006, Earth and Planetary Science Letters, 245, 278

\bibitem[{{Lichtenberg} {et~al.}(2021){Lichtenberg}, {Bower}, {Hammond}, {Boukrouche}, {Sanan}, {Tsai}, \& {Pierrehumbert}}]{Lichtenberg2021}
{Lichtenberg}, T., {Bower}, D.~J., {Hammond}, M., {et~al.} 2021, Journal of Geophysical Research (Planets), 126, e06711

\bibitem[{{Lim} {et~al.}(2023){Lim}, {Benneke}, {Doyon}, {MacDonald}, {Piaulet}, {Artigau}, {Coulombe}, {Radica}, {L'Heureux}, {Albert}, {Rackham}, {de Wit}, {Salhi}, {Roy}, {Flagg}, {Fournier-Tondreau}, {Taylor}, {Cook}, {Lafreni{\`e}re}, {Cowan}, {Kaltenegger}, {Rowe}, {Espinoza}, {Dang}, \& {Darveau-Bernier}}]{Lim2023}
{Lim}, O., {Benneke}, B., {Doyon}, R., {et~al.} 2023, \apjl, 955, L22

\bibitem[{{Lincowski} {et~al.}(2023){Lincowski}, {Meadows}, {Zieba}, {Kreidberg}, {Morley}, {Gillon}, {Selsis}, {Agol}, {Bolmont}, {Ducrot}, {Hu}, {Koll}, {Lyu}, {Mandell}, {Suissa}, \& {Tamburo}}]{Lincowski2023}
{Lincowski}, A.~P., {Meadows}, V.~S., {Zieba}, S., {et~al.} 2023, \apjl, 955, L7

\bibitem[{{Liu} {et~al.}(2005){Liu}, {Zhang}, \& {Behrens}}]{Liu2005}
{Liu}, Y., {Zhang}, Y., \& {Behrens}, H. 2005, Journal of Volcanology and Geothermal Research, 143, 219

\bibitem[{Lodders(2003)}]{Lodders2003}
Lodders, K. 2003, The Astrophysical Journal, 591, 1220

\bibitem[{{Luger} \& {Barnes}(2015)}]{Luger2015}
{Luger}, R. \& {Barnes}, R. 2015, Astrobiology, 15, 119

\bibitem[{{Marcq} {et~al.}(2017){Marcq}, {Salvador}, {Massol}, \& {Davaille}}]{Marcq2017}
{Marcq}, E., {Salvador}, A., {Massol}, H., \& {Davaille}, A. 2017, Journal of Geophysical Research (Planets), 122, 1539

\bibitem[{{Mason} \& {Marrero}(1970)}]{Mason1970}
{Mason}, E.~A. \& {Marrero}, T.~R. 1970, Advances in Atomic and Molecular Physics, 6, 155

\bibitem[{{Maurice} {et~al.}(2017){Maurice}, {Tosi}, {Samuel}, {Plesa}, {H{\"u}ttig}, \& {Breuer}}]{Maurice2017}
{Maurice}, M., {Tosi}, N., {Samuel}, H., {et~al.} 2017, Journal of Geophysical Research (Planets), 122, 577

\bibitem[{{Miguel} {et~al.}(2020){Miguel}, {Cridland}, {Ormel}, {Fortney}, \& {Ida}}]{Miguel2020}
{Miguel}, Y., {Cridland}, A., {Ormel}, C.~W., {Fortney}, J.~J., \& {Ida}, S. 2020, \mnras, 491, 1998

\bibitem[{Mlawer {et~al.}(2012)Mlawer, Payne, Moncet, Delamere, Alvarado, \& Tobin}]{12MlPaMo}
Mlawer, E.~J., Payne, V.~H., Moncet, J.-L., {et~al.} 2012, Philosophical Transactions of the Royal Society A: Mathematical, Physical and Engineering Sciences, 370, 2520

\bibitem[{{Molli{\`e}re} {et~al.}(2019){Molli{\`e}re}, {Wardenier}, {van Boekel}, {Henning}, {Molaverdikhani}, \& {Snellen}}]{Molliere2019}
{Molli{\`e}re}, P., {Wardenier}, J.~P., {van Boekel}, R., {et~al.} 2019, \aap, 627, A67

\bibitem[{{Moore} \& {Carmichael}(1998)}]{Moore1998}
{Moore}, G. \& {Carmichael}, I.~S.~E. 1998, Contributions to Mineralogy and Petrology, 130, 304

\bibitem[{{Moore} {et~al.}(2023){Moore}, {Cowan}, \& {Boukar{\'e}}}]{Moore2023}
{Moore}, K., {Cowan}, N.~B., \& {Boukar{\'e}}, C.-{\'E}. 2023, \mnras, 526, 6235

\bibitem[{{Mousis} {et~al.}(2021){Mousis}, {Aguichine}, {Bouquet}, {Lunine}, {Danger}, {Mandt}, \& {Luspay-Kuti}}]{Mousis2021}
{Mousis}, O., {Aguichine}, A., {Bouquet}, A., {et~al.} 2021, The Planetary Science Journal, 2, 72

\bibitem[{{M{\"u}ller} {et~al.}(2024{\natexlab{a}}){M{\"u}ller}, {Bitsch}, \& {Schneider}}]{Muller2024}
{M{\"u}ller}, J., {Bitsch}, B., \& {Schneider}, A.~D. 2024{\natexlab{a}}, \aap, 688, A139

\bibitem[{{M{\"u}ller} {et~al.}(2024{\natexlab{b}}){M{\"u}ller}, {Baron}, {Helled}, {Bouchy}, \& {Parc}}]{Baron2024}
{M{\"u}ller}, S., {Baron}, J., {Helled}, R., {Bouchy}, F., \& {Parc}, L. 2024{\natexlab{b}}, \aap, 686, A296

\bibitem[{{Mysen} {et~al.}(1975){Mysen}, {Arculus}, \& {Eggler}}]{Mysen1975}
{Mysen}, B.~O., {Arculus}, R.~J., \& {Eggler}, D.~H. 1975, Contributions to Mineralogy and Petrology, 53, 227

\bibitem[{{Nikolaou} {et~al.}(2019){Nikolaou}, {Katyal}, {Tosi}, {Godolt}, {Grenfell}, \& {Rauer}}]{Niko2019}
{Nikolaou}, A., {Katyal}, N., {Tosi}, N., {et~al.} 2019, \apj, 875, 11

\bibitem[{{Noack} {et~al.}(2016){Noack}, {H{\"o}ning}, {Rivoldini}, {Heistracher}, {Zimov}, {Journaux}, {Lammer}, {Van Hoolst}, \& {Bredeh{\"o}ft}}]{Noack2016}
{Noack}, L., {H{\"o}ning}, D., {Rivoldini}, A., {et~al.} 2016, \icarus, 277, 215

\bibitem[{Noack \& Lasbleis(2020)}]{Noack2020}
Noack, L. \& Lasbleis, M. 2020, Astronomy \& Astrophysics, 638, A129

\bibitem[{Noack {et~al.}(2017)Noack, Rivoldini, \& Van~Hoolst}]{Noack2017}
Noack, L., Rivoldini, A., \& Van~Hoolst, T. 2017, Physics of the Earth and Planetary Interiors, 269, 40

\bibitem[{Odintsova {et~al.}(2020)Odintsova, Tretyakov, Simonova, Ptashnik, Pirali, \& Campargue}]{20OdTrSi}
Odintsova, T.~A., Tretyakov, M.~Y., Simonova, A.~A., {et~al.} 2020, Journal of Molecular Structure, 1210, 128046

\bibitem[{{O'Neill} \& {Palme}(1998)}]{O_neill_1998}
{O'Neill}, H. \& {Palme}, H. 1998, The Earth’s Mantle: Composition, Structure, and Evolution, ed. I.~{Jackson} (Cambridge University Press)

\bibitem[{{Ormel} {et~al.}(2017){Ormel}, {Liu}, \& {Schoonenberg}}]{Ormel2017}
{Ormel}, C.~W., {Liu}, B., \& {Schoonenberg}, D. 2017, \aap, 604, A1

\bibitem[{{O'Rourke}(2020)}]{Orourke2020}
{O'Rourke}, J.~G. 2020, \grl, 47, e86126

\bibitem[{{Ortenzi} {et~al.}(2020){Ortenzi}, {Noack}, {Sohl}, {Guimond}, {Grenfell}, {Dorn}, {Schmidt}, {Vulpius}, {Katyal}, {Kitzmann}, \& {Rauer}}]{Ortenzi2020}
{Ortenzi}, G., {Noack}, L., {Sohl}, F., {et~al.} 2020, Scientific Reports, 10, 10907

\bibitem[{{Pahlevan} {et~al.}(2022){Pahlevan}, {Schaefer}, {Elkins-Tanton}, {Desch}, \& {Buseck}}]{Pahlevan2022}
{Pahlevan}, K., {Schaefer}, L., {Elkins-Tanton}, L.~T., {Desch}, S.~J., \& {Buseck}, P.~R. 2022, Earth and Planetary Science Letters, 595, 117772

\bibitem[{{Pan} {et~al.}(1991){Pan}, {Holloway}, \& {Hervig}}]{Pan1991}
{Pan}, V., {Holloway}, J.~R., \& {Hervig}, R.~L. 1991, \gca, 55, 1587

\bibitem[{{Papale}(1997)}]{Papale1997}
{Papale}, P. 1997, Contributions to Mineralogy and Petrology, 126, 237

\bibitem[{Paynter {et~al.}(2009)Paynter, Ptashnik, Shine, Smith, McPheat, \& Williams}]{09PaPtSh}
Paynter, D.~J., Ptashnik, I.~V., Shine, K.~P., {et~al.} 2009, Journal of Geophysical Research: Atmospheres, 114

\bibitem[{{Pichierri} {et~al.}(2024){Pichierri}, {Morbidelli}, {Batygin}, \& {Brasser}}]{Pichierri2024}
{Pichierri}, G., {Morbidelli}, A., {Batygin}, K., \& {Brasser}, R. 2024, Nature Astronomy, 8, 1408

\bibitem[{{Pierrehumbert}(2010)}]{Pierrehumbert}
{Pierrehumbert}, R.~T. 2010, {Principles of Planetary Climate}

\bibitem[{{Piso} {et~al.}(2015){Piso}, {{\"O}berg}, {Birnstiel}, \& {Murray-Clay}}]{Piso2015}
{Piso}, A.-M.~A., {{\"O}berg}, K.~I., {Birnstiel}, T., \& {Murray-Clay}, R.~A. 2015, \apj, 815, 109

\bibitem[{{Pluriel} {et~al.}(2019){Pluriel}, {Marcq}, \& {Turbet}}]{Pluriel2019}
{Pluriel}, W., {Marcq}, E., \& {Turbet}, M. 2019, \icarus, 317, 583

\bibitem[{Ptashnik {et~al.}(2011)Ptashnik, McPheat, Shine, Smith, \& Williams}]{11PtMcSh}
Ptashnik, I.~V., McPheat, R.~A., Shine, K.~P., Smith, K.~M., \& Williams, R.~G. 2011, Journal of Geophysical Research: Atmospheres, 116

\bibitem[{{Raymond} {et~al.}(2022){Raymond}, {Izidoro}, {Bolmont}, {Dorn}, {Selsis}, {Turbet}, {Agol}, {Barth}, {Carone}, {Dasgupta}, {Gillon}, \& {Grimm}}]{Raymond2022}
{Raymond}, S.~N., {Izidoro}, A., {Bolmont}, E., {et~al.} 2022, Nature Astronomy, 6, 80

\bibitem[{{Ribas} {et~al.}(2005){Ribas}, {Guinan}, {G{\"u}del}, \& {Audard}}]{Ribas2005}
{Ribas}, I., {Guinan}, E.~F., {G{\"u}del}, M., \& {Audard}, M. 2005, Astrophys.~J., 622, 680

\bibitem[{{Richey-Yowell} {et~al.}(2022){Richey-Yowell}, {Shkolnik}, {Loyd}, {Jackman}, {Schneider}, {Ag{\"u}eros}, {Barman}, {Meadows}, {Gibson}, \& {Douglas}}]{RicheyYowell2022}
{Richey-Yowell}, T., {Shkolnik}, E.~L., {Loyd}, R.~O.~P., {et~al.} 2022, \apj, 929, 169

\bibitem[{{Ros} \& {Johansen}(2013)}]{Ros2013}
{Ros}, K. \& {Johansen}, A. 2013, \aap, 552, A137

\bibitem[{{Salvador} \& {Samuel}(2023)}]{Salvador2023}
{Salvador}, A. \& {Samuel}, H. 2023, \icarus, 390, 115265

\bibitem[{{Samuel} {et~al.}(2023){Samuel}, {Drilleau}, {Rivoldini}, {Xu}, {Huang}, {Garcia}, {Leki{\'c}}, {Irving}, {Badro}, {Lognonn{\'e}}, {Connolly}, {Kawamura}, {Gudkova}, \& {Banerdt}}]{Samuel2023}
{Samuel}, H., {Drilleau}, M., {Rivoldini}, A., {et~al.} 2023, \nat, 622, 712

\bibitem[{{Schaefer} {et~al.}(2016){Schaefer}, {Wordsworth}, {Berta-Thompson}, \& {Sasselov}}]{Schaefer2016}
{Schaefer}, L., {Wordsworth}, R.~D., {Berta-Thompson}, Z., \& {Sasselov}, D. 2016, \apj, 829, 63

\bibitem[{{Schlecker} {et~al.}(2024){Schlecker}, {Apai}, {Lichtenberg}, {Bergsten}, {Salvador}, \& {Hardegree-Ullman}}]{Schlecker2024}
{Schlecker}, M., {Apai}, D., {Lichtenberg}, T., {et~al.} 2024, The Planetary Science Journal, 5, 3

\bibitem[{{Schneider} \& {Bitsch}(2021)}]{Schneider2021}
{Schneider}, A.~D. \& {Bitsch}, B. 2021, \aap, 654, A72

\bibitem[{{Schoonenberg} {et~al.}(2019){Schoonenberg}, {Liu}, {Ormel}, \& {Dorn}}]{Schoonenberg2019}
{Schoonenberg}, D., {Liu}, B., {Ormel}, C.~W., \& {Dorn}, C. 2019, \aap, 627, A149

\bibitem[{{Sergeev} {et~al.}(2024){Sergeev}, {Boutle}, {Lambert}, {Mayne}, {Bendall}, {Kohary}, {Olivier}, \& {Shipway}}]{Sergeev2024}
{Sergeev}, D.~E., {Boutle}, I.~A., {Lambert}, F.~H., {et~al.} 2024, \apj, 970, 7

\bibitem[{{Sergeev} {et~al.}(2022){Sergeev}, {Fauchez}, {Turbet}, {Boutle}, {Tsigaridis}, {Way}, {Wolf}, {Domagal-Goldman}, {Forget}, {Haqq-Misra}, {Kopparapu}, {Lambert}, {Manners}, \& {Mayne}}]{Sergeev2022}
{Sergeev}, D.~E., {Fauchez}, T.~J., {Turbet}, M., {et~al.} 2022, The Planetary Science Journal, 3, 212

\bibitem[{Shine {et~al.}(2016)Shine, Campargue, Mondelain, McPheat, Ptashnik, \& Weidmann}]{16ShCaMo}
Shine, K.~P., Campargue, A., Mondelain, D., {et~al.} 2016, Journal of Molecular Spectroscopy, 327, 193

\bibitem[{{Silver} {et~al.}(1990){Silver}, {Ihinger}, \& {Stolper}}]{Silver1990}
{Silver}, L.~A., {Ihinger}, P.~D., \& {Stolper}, E. 1990, Contributions to Mineralogy and Petrology, 104, 142

\bibitem[{Stixrude(2014)}]{Stixrude2014}
Stixrude, L. 2014, Philosophical Transactions of the Royal Society A: Mathematical, Physical and Engineering Sciences, 372, 20130076

\bibitem[{{Stolper} \& {Holloway}(1988)}]{Stolper1988}
{Stolper}, E. \& {Holloway}, J.~R. 1988, Earth and Planetary Science Letters, 87, 397

\bibitem[{{St{\"u}eken} {et~al.}(2020){St{\"u}eken}, {Som}, {Claire}, {Rugheimer}, {Scherf}, {Spro{\ss}}, {Tosi}, {Ueno}, \& {Lammer}}]{Stueken2020}
{St{\"u}eken}, E.~E., {Som}, S.~M., {Claire}, M., {et~al.} 2020, \ssr, 216, 31

\bibitem[{{Tian} \& {Ida}(2015)}]{Tian2015}
{Tian}, F. \& {Ida}, S. 2015, Nature Geoscience, 8, 177

\bibitem[{{Turbet} {et~al.}(2021){Turbet}, {Bolmont}, {Chaverot}, {Ehrenreich}, {Leconte}, \& {Marcq}}]{Turbet2021}
{Turbet}, M., {Bolmont}, E., {Chaverot}, G., {et~al.} 2021, \nat, 598, 276

\bibitem[{{Turbet} {et~al.}(2019){Turbet}, {Ehrenreich}, {Lovis}, {Bolmont}, \& {Fauchez}}]{Turbet2019}
{Turbet}, M., {Ehrenreich}, D., {Lovis}, C., {Bolmont}, E., \& {Fauchez}, T. 2019, \aap, 628, A12

\bibitem[{{Unterborn} {et~al.}(2018){Unterborn}, {Hinkel}, \& {Desch}}]{Unterborn2018}
{Unterborn}, C.~T., {Hinkel}, N.~R., \& {Desch}, S.~J. 2018, Research Notes of the American Astronomical Society, 2, 116

\bibitem[{{Watson} {et~al.}(1981){Watson}, {Donahue}, \& {Walker}}]{Watson1981}
{Watson}, A.~J., {Donahue}, T.~M., \& {Walker}, J.~C.~G. 1981, Icarus, 48, 150

\bibitem[{{Way} {et~al.}(2022){Way}, {Ernst}, \& {Scargle}}]{Way2022}
{Way}, M.~J., {Ernst}, R.~E., \& {Scargle}, J.~D. 2022, The Planetary Science Journal, 3, 92

\bibitem[{Williams \& Cieza(2011)}]{Williams2011}
Williams, J.~P. \& Cieza, L.~A. 2011, Annual Review of Astronomy and Astrophysics, 49, 67

\bibitem[{{Wordsworth} {et~al.}(2018){Wordsworth}, {Schaefer}, \& {Fischer}}]{Wordsworth2018}
{Wordsworth}, R.~D., {Schaefer}, L.~K., \& {Fischer}, R.~A. 2018, \aj, 155, 195

\bibitem[{{Yamashita}(1999)}]{Yamashita1999}
{Yamashita}, S. 1999, Journal of Petrology, 40, 1497

\bibitem[{{Yang} {et~al.}(2014){Yang}, {Bou{\'e}}, {Fabrycky}, \& {Abbot}}]{Yang2014}
{Yang}, J., {Bou{\'e}}, G., {Fabrycky}, D.~C., \& {Abbot}, D.~S. 2014, \apjl, 787, L2

\bibitem[{{Young} {et~al.}(2023){Young}, {Shahar}, \& {Schlichting}}]{Young2023}
{Young}, E.~D., {Shahar}, A., \& {Schlichting}, H.~E. 2023, \nat, 616, 306

\bibitem[{{Zahnle} \& {Kasting}(1986)}]{Zahnle1986}
{Zahnle}, K.~J. \& {Kasting}, J.~F. 1986, \icarus, 68, 462

\bibitem[{{Zahnle} {et~al.}(2020){Zahnle}, {Lupu}, {Catling}, \& {Wogan}}]{Zahnle2020}
{Zahnle}, K.~J., {Lupu}, R., {Catling}, D.~C., \& {Wogan}, N. 2020, The Planetary Science Journal, 1, 11

\bibitem[{{Zahnle} {et~al.}(2015){Zahnle}, {Lupu}, {Dobrovolskis}, \& {Sleep}}]{Zahnle2015}
{Zahnle}, K.~J., {Lupu}, R., {Dobrovolskis}, A., \& {Sleep}, N.~H. 2015, Earth and Planetary Science Letters, 427, 74

\bibitem[{{Zieba} {et~al.}(2023){Zieba}, {Kreidberg}, {Ducrot}, {Gillon}, {Morley}, {Schaefer}, {Tamburo}, {Koll}, {Lyu}, {Acu{\~n}a}, {Agol}, {Iyer}, {Hu}, {Lincowski}, {Meadows}, {Selsis}, {Bolmont}, {Mandell}, \& {Suissa}}]{Zieba2023}
{Zieba}, S., {Kreidberg}, L., {Ducrot}, E., {et~al.} 2023, \nat, 620, 746

\end{thebibliography}
\begin{appendix}

\section{Derivation of the coupled \ce{H2O}-\ce{CO2} outgassing differential equations}
\label{sec: Derivation melt fraction}

We present here an improvement of the volatile model used by \citet{Barth2021} for \magmoc{1.0}, which originally adopted the magma ocean model including outgassing of \citet{Schaefer2016}. To derive the coupled differential equations that drive multi-component outgassing, we use the mass balance in the magma ocean and atmosphere as a starting point as by \citet[][Eq.~12]{Barth2021} (See also Eq~\ref{eq: mass balance}).

For clarity, we outline again the set-up of the volatile reservoirs. Each volatile $i$ (here \ce{H2O} and \ce{CO2} ) is stored across the magma ocean and atmosphere system (MOA) as follows (Fig.~\ref{fig: Volatile_Flowchart}): a) in the crystallized portion of the magma ocean ($M_i^{\mathrm{crystal}}$), b)in the liquid magma phase ($M_i^{\mathrm{liq}}$), c) and in the atmosphere ($M_i^{\mathrm{atm}}$). Additionally, we define, $F_i$, as the mass fraction of the volatile $i$ solved in the melt. We also assume a constant, mantle-averaged partition coefficient $k_i$ between the melt and the crystal phase. The latter is assumed to sink to the bottom and form the solidified mantle. For the partition coefficient, we adopt  $k_{\ce{H2O}}$ from \citet{Schaefer2016} and $k_{\ce{CO2}}$ from \citet{Lebrun2013} (Table~\ref{Tab_geo}).

Lastly, the mass of volatile $i$ in the atmosphere is calculated via:
\begin{equation}
 M_i^{\mathrm{atm}} =  \frac{4 \pi r_\mathrm{p}^2}{g} p_{i,mass}
\end{equation}
where $p_{i,\rm{mass}}$ is mass weighted pressure with respect to the partial pressure $p_{i,\rm{part}}$ of volatile $i$ (see e.g. \citet{Bower2019}):
\begin{eqnarray}
 p_{i,\rm{mass}} &=& \frac{\mu_i} {\overline{\mu}_{\rm{atm}}} p_{i,\rm{part}}
 \label{eq: mass pressure}
\end{eqnarray}
where $\mu_i$ represents the molecular mass of the specific volatile and $\overline{\mu}_{atm}$ stands for the mean atmospheric molecular mass, considering the combination of all outgassed volatiles. It is worth noting that in the water-dominated steam atmosphere set-up of \magmoc{1.0} \citep{Barth2021}, the authors assumed $p_{\ce{H2O},mass}= p_{\ce{H2O}}$ because it held true for most of the simulation time that $\mu_{\ce{H2O}} \approx \overline{\mu}_{atm}$. This assumption no longer holds in cases where a \ce{H2O} dominated atmosphere evolves into \ce{CO2} dominated atmosphere as simulated in this work. We also neglect here \ce{O2} because atmospheric \ce{O2} is in this model typically produced after the end of the magma ocean and therefore does not enter the mass balance equations describing balance between outgassing and dissolved volatile mass in the magma ocean phase. 

Thus, the volatile mass balance can be summarized as (See Table~\ref{Tab_Volat_Model} for an overview of all the components of the volatile model.):

\begin{align}
		M_i^{\mathrm{moa}} &= M_i^{\mathrm{crystal}} + M_i^{\mathrm{liq}} + M_i^{\mathrm{atm}} \nonumber\\
		&= k_i F_i M^{\mathrm{crystal}} + F_i M^{\mathrm{liq}} + \frac{4 \pi r_\mathrm{p}^2}{g} p_{i,mass},
\label{eq: mass balance2}
\end{align}
where $M^{\mathrm{crystal}}$ denotes the mass of the crystallized part of the magma ocean and $M^{\mathrm{liq}}$ is the mass of the liquid phase of the magma ocean. We further assume that the partial pressure of the volatile at the surface is determined by the fraction of the volatile in the melt $F_i$ \citep{Elkins-Tanton2008,Schaefer2016,Lichtenberg2021}. 

In \citet{Barth2021}, the water outgassing law of \citet{Schaefer2016}, based on the laboratory data of \citet{Papale1997}, was used. Therefore, we employ for the pure water outgassing the same relation to compare with \magmoc{1.0}:

\begin{equation}
 p_{\ce{H2O}} =\left(\frac{F_{\ce{H2O}}}{3.44\times 10^{-8}}\right)^{1/0.74} \, \rm{[Pa]}\label{eq: H2O_Schaefer} .      
\end{equation}

For a mixed atmosphere model, \citet{Elkins-Tanton2008} assumed the following solubility laws for \ce{H2O} and \ce{CO2} that were also based \citet{Papale1997} but have a form that suppresses outgassing below a certain volatile melt fraction:

\begin{align}
p_{\ce{H2O}} &=\left(\frac{F_{\ce{H2O}}-3 \times 10^{-3}}{2.08\times 10^{-6}}\right)^{1/0.52}\, \rm{[Pa]}\\
p_{\ce{CO2}} &=\left(\frac{F_{\ce{CO2}}-5\times 10^{-4}}{2.08\times 10^{-6}}\right)^{1/0.45}\, \rm{[Pa]} \label{eq: mixed_Papale}.
\end{align}

\begin{figure}
    \centering
    \includegraphics[width=0.49\textwidth]{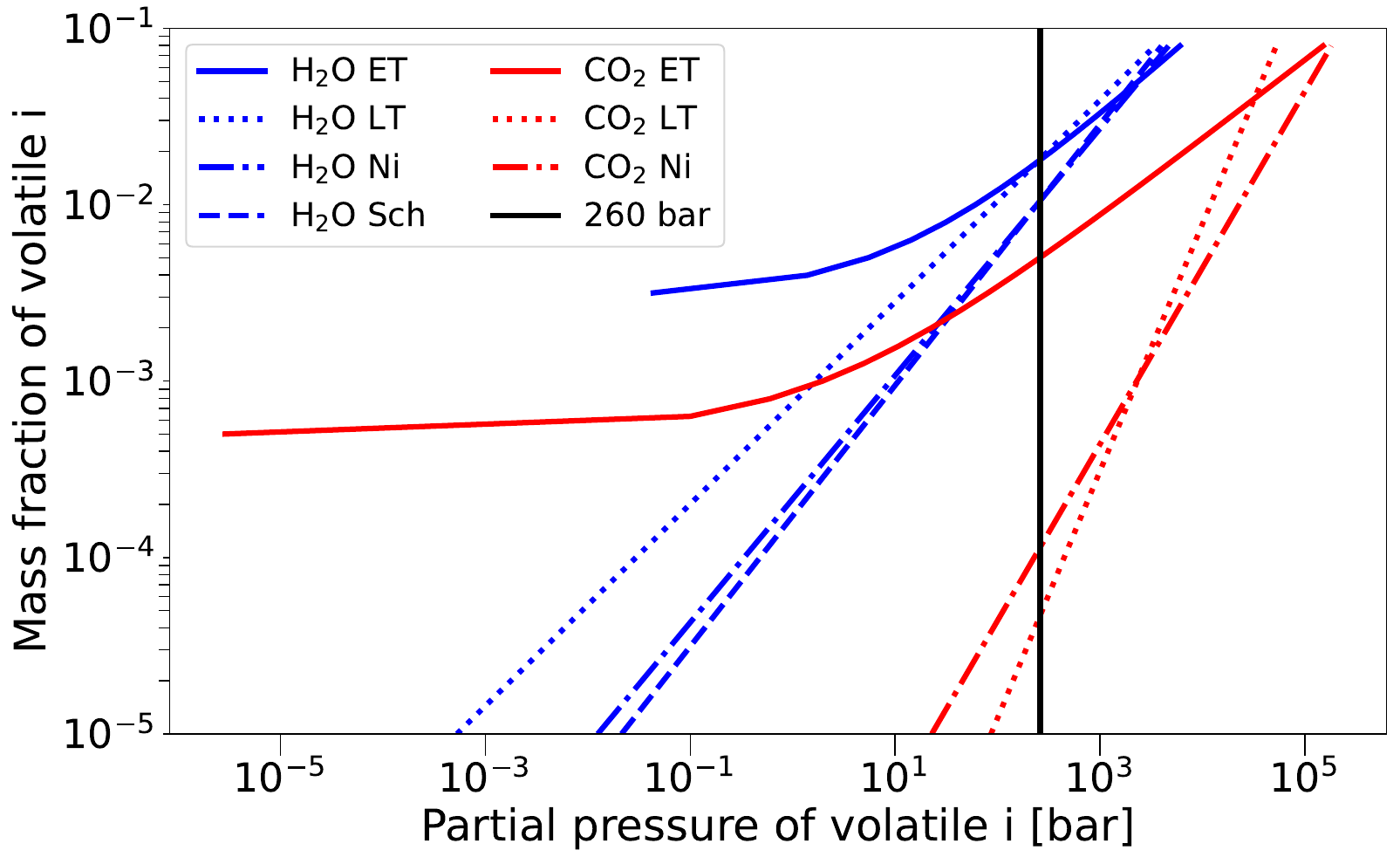}
    \caption{Outgassing of \ce{H2O} (blue) and \ce{CO2} (red) from different publications (ET: \citet{Elkins-Tanton2008}, Sch: \citet{Schaefer2016}, LT: \citet{Lichtenberg2021}, Ni: \citet{Niko2019}). The data reference for ET and Sch is \citet{Papale1997}, for Ni these are \citet{Caroll1994} for \ce{H2O} and \citet{Pan1991} for \ce{CO2}, for LT these are \citet{Silver1990,Holtz1995,Moore1998,Yamashita1999,Gardner1999,Liu2005} for \ce{H2O} and \citet{Mysen1975,Stolper1988,Pan1991,Blank1993,Dixon1995} for \ce{CO2}.}
    \label{fig: OutgasLaws}
\end{figure}

Figure~\ref{fig: OutgasLaws} provides an overview of solubility laws used in the literature. The \ce{H2O} solubility laws agree with each other within one order of magnitude for $260$~bar$\leq p \leq 10^{4}$~bar. Consequently, it is not surprising that magma oceans with \ce{H2O} outgassing yield generally similar results (See Sect.~\ref{sec:Earth}). However, significant deviations are apparent between the \ce{CO2} solubility law used by \citet{Elkins-Tanton2008} and those used by \citet{Lichtenberg2021} and \citet{Niko2019}. Conversely, the \ce{CO2}laws used by \citet{Lichtenberg2021} and \citet{Niko2019} generally agree with each other. 

 In  \magmoc{1,0} the mass balance equations (\ref{eq: mass balance2}) were solved for $F_{\ce{H2O}}$ at every time step using a root finding method. Using this approach for two or more volatiles, however, is cumbersome and numerically costly.
 
 In the new model \magmoc{2.0}, we opt instead to solve by numerical integration:
\begin{align}
F_{\ce{H2O}}(t) &= F_{\ce{H2O}}(0)+\frac{d  F_{\ce{H2O}}(t)}{dt} \label{eq: F0+dFH2O}\Delta t\\
F_{\ce{CO2}}(t) &= F_{\ce{CO2}}(0)+\frac{d  F_{\ce{CO2}}(t)}{dt} \Delta t
\label{eq: F0+dFCO2}
\end{align}
where the starting points of the integration, $F_{\ce{H2O}}(0)$ and $F_{\ce{CO2}}(0)$, are input parameters for \magmoc{2.0} and are derived via root finding only once when preparing the simulations. 

The derivatives of the melt fractions $F'_i$ can be found by differentiating Equations \ref{eq: mass balance2}, assuming that the total volatile mass in the system, $M_i^{moa}$, has to remain constant when a new equilibrium in mass for \ce{H2O} and \ce{CO2}  is established between all reservoirs. 

For mixed \ce{H2O}-\ce{CO2} outgassing, we also account for changes in the mean molecular mass of the atmosphere $\overline{\mu}_{atm}$, given by:
 \begin{align}
\overline{\mu}_{atm} =\frac{p_{\ce{H2O}}\cdot \mu_{\ce{H2O}}+ p_{\ce{CO2}}\cdot \mu_{\ce{CO2}} }{\underbrace{p_{\ce{H2O}}+p_{\ce{CO2}}}_{p_{surf}}}, 
 \end{align}
where $p_{\ce{H2O}}+p_{\ce{CO2}}$ is the surface pressure of the complete atmosphere $p_{surf}$.

\begin{table*}[h]
    \centering
    \caption{Placeholder for differentiation of mass balance equations. }
    \begin{tabular}{|p{0.3cm} l|p{0.3cm} l|}
    \hline
      & mass balance \ce{H2O}  &  & mass balance \ce{CO2}\\
      & (volatile 1)  &  & (volatile 2)\\
      \hline
       $g$&$=F_{\ce{H2O}}$ & $l$&$= F_{\ce{CO2}}$\\
       $a_1$&$=-M_{\ce{H2O}}^{\mathrm{moa}}$ (1) &$a_2$&$=-M_{\ce{CO2}}^{\mathrm{moa}}$ (1)\\
       $b_1$&$=k_{\ce{H2O}} \cdot M^{\mathrm{crystal}}$ & $b_2$&$=k_{\ce{CO2}} \cdot M^{\mathrm{crystal}}$\\
       $f$&$=3.44 \times 10^{-8}$ (Schaefer) or $2.08 \times 10^{-6}$ (E-T)&$v$&$=2.08 \times 10^{-6}$ (E-T)\\
       $e$&= 0 (Schaefer) or $-3\times 10^{-3}$ (E-T) &$u$&$=-5\times 10^{-4}$ (E-T)\\
       $exp_1$ &= 1/0.74 (Schaefer) or 1/0.52 (E-T) &$exp_2$ & = 1/0.45 (E-T)\\
       $w_1$&$=\mu_{\ce{CO2}}/\mu_{\ce{H2O}}$ &$w_2$&$=\mu_{\ce{H2O}}/\mu_{\ce{CO2}}$\\
         \hline
           \multicolumn{4}{|c|}{Both volatile systems}\\
           \hline
                \multicolumn{4}{|c|}{$h=M^{\mathrm{liq}}$}\\
                \multicolumn{4}{|c|}{$d=\frac{4\pi R_{Pl}^2}{g}$}\\
                 \multicolumn{4}{|c|}{$m=p_{surf}=p_{\ce{H2O}}+p_{\ce{H2O}}$}\\
                 \hline
    \end{tabular}
    \newline (1) Due to numerical reasons, we add in the code implementation mass budget correction terms to $a_1$ and $a_2$. See Sect.~\ref{sec: Stability} for a description and discussion of code stability and performance.

    \label{tab: Placeholders}
\end{table*}

To make the derivation more manageable, we introduce placeholders for several terms of the mass balance equation, which are listed in Table~\ref{tab: Placeholders}. The derivatives of the mass balance equations for \ce{H2O} and \ce{CO2} are then expressed by:
\begin{align}
\frac{d}{dt}\left(a_1+b_1\cdot g+ h\cdot g+\frac{d\cdot m}{w_1 \left(\frac{l+u}{v}\right)^{exp_2} \cdot \left(\frac{g+e}{f}\right)^{-exp_1}+1}\right) &= 0  \\ 
\frac{d}{dt}\left(a_2+b_2\cdot l+ h\cdot l+\frac{d\cdot m}{w_2 \left(\frac{g+e}{f}\right)^{exp_1}\cdot \left(\frac{l+u}{v}\right)^{-exp_2}+1}\right) &= 0
\end{align}

We point out that $m'$ is a place holder for the derivative of the total surface pressure $p_{surf}$ of the \ce{H2O}- \ce{CO2} atmosphere and thus depends on both, $F'_{\ce{H2O}}=g'$ and  $F'_{\ce{CO2}}=l'$. Its derivative is (according to the substitution we chose in Table~\ref{tab: Placeholders}):\begin{align}
   \frac{d m}{dt} & =  \frac{d}{dt} \left(\frac{g+e}{f} \right)^{exp_1}+\frac{d}{dt}\left(\frac{l+u}{v}\right)^{exp_2} \nonumber\\
   &=exp_1\, \left(\frac{g'}{f}\right) \left(\frac{g+e}{f} \right)^{exp_1-1}+exp_2\, \left(\frac{l'}{v}\right) \left(\frac{l+u}{v}\right)^{exp_2-1}.
\end{align}

We now replace $m'$ in the derivation of the mass balance equations with this resolved expression and rearrange the two mass balance equations to gather all terms that contain $F'_{\ce{H2O}}=g'$ and $F'_{\ce{CO2}}=l'$ on the right hand side and all other terms on the left hand side. We thus derive for \ce{H2O}:

\begin{align}
g' &\bigg[b_1 + h + d \cdot m \cdot 
    \frac{ \exp_1 w_1 \frac{1}{f} \cdot 
    \left(\frac{l+u}{v}\right)^{\exp_2}}
    {\left(\frac{g+e}{f}\right)^{\exp_1+1}}
    \cdot \left(\frac{w_1 \left(\frac{l+u}{v}\right)^{\exp_2}}
    {\left(\frac{g+e}{f}\right)^{\exp_1}} + 1\right)^{-2} \nonumber\\
&+ \frac{d \cdot \exp_1 \frac{1}{f} \cdot 
    \left(\frac{g+e}{f}\right)^{\exp_1-1}}
    {w_1 \left(\frac{l+u}{v}\right)^{\exp_2} 
    \left(\frac{g+e}{f}\right)^{-\exp_1} + 1}
    \bigg] \nonumber\\
&+ l' \bigg[- d \cdot m \cdot 
    \frac{\exp_2 w_1 \frac{1}{v} \cdot 
    \left(\frac{l+u}{v}\right)^{\exp_2-1}}
    {\left(\frac{g+e}{f}\right)^{\exp_1}} 
    \cdot \left(\frac{w_1 \left(\frac{l+u}{v}\right)^{\exp_2}}
    {\left(\frac{g+e}{f}\right)^{\exp_1}} + 1\right)^{-2} \nonumber\\
&+ \frac{d \cdot \exp_2 \frac{1}{v} 
    \left(\frac{l+u}{v}\right)^{\exp_2-1}}
    {w_1 \left(\frac{l+u}{v}\right)^{\exp_2} 
    \left(\frac{g+e}{f}\right)^{-\exp_1} + 1} 
    \bigg] \nonumber\\
&= -[a'_1 + g \cdot b'_1 + g \cdot h'].
\end{align}
 
and accordingly for \ce{CO2}:

\begin{align}
l' &\bigg[b_2 + h + d \cdot m \cdot 
    \frac{\exp_2 w_2 \left(\frac{1}{v}\right) 
    \left(\frac{g+e}{f}\right)^{\exp_1}}
    {\left(\frac{l+u}{v}\right)^{\exp_2+1}}
    \cdot \left(\frac{w_2 \left(\frac{g+e}{f}\right)^{\exp_1}}
    {\left(\frac{l+u}{v}\right)^{\exp_2}} + 1 \right)^{-2} \nonumber\\
&+ \frac{d \cdot \exp_2 \left(\frac{1}{v}\right) 
    \left(\frac{l+u}{v}\right)^{\exp_2-1}}
    {w_2 \left(\frac{g+e}{f}\right)^{\exp_1} 
    \left(\frac{l+u}{v}\right)^{-\exp_2} + 1} 
    \bigg] \nonumber\\
&+ g' \bigg[-d \cdot m \cdot 
    \frac{\exp_1 w_2 \frac{1}{f} 
    \left(\frac{g+e}{f}\right)^{\exp_1-1}}
    {\left(\frac{l+u}{v}\right)^{\exp_2}}
    \cdot \left(\frac{w_2 \left(\frac{g+e}{f}\right)^{\exp_1}}
    {\left(\frac{l+u}{v}\right)^{\exp_2}} + 1\right)^{-2} \nonumber\\
&+ \frac{d \cdot \exp_1 \frac{1}{f} 
    \left(\frac{g+e}{f}\right)^{\exp_1-1}}
    {w_2 \left(\frac{g+e}{f}\right)^{\exp_1} 
    \left(\frac{l+u}{v}\right)^{-\exp_2} + 1} 
    \bigg] \nonumber\\
&= -[a'_2 + l \cdot b_2' + l \cdot h'].
\end{align}

These two equations establish a set of linear equations with respect to  $g'=F'_{\ce{H2O}}$ and $l'=F'_{\ce{CO2}}$ of the form:
\begin{align}
A_{\ce{H2O}} l' +  B_{\ce{H2O}} g' &= - C_{\ce{H2O}} \\
A_{\ce{CO2}} l' + B_{\ce{CO2}} g' &=-C_{\ce{CO2}}.
\end{align}

Therefore, we can use Cramer's rule to derive unique solutions for $g'$ and $l'$ as long as $A_{\ce{H2O}}B_{\ce{CO2}} \neq A_{\ce{CO2}}B_{\ce{H2O}}$ :
\begin{align}
g'= \frac{d F_{\ce{H2O}}}{dt} &= \frac{C_{\ce{H2O}}A_{\ce{CO2}} -C_{\ce{CO2}} A_{\ce{H2O}}}{A_{\ce{H2O}}B_{\ce{CO2}}-A_{\ce{CO2}}B_{\ce{H2O}}} \label{eq: dFH2O} \\
l' = \frac{d F_{\ce{CO2}}}{dt} & = \frac{C_{\ce{CO2}}B_{\ce{H2O}} -C_{\ce{H2O}} B_{\ce{CO2}}}{A_{\ce{H2O}}B_{\ce{CO2}}-A_{\ce{CO2}}B_{\ce{H2O}}}  \label{eq: dFCO2}. 
\end{align}

Thus, we demonstrate that while the mass balance equation with the parametric equation for volatile outgassing cannot be solved for analytically with respecz to $F_{\ce{H2O}}$ and $F_{\ce{CO2}}$, their derivatives can. Therefore in \magmoc{2.0}, we add the two differential equations~(\ref{eq: F0+dFH2O}) and (\ref{eq: F0+dFCO2}) with equations~\ref{eq: dFH2O} and \ref{eq: dFCO2}. We note that the last term of the sum comprising $A_{\ce{H2O}}, \dots, B_{\ce{H2O}}$ encapsulates changes in surface pressure and also in mean molecular weight (see also Eq.~\ref{eq: mass pressure}). 

We further emphasize, similar to \citet{Bower2019}, that the derivation outlined in this section can in principle be generalized to more than two volatiles as long as it is possible to derive a set of $n$ independent linear equations for $n$ volatiles that can be solved for time derivatives of all relevant volatile mass fractions $F_{1...n}$. 
\section{A corrected gray atmosphere model for a two component H2O-CO2 atmosphere}
\label{sec: grey}

As outlined in Sect.~\ref{sec: RT}, the thermal evolution of the planet is determined by the net outgoing flux $F_{\rm net}$, that is, the difference between absorbed stellar radiation (ASR) and the thermal flux emitted by the planet also known as outgoing longwave radiation (OLR) on top of the atmosphere. 

In \magmoc{1.0} \citep{Barth2021}, the outgoing longwave radiation was calculated with a gray atmosphere model:
\begin{equation}
 F_{OLR}= \epsilon_{\ce{H2O}} \sigma T_{surf}^4,
\end{equation}
where $\sigma$ is the Stefan-Boltzmann constant, and $ \epsilon_{\ce{H2O}}$ is the emissivity of the \ce{H2O} atmosphere.  

The emissivity of a volatile $i$ can be derived from the optical depth of an atmosphere compised of volatile $i$ with the Rosseland mean opacity over the infrared wavelength range $\kappa_0(i)$ at reference pressure $p_0$. The optical depth $\tau_{i}$ further depends on surface gravity $g$ and partial pressure $p_i$ at surface as \citep[see e.g.][]{Carone2014,Elkins-Tanton2008,CatlingBook,Pierrehumbert} :
\begin{equation}
\tau_{i}=p_{i} \sqrt{0.75 \cdot\kappa_0(i)/(g\cdot p_0)} \label{eq: tau}
\end{equation}

From this relation,the emissivity $\epsilon_{i}$ follows as:
\begin{equation}
 \epsilon_i=\frac{2}{\tau_i+2} \label{eq: emissivity}.  
\end{equation}

We use this gray formalism as a bases to compare to calculations with full radiative transfer in a vertically extended atmosphere and to derive a parametric fit.

\subsection{Pure H2O atmosphere for Earth gravity}
\begin{figure}
    \centering
    \includegraphics[width=0.49\textwidth]{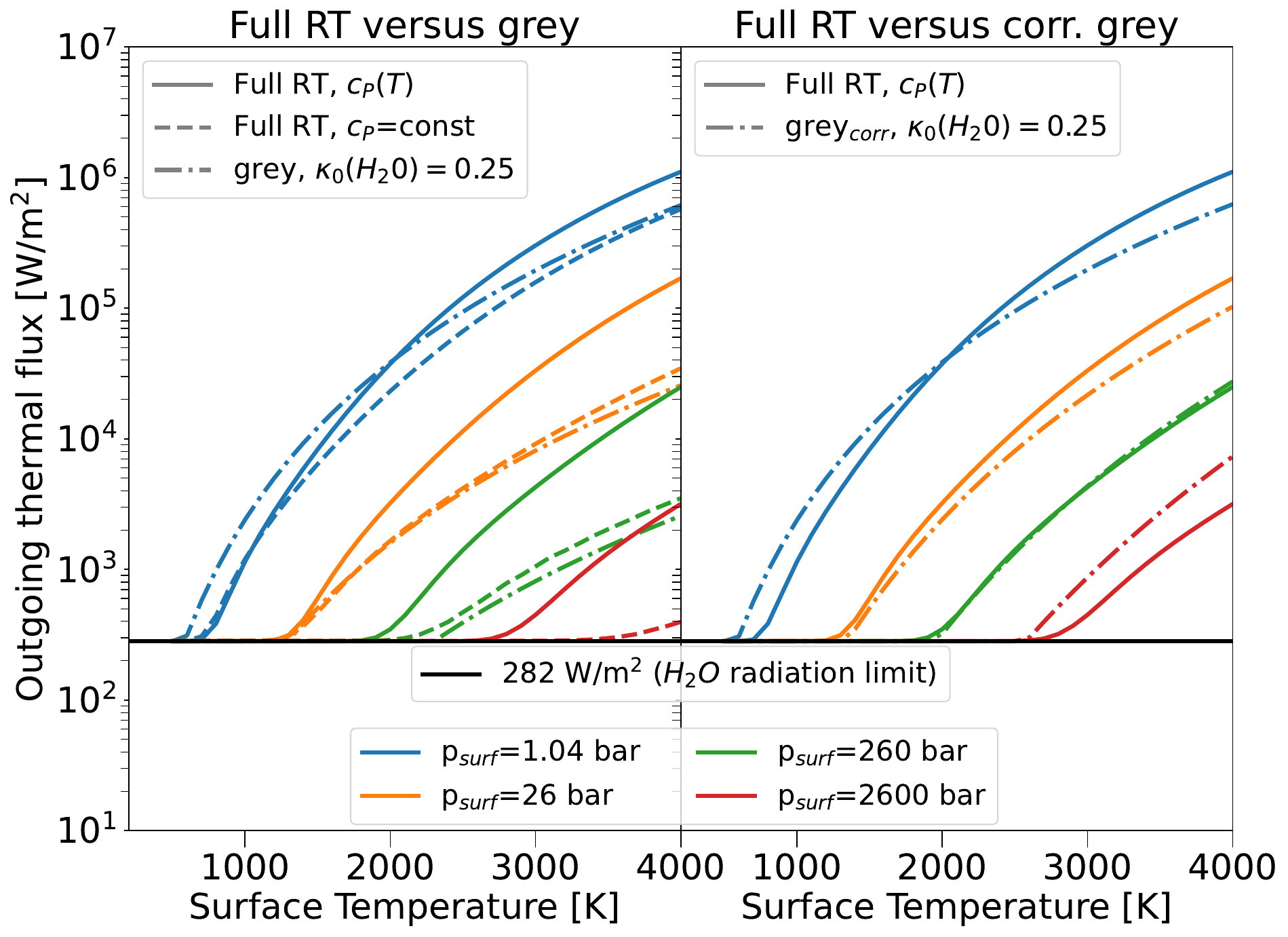}
    \caption{Outgoing thermal flux for a pure \ce{H2O} atmosphere on an Earth gravity planet ($g=9.81$~m/s${}^2$) based on the analytic gray atmosphere and full radiative transfer calculations under different assumptions. The results of the full radiative transfer calculations in a vertically extended atmosphere (Full RT) with heat capacity varying with temperature, $c_P(T)$, are used for benchmarking (solid lines). The models are compared for the same surface temperatures and pressures, where different surface pressures are indicated by variations in color. Left panel: The results of full radiative transfer calculations with temperature-dependent $c_P$ (solid lines), full radiative transfer withconstant $c_P$ (dashed lines), and the gray atmosphere model (dashed dotted lines). Right panel: The results of the corrected gray atmosphere (dashed dotted lines) compared to the full RT. }
    \label{fig: Grey_H2O}
\end{figure}

For a pure \ce{H2O} atmosphere, the gray atmosphere analytical model with $\kappa_0(\ce{H2O})=0.25$~m${}^2$/kg at $p_0=1.013$~bar and the runaway greenhouse radiation limit of $OLR_{lim}(\ce{H2O})=282$~W/m${}^2$  by prescribing (Fig.~\ref{fig: Grey_H2O}, left):
\begin{equation}
F_{OLR}(\ce{H2O}, c_P=\textrm{const})=\mathrm{max}\left(\textrm{OLR}_{lim}(\ce{H2O}),\epsilon_{\ce{H2O}} \sigma T_{surf}^4\right),    
\end{equation}
yields agreement for $T_{surf}=500 -4000$~K and $p_{surf}=1 -26000$~bar within one order of magnitude when compared to thermal emission from full radiative transfer (full RT, Sect.~\ref{sec: RT}) in a vertically extended atmosphere (Sect.~\ref{sec: PT}) with constant dry adiabatic lapse rate of $c_{p,\ce{H2O}}=37$~J mole${}^{-1}$ K${}^{-1}$.\footnote{Using $\kappa_0(\ce{H2O})=0.01$~m${}^2$/kg as in \citet{Elkins-Tanton2008,Barth2021} yields larger deviations (not shown).}

\begin{figure}
    \centering
    \includegraphics[width=0.49\textwidth]{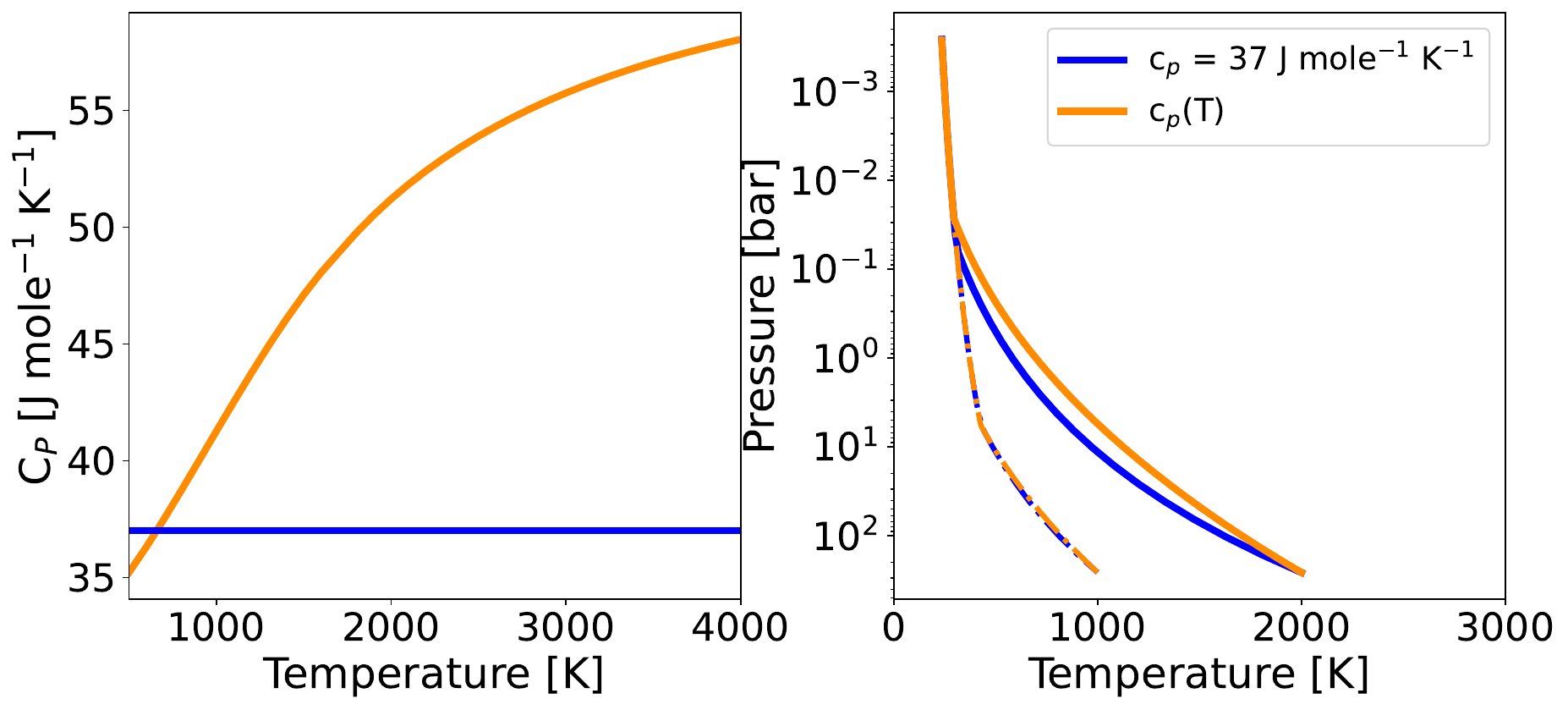}
    \caption{Left: Specific heat capacity versus temperature for constant $c_P$ (blue) and $c_P(T)$ (dark orange). Right: \ce{H2O} pressure-temperature profiles for $p_{surf}=260$~bar and $T_{surf}=1000$~K (dashed-dotted) and 2000~K (solid). Profiles assuming constant $c_P$ are shown in blue and profiles assuming temperature dependent $c_P$ are shown in dark orange. Note that for a 1000 K surface, the difference between the profiles is negligible. }
    \label{fig: H2O_PT}
\end{figure}

The heat capacity of water is, however, not constant but increases strongly for $T>1000$~K, which results in a steeper gradient for temperature profiles for the early stages of the magma ocean evolution (Fig.~\ref{fig: H2O_PT}). The steeper temperature-pressure gradient results in hotter upper atmospheric layers and consequently higher outgoing thermal fluxes (Fig.~\ref{fig: Grey_H2O}, left). We can account to first order for increased thermal emission with larger surface temperatures and pressures (Fig.~\ref{fig: Grey_H2O}, right) by adopting an emission correction term:
\begin{equation}
E^*(\ce{H2O}) =  \left(\frac{T_{surf}}{1500 K}\right)^{log_{10}\left(\frac{p_{\ce{H2O}}}{1bar}\right)}. 
\end{equation}

The corrected gray atmosphere model for a pure \ce{H2O} is then described with:
\begin{equation}
F_{OLR}(\ce{H2O})=\mathrm{max}\left(\textrm{OLR}_{lim}(\ce{H2O}),\epsilon_{\ce{H2O}} \sigma T_{surf}^4 E^*(\ce{H2O}) \right) \label{eq: H2O grey corr}.    
\end{equation}

\subsection{Pure CO2 atmosphere for Earth gravity}

\begin{figure}
    \centering
\includegraphics[width=0.49\textwidth]{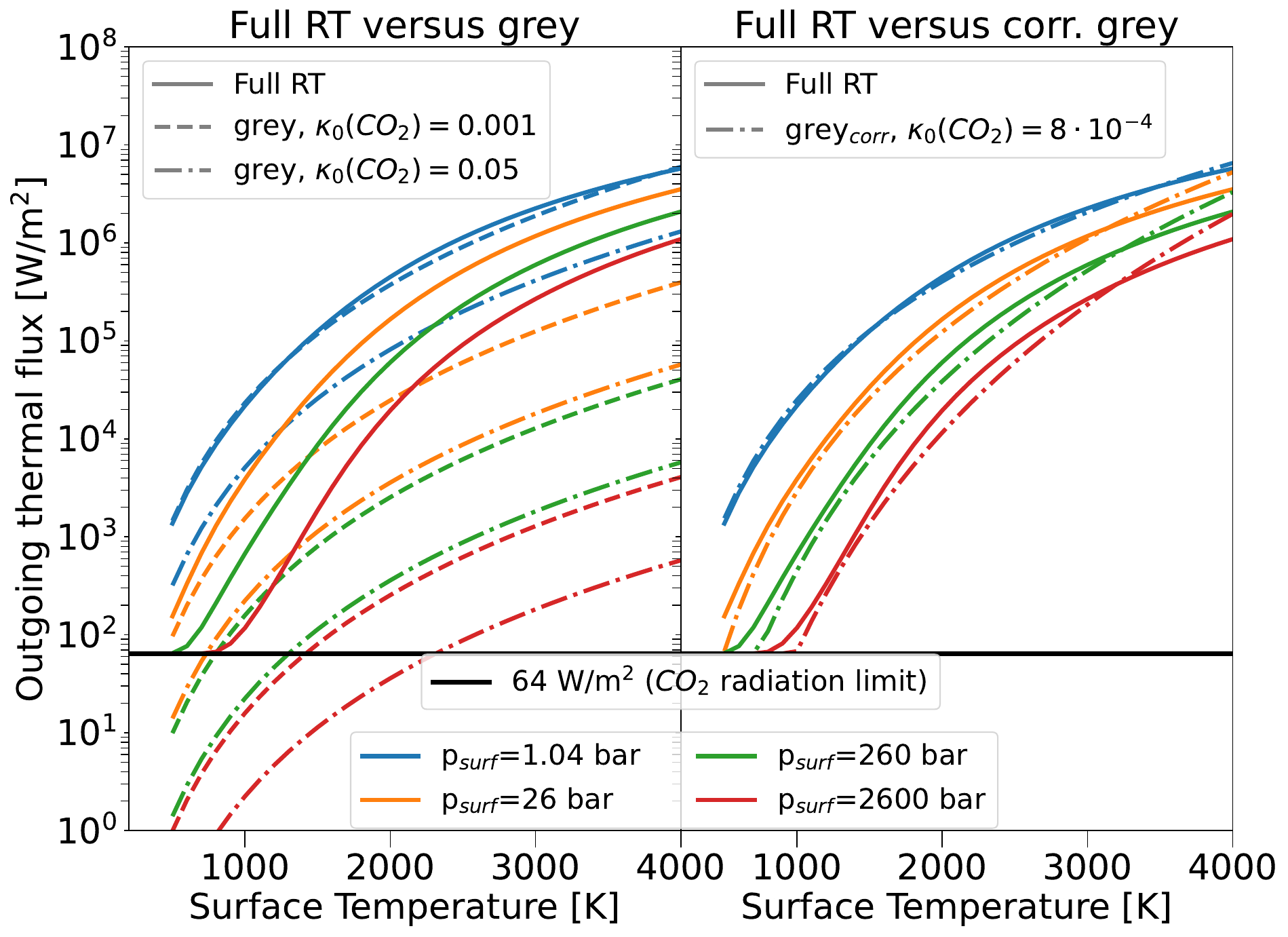}
    \caption{Outgoing thermal flux for a pure \ce{CO2} atmosphere on an Earth gravity planet ($g=9.81$~m/s${}^2$) are shown for gray atmosphere model with different assumptions.  The results of the full radiative transfer in a vertically extended atmosphere (``Full RT'') is used for benchmarking (solid lines). The models are compared for the same surface temperatures and pressures, where different surface pressures are indicated by variations in color.  Left panel: The results of the gray atmosphere model is shown for $\kappa_0(\ce{CO2})=0.001$~m${}^2$/kg (dashed lines) and $\kappa_0(\ce{CO2})=0.05$~m${}^2$/kg (dashed-dotted lines). Right panel: The results of the corrected gray atmosphere is shown (dashed dotted lines) in comparison to the Full RT calculations.  }
    \label{fig: Grey_CO2}
\end{figure}

For a pure \ce{CO2} atmosphere, we find even larger disagreement between thermal emission based on the gray atmosphere model and the full radiative transfer calculations (Fig.~\ref{fig: Grey_CO2}, left). The gray atmosphere model ($\kappa_0(\ce{CO2})=0.001$~m${}^2$/kg  at $p_0=1.013$~bar) yields only good agreement for $p_{surf}=1$~bar and underestimates the OLR by orders of magnitudes for higher surface pressures. Assuming $\kappa_0(\ce{CO2})=0.005$~m${}^2/$kg at $p_0=1.013$~bar, as used in \citet{Lebrun2013,Elkins-Tanton2008}, yields even worse results and underestimates the OLR by at least one order of magnitude also for $p_{surf}=1$~bar.

The following analytical prescription is, however, able to mimic the outgoing long wave radiation (OLR) from the full radiative transfer calculations in a vertically extended pure \ce{CO2} atmosphere to good accuracy for surface pressures between $1-26000$~bar and $T_{surf}=500 -4000$~K. Here, we set the \ce{CO2} radiation limit $OLR_{lim}(\ce{CO2})=64$~W/m${}^2$ (Fig.~\ref{fig: Grey_CO2}, right):
\begin{equation}
F_{OLR}(\ce{CO2})=\mathrm{max}\left(OLR_{lim}(\ce{CO2}),\epsilon^{*}_{\ce{CO2}} \sigma T_{surf}^4\cdot E^{*}\right)\label{eq: CO2_emission},        
\end{equation}

and use an adapted optical depth for \ce{CO2} and partial pressure $p_{\ce{CO2}}$:
\begin{equation}
\tau^{*}_{\ce{CO2}}=\left(\frac{p_{\ce{CO2}}}{p_{ref}(\ce{CO2})}\right)^{0.75}\cdot p_{ref}(\ce{CO2}) \sqrt{0.75 \cdot\kappa_0(\ce{CO2})/(g\cdot p_0)}, \label{eq: tau_CO2}
\end{equation}
where  $p_{ref}(\ce{CO2})$=1~bar is the reference pressure for the fit. We then derive $\epsilon_{\ce{CO2}}^{*}$  via Eq.~(\ref{eq: tau}) by substituting $\tau_{\ce{CO2}}$ with $\tau^{*}_{\ce{CO2}}$.

As in the pure \ce{H2O}-atmosphere case, the steepness of the pressure temperature profile for large surface temperatures in a \ce{CO2} atmosphere requires a correction term for the emission, which we fit as:
\begin{equation}
 E^{*}(\ce{CO2})= \left(\frac{T_{surf}}{1200 K}\right)^{log_{10}\left(\frac{p_{\ce{CO2}}}{1bar}\right)}.
\end{equation}

\subsection{Mixed H2O-CO2 atmosphere for Earth gravity}
\begin{figure}
    \centering
    \includegraphics[width=0.49\textwidth]{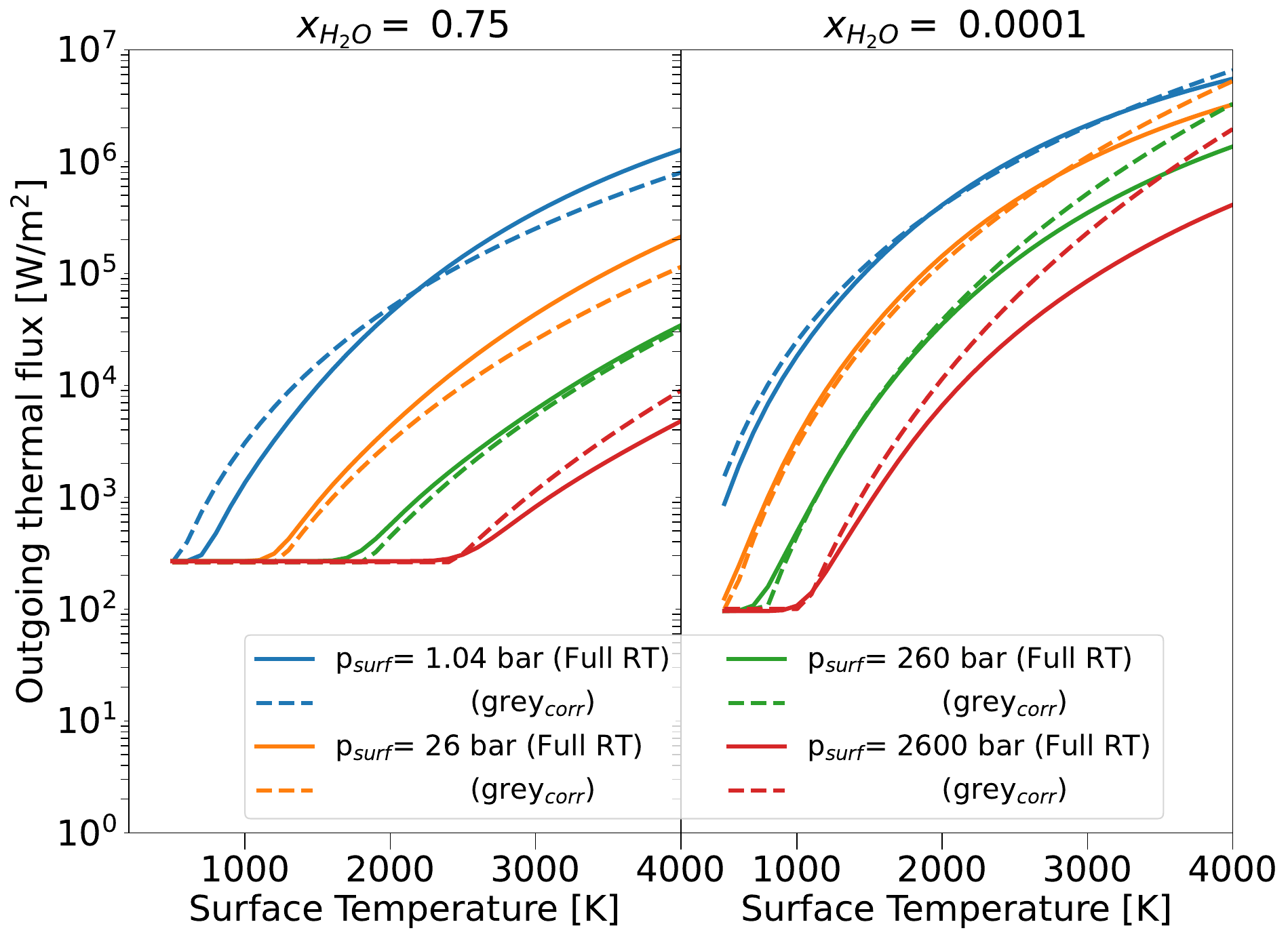}
    \caption{ Outgoing thermal flux for a mixed \ce{H2O}-\ce{CO2} atmosphere on an Earth gravity planet ($g=9.81$~m/s${}^2$)  based on the corrected gray atmosphere (dashed lines) and full radiative transfer calculations  with vertically extended atmosphere (solid lines) for the same surface temperatures and pressures, respectively. Different surface pressures are indicated by variations in color. Results are shown for two examples of a two component atmosphere with water volume mixing ratio  $x_{\ce{H2O}}=0.75$ (left panel) and $x_{\ce{H2O}}=10^{-4}$ (right panel), respectively.}
    \label{fig: Mixed_Emission}
\end{figure}

For mixed \ce{H2O}-\ce{CO2} atmospheres, the following parametric  approximation yields thermal emission within one order of magnitude compared to the full radiative transfer model (Fig.~\ref{fig: Mixed_Emission}) for $T_{surf}=500$~K~-~$ 4000$~K and $p_{surf}=1$~bar~-~$2600$~bar using: 
\begin{equation}
F_{\rm{OLR}}(x_{\ce{H2O}})=\mathrm{max}\left(\rm{OLR}_{lim}(x_{\ce{H2O}}),\epsilon^{*}_{Mix} \sigma T_{surf}^4\cdot E_{Mix}^{*}\right)\label{eq: Mixed_emission},        
\end{equation}
with
\begin{equation}
\epsilon^{*}_{Mix} = \frac{2}{\tau_{Mix}+2},
\end{equation}

where $\tau_{Mix}$ is the sum of the modified optical depths derived in the previous two subsections:
\begin{equation}
\tau_{Mix} =\tau_{\ce{H2O}}+\tau^{*}_{\ce{CO2}}.
\end{equation}

\begin{figure}
    \centering
    \includegraphics[width=0.49\textwidth]{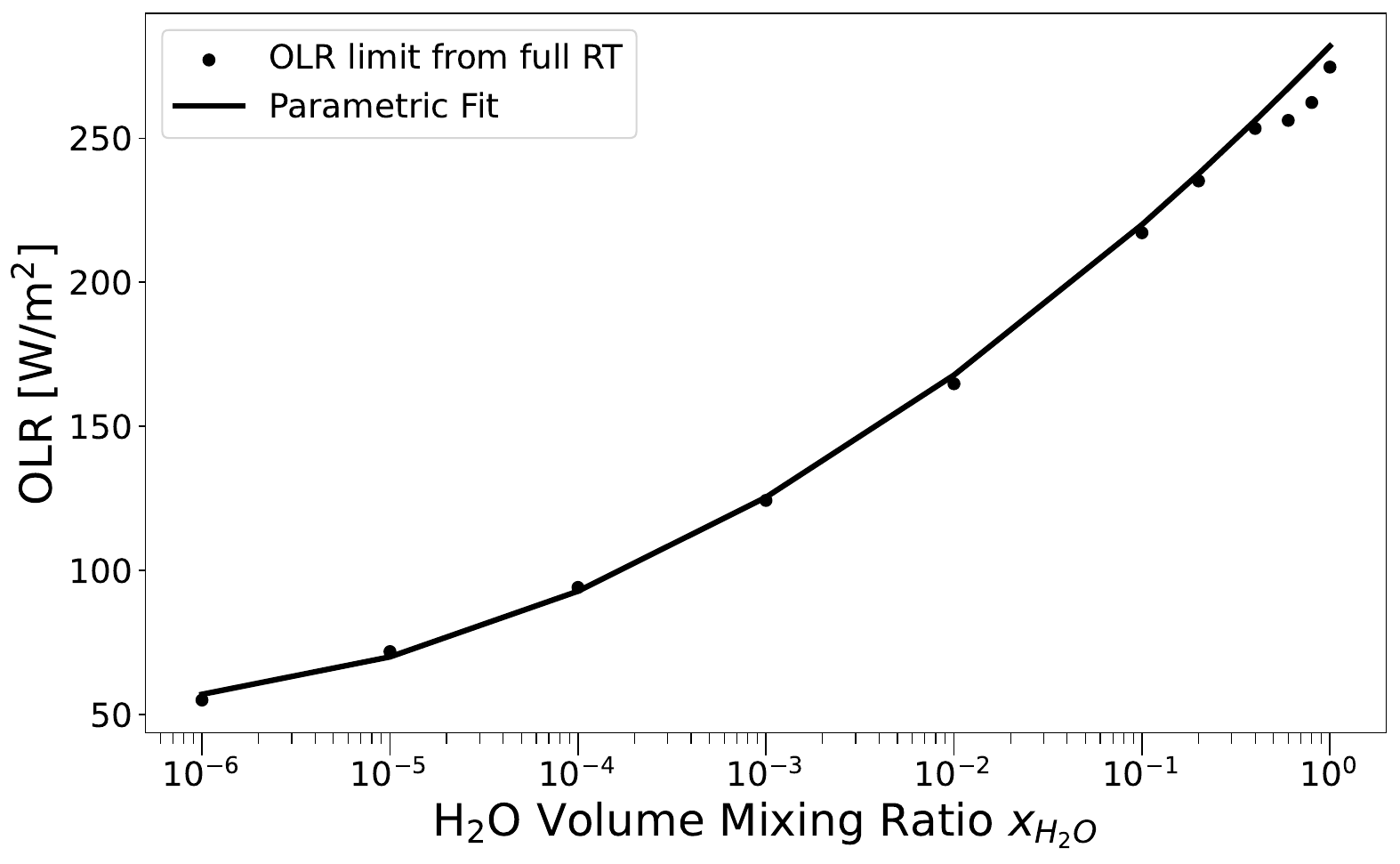}
    \caption{OLR limit for Earth with a two component \ce{H2O}-\ce{CO2} atmosphere for different \ce{H2O} volume mixing ratios using full radiative transfer (circles) versus the parametric fit as defined in Equation~(\ref{eq: Param_Fit}) (solid line). }
    \label{fig: OLRLimit}
\end{figure}

The runaway greenhouse radiation radiation, or \ce{H2O} OLR limit, decreases when \ce{CO2} is added to a \ce{H2O}-dominated atmosphere,  as already pointed out by \citet{Goldblatt2013}. We systemtically investigate this effect for a vertically extended \ce{H2O}-\ce{CO2} atmosphere. We find that the following fit to the \ce{H2O} OLR limit ${OLR}_{Lim}(x_{\ce{H2O}})$ with respect to volume mixing ratio ($x_{\ce{H2O}}$) and for Earth surface gravity ($g=9.81$~m/s${}²$) yields agreement to first order (Fig.~\ref{fig: OLRLimit}):
\begin{equation}
   \rm{OLR}_{Lim}(x_{\ce{H2O}})= 4.9\cdot \log_{10}(x_{\ce{H2O}})^2+66.9\cdot \log_{10}(x_{\ce{H2O}})+282, \label{eq: Param_Fit}
\end{equation}
where $x_{\ce{H2O}}\leq 10^{-6}$ is set to be identical to a pure \ce{CO2} atmosphere. That is, we assume for such low volume mixing ratios of water in a \ce{CO2}-dominated atmosphere that the water contribution can be neglected (within 10\% accuracy) when simulating the thermal evolution of the magma ocean. See also Fig.~\ref{fig: PT_profiles} which shows that both, the pressure-temperature profile as well as the emission are very close to that of a 100\% \ce{CO2} atmosphere for $x_{\ce{H2O}}=10^{-5}$.

We adopt the following correction in emission, using the volume fraction of \ce{H2O}, $x_{\ce{H2O}}=p_{\ce{H2O}}/p_{surf}$, in a two-component atmosphere set-up with $p_{\ce{H2O}}+p_{\ce{CO2}}=p_{surf}$:
\begin{align}
E_{Mix}^{*}= &\left(\frac{T_{surf}}{1200 K}\right)^{log_{10}\left(\left(\frac{p_{surf}}{1bar}\right)\cdot (1-x_{\ce{H2O}})+x_{\ce{H2O}}\right)}\cdot (1-x_{\ce{H2O}}) +\nonumber\\
        &  \left(\frac{T_{surf}}{1500 K}\right)^{log_{10}\left(\left(\frac{p_{surf}}{1bar}\right)\cdot x_{\ce{H2O}}+ (1-x_{\ce{H2O})}\right)}\cdot x_{\ce{H2O}}
\end{align}

Two examples of the resulting grid emission and corrected gray emission are shown in (Fig.~\ref{fig: Mixed_Emission}), using $\kappa_0(\ce{CO2})=0.0008$~m${}^2/$kg and $\kappa_0(\ce{H2O})=0.25$~m${}^2/$kg at $p_0=1.013$~bar. These show that the corrected gray atmosphere model yields agreement to first order to the full radiative transfer calculations even for very different atmosphere mixing. We note that this two-component \ce{H2O}-\ce{CO2} atmosphere model also encapsulates 100\% \ce{H2O} and \ce{CO2} compositions by setting $x_{\ce{H2O}}$ equal to $1$ or $0$, respectively. 

\subsection{Generalization of gray atmosphere for rocky exoplanets}
\label{sec: Grey_exo}

So far, we have applied the radiative transfer and corrected gray atmosphere model to Earth. In principle, the latter can also be applied to any rocky planet with a different surface gravity $g$.

\begin{figure}
    \centering
    \includegraphics[width=0.49\textwidth]
    {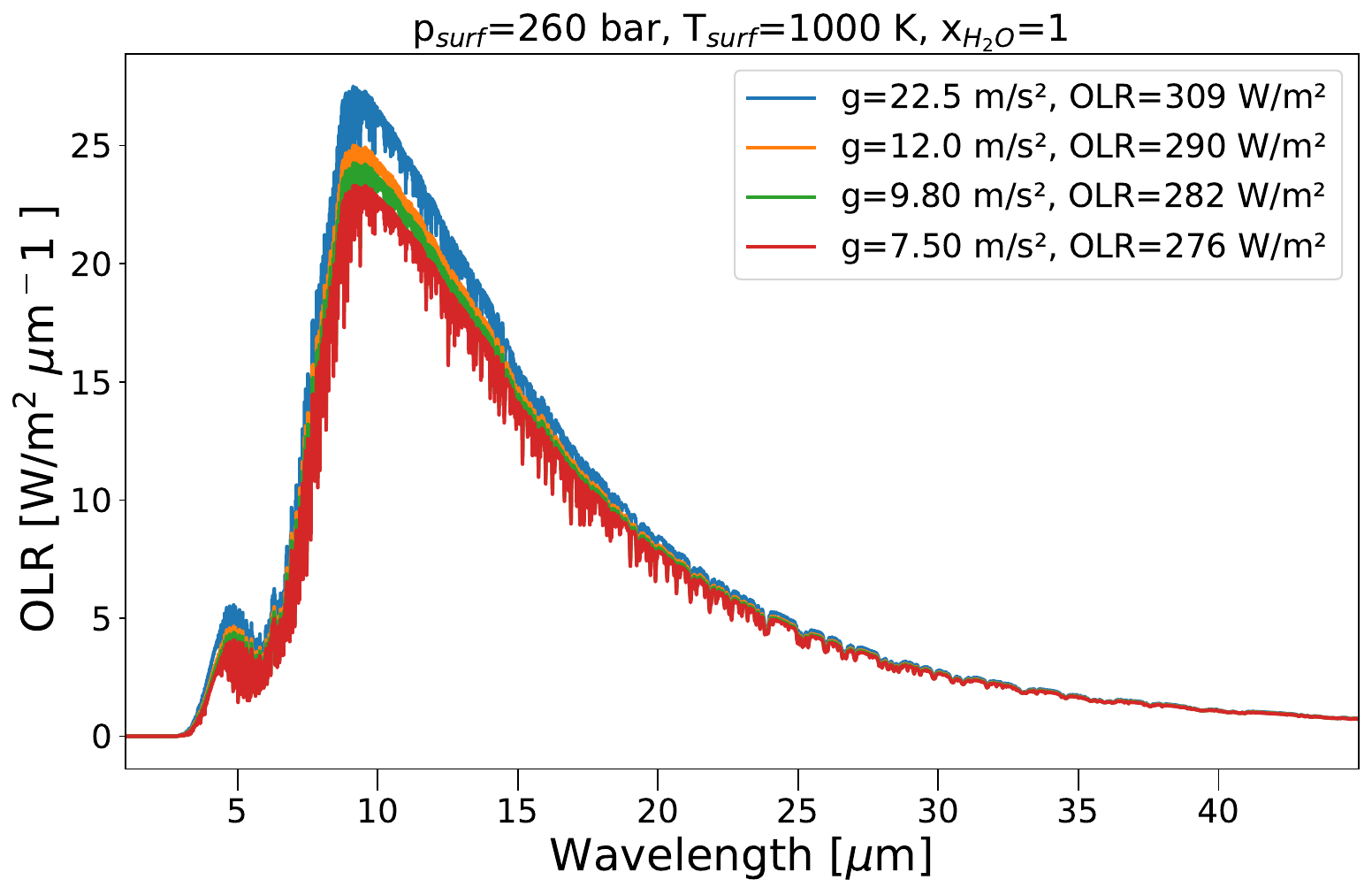}\\
        \includegraphics[width=0.49\textwidth]{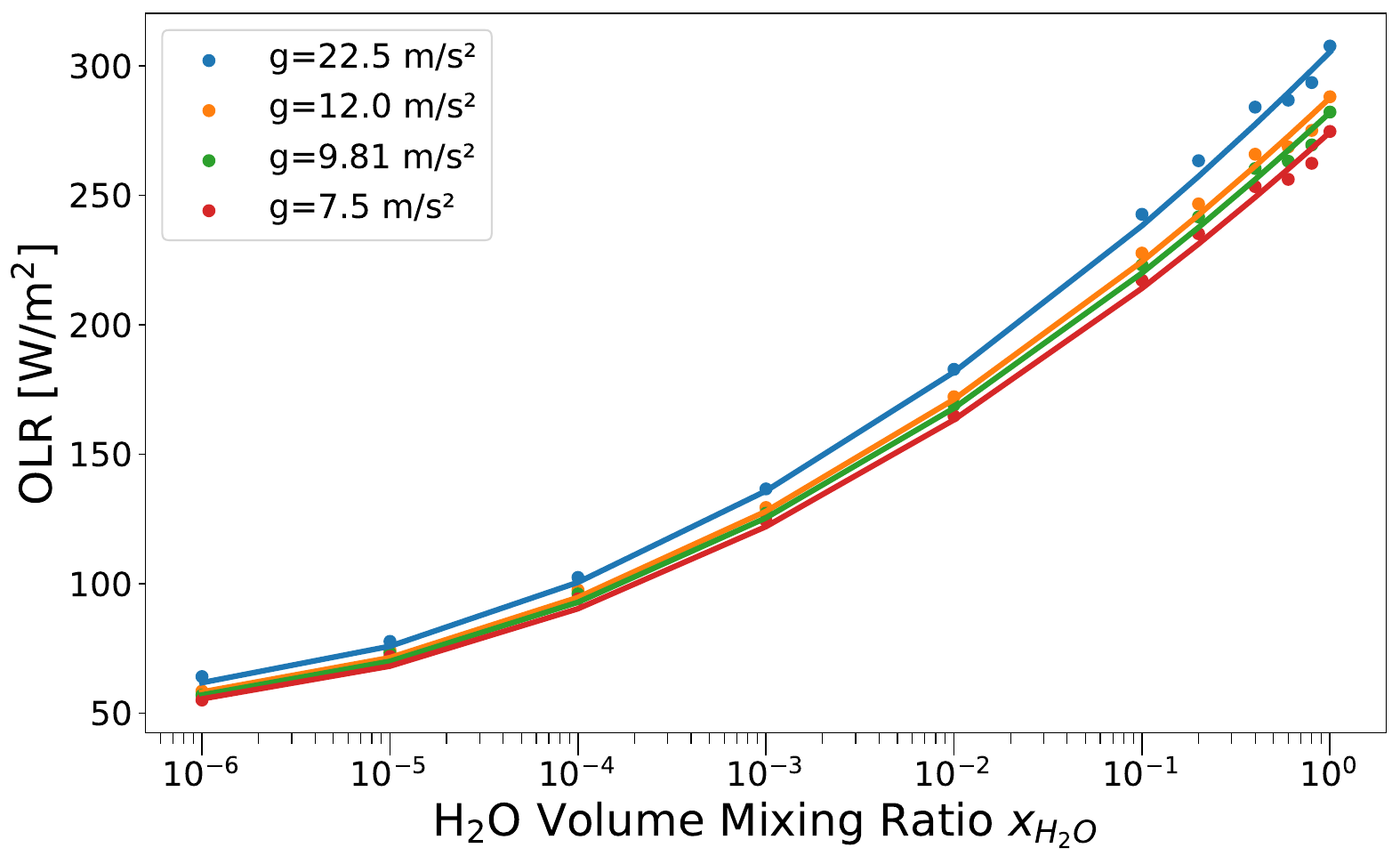}
    \caption{Left panel: Emission versus wavelength for a pure \ce{H2O} atmosphere in the runaway greenhouse limit for different surface gravities. Right panel: OLR limit for a two component \ce{H2O}-\ce{CO2} atmosphere and different surface gravities (circles) including the adjusted parametric fits (solid lines) using Equation~(\ref{eq: Paramfit_g}). }
    \label{fig: H2O emission gravity}
\end{figure}

The pressure-temperature profiles will not change with different surface gravities for a given atmospheric composition and surface temperature and pressure set by the hot magma ocean, because the atmospheric lapse rate is determined either by $R/c_{p,i}$ when the vertical coordinate is chosen, where $R$ is the ideal gas constant and $c_{p}(T)$ is the combined specific heat capacity of the volatiles in the atmosphere or by latent heat release $L_i$. Thus, pressure-temperature profiles for rocky planets with the same atmospheric composition, surface temperature, and surface pressure are in this set-up identical for different surface gravities. The emission arising from the top of the atmosphere, however, will still change because the planet's opacity depends on surface gravity (Eq.~\ref{eq: tau}). Consequently, also the runaway greenhouse limit changes with surface gravity (Fig.~\ref{fig: H2O emission gravity}, left).

We find that the runaway greenhouse radiation limit for different surface gravities can be parameterized with a simple adjustment to Equation~(\ref{eq: Param_Fit})  (Fig.~\ref{fig: H2O emission gravity}, right):

\begin{equation}
{\rm{OLR}}_{\rm{Lim}}(x_{\ce{H2O}},\rm{g})={\rm{OLR}}_{\rm{Lim}}(x_{\ce{H2O}})\cdot \left(1+\ln\left(\frac{g}{g_{\rm{Earth}}}\right)\cdot 0.1\right).\label{eq: Paramfit_g}
\end{equation}

\section{Validation of \magmoc{2.0} for an oxidized Earth}
\label{sec:Earth}

We apply our magma ocean to the Earth regime to compare with previous results of \citet{Barth2021} who also benchmarked \magmoc{1.0} for Earth-like conditions for comparison with \citet{Hamano2013} and \citet{Elkins-Tanton2008}. We further exploit the versatility of \magmoc{2.0} to test three different atmosphere models and their impact on the thermal evolution of the mag,a ocean: the gray formalism as used in \magmoc{1.0}, an improved gray formalism, and a model based on full radiative transfer calculations in a vertically extended atmosphere. Henceforth, we refer to the first model as the "gray" model, to the second as "corrected gray" model and the third as the "RT" model. Moreover, we investigate the impact of various \ce{H2O} and \ce{CO2} outgassing laws. Specifically, we compare  the laws used by \citet{Elkins-Tanton2008} and \citet{Niko2019}, respectively. Finally, we assess the impact of adding \ce{CO2} to a water-dominated atmosphere for our simulations of the oxidized Earth magma ocean.

In the following, we begin by evaluation the role of the different atmosphere models for pure \ce{H2O} and \ce{CO2} atmospheres, respectively. Subsequently, we compare the outcome of simulations using the different atmosphere models for mixed \ce{H2O}-\ce{CO2} atmospheres. For these mixed scenarios, we adopt as in \citet{Barth2021}, an initial magma ocean depth of 2000~km, consistent with \citet[][]{Elkins-Tanton2008}, one of the first models to investigate magma oceans with outgassing of both, \ce{H2O} and \ce{CO2}.  Next, we assess the feedback between \ce{H2O} and \ce{CO2} outgassing and compare to the results of \citet{Bower2019}.

\subsection{Pure H2O atmosphere}
\label{sec: Earth - pure H2O}

For the pure \ce{H2O} atmosphere, we follow \citet{Barth2021} by simulating the evolution of the Earth magma ocean with an initial water content of 5~TO for comparison with \citet{Hamano2013}. We use the \ce{H2O} solubility law of \citet{Schaefer2016} and begin with fully molten mantle down to the core ($R_{Core}=3400$~km), which gives an initial magma ocean depth of $2978$~km. We test \magmoc{2.0} with the gray atmosphere model and  $\kappa_0(\ce{H2O})=0.01$~m${}^2$/kg, the corrected gray atmosphere model (Sect.~\ref{sec: grey} and $\kappa_0(\ce{H2O})=0.25$~m${}^2$/kg), and the RT atmosphere model (Sect.~\ref{sec: RT}). 

\begin{figure*}
    \centering
    \includegraphics[width=0.95 \textwidth]{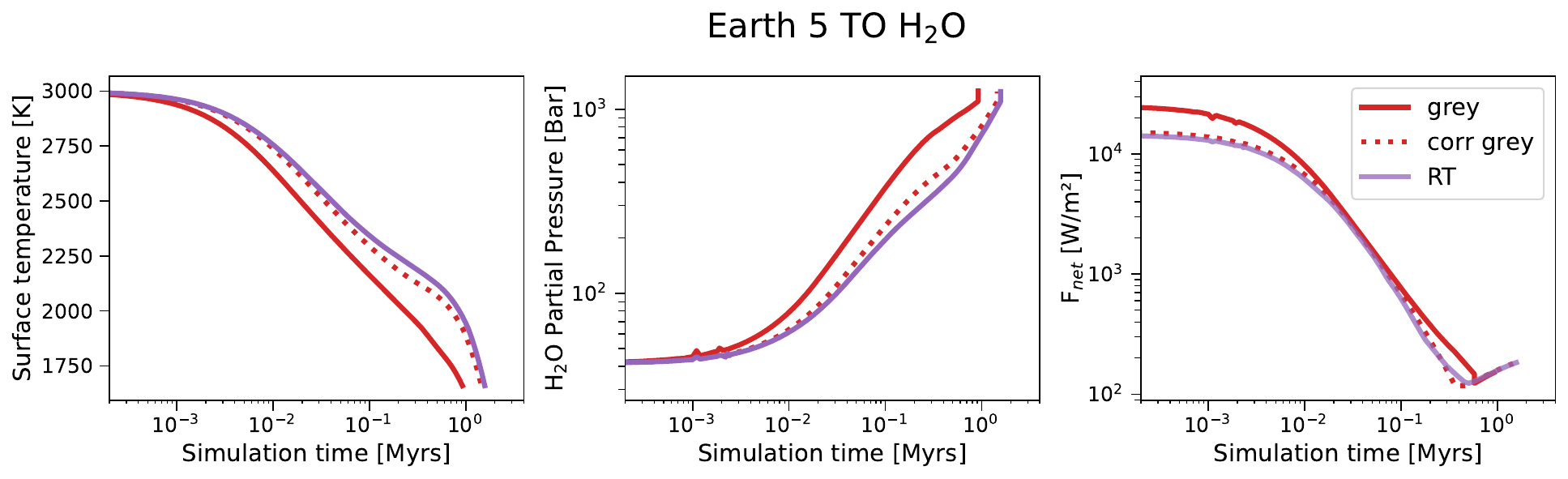}
    \includegraphics[width=0.95 \textwidth]{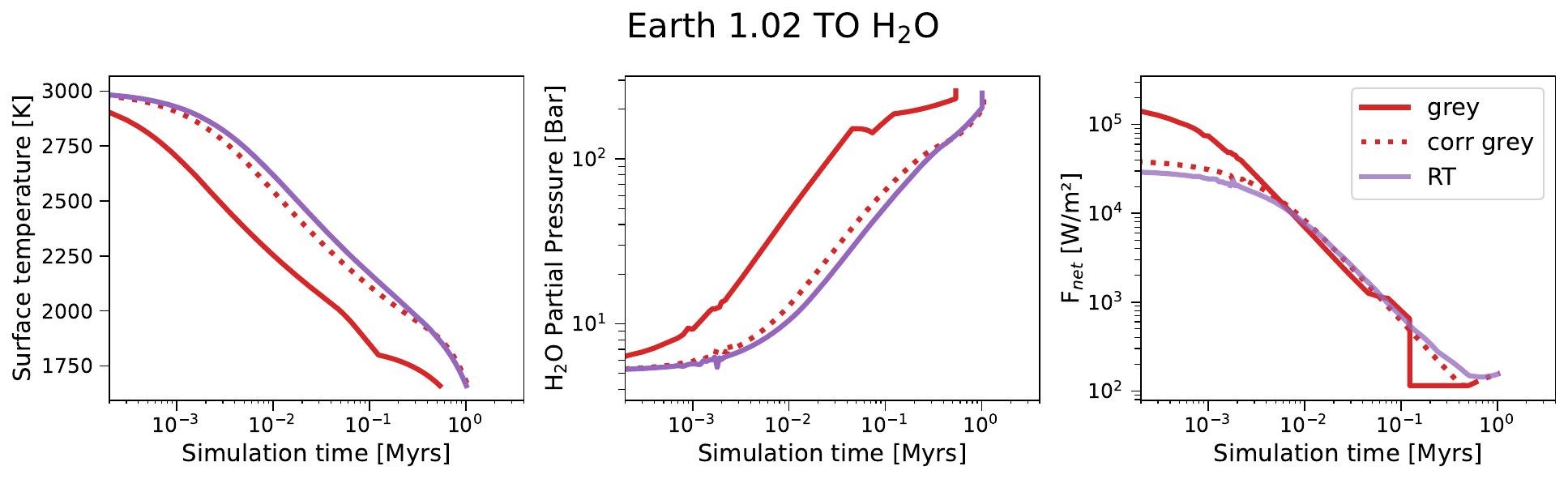}
        \caption{\magmoc{2.0} simulations for Earth and an initial water content of 5~TO (top panel) and 1.02~TO (bottom panel). Depicted from left to right in each panel are: Surface temperature, \ce{H2O} atmosphere surface pressure and net flux at the top of the atmosphere ($F_{OLR}-F_{ASR}$).  Three different atmosphere models are used: The gray atmosphere model (solid red line), the corrected gray model (dotted red line), and the RT model (solid purple line).}
    \label{fig: Earth_5TO}
\end{figure*}

Figure~\ref{fig: Earth_5TO} (top) illustrates that the gray atmosphere model yields solidification after 0.9~Myrs as expected \footnote{We have verified that \magmoc{1.0} and \magmoc{2.0} always yield the same results for the pure \ce{H2O} outgassing mode and the gray atmosphere model.} from \citet{Barth2021}. The solidification time of 0.9 Myrs is notably shorter than the Earth magma ocean duration of 4~Myrs reported by \citet{Hamano2013} for the same initial water mass. The simulations with the corrected gray and full radiative transfer (RT) atmosphere model yield longer solidification times of 1.5~Myrs, aligning more closely to \citet{Hamano2013}. The magma ocean simulations reach across all three atmosphere models the runaway greenhouse radiation limit, $OLR_{lim}(1)=282$~W/m$^{2}$, towards the end of the magma ocean evolution. As the simulations progress further, the net flux increases because the Sun's bolometric flux and thus the absorbed stellar radiation (ASR) has substantially decreased after 1~Myrs, whereas the planet's OLR does not change significantly. We note that the gray atmosphere model consistently assumes $OLR_{lim}(1)=282$~W/m$^{2}$ for $T_{surf} \leq 1800$~K, irrespective of surface pressure and prior OLR evolution. This assumption can result in an abrupt drop in net flux, as is evident in Fig.~\ref{fig: Earth_5TO} (bottom panel).

We further compare our results to the Earth magma ocean simulations of \citet{Niko2019}, who also investigated differences between a gray atmosphere model and a full radiative transfer model for the pure \ce{H2O} atmosphere. However, these authors adopted an initial water content of about 1.0~TO \ce{H2O}. For better comparison with this work and that of \citet{Elkins-Tanton2008}, we adopt an initial \ce{H2O} melt fraction of $0.05$ in a 2000~km deep magma ocean, which is equivalent to 1.02~TO initial \ce{H2O} mass and aligns with the magma ocean depth of the model by \citet{Elkins-Tanton2008}. These simulations  yield shorter solidification times ranging between 0.8 and 1~Myrs compared to the 5~TO \ce{H2O} simulation, where once again the gray model results again in the shortest solidification time (Fig.~\ref{fig: Earth_5TO}, bottom).

In both the 5~TO and 1.02~TO \ce{H2O} cases, we consistently find that the OLR of the gray model is higher during the earliest evolution stages (within the range of $10^3-10^4$~years)  compared to the corrected gray and RT atmosphere model. The higher inital OLR of the gray atmosphere model leads to faster overall cooling, thereby explaining the shorter solidification times. A similar deviation in OLR evolution between a gray and RT atmosphere model was reported by \citet[][Figure~3]{Niko2019}. Consequently, they found that in their simulation with a gray atmosphere model that the magma ocean solidified after several 0.1~Myrs, whereas the simulation using the line-by-line radiative transfer model of \citet{Katyal2019} yielded a solidification time of 1 Myrs, in good agreement with our 1.02~TO \ce{H2O} simulations. These results confirm that the magma ocean lifetime for a pure \ce{H2O} atmosphere can differ by several 0.1~million years across different models and  atmosphere assumptions \citep{Niko2019}. 

However, our simulations show here that the corrected gray and full radiative transfer yield evolution tracks that are in good agreement with each other in our magma ocean model. We further find agreement between the OLR of our RT model and the radiative transfer calculations of \citet{Lichtenberg2021} for a pure \ce{H2O} atmosphere with $T_{surf}=500 - 3000$~K and $p_{surf}=1$~bar and 260~bar. These authors similarly reproduce a magma ocean lifetime of about 1~Myrs with a pure \ce{H2O} atmosphere, where we note different definitions for solidification times (see Sect.~\ref{sec: Simulations}). Furthermore,  \citet{Lichtenberg2021} have a cooler initial temperature of 3000~K compared to our initial temperature of 4000~K \citep{Hamano2013,Schaefer2016,Niko2019,Barth2021}. The general agreement between our pure \ce{H2O} simulations and that of \citet{Lichtenberg2021} is not surprising, given the usage of very similar \ce{H2O} opacities (Table~\ref{t:opacities}) and vertically extended atmospheres for the RT calculations. 

\subsection{Pure CO2 atmosphere}
\label{sec: Earth - pure CO2}

We simulate the pure \ce{CO2} outgassing scenario for Earth following \citet{Elkins-Tanton2008}. Consequently, we assume an initial magma ocean depth of 2000~km and a \ce{CO2} outgassing law with initial saturation in the melt (Sect.~\ref{sec: outgas}). The simulation assumes an initial \ce{CO2} melt fraction of 0.6, which corresponds to 14.7 TO \ce{CO2}.

\begin{figure*}
    \centering
    \includegraphics[width=0.95\textwidth]{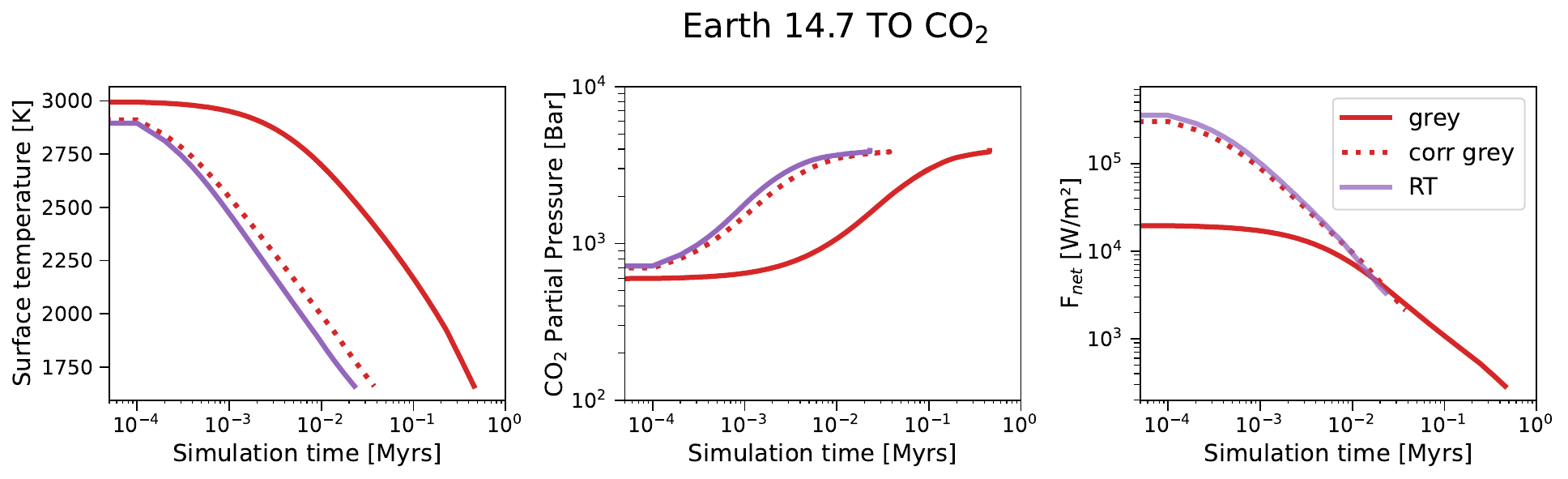}
    \caption{Simulation of \texttt{MagmOcV2.0} for Earth and 14.7~TO \ce{CO2} content for a 2000~km deep magma ocean. Depicted from left to right: Surface temperature, \ce{H2O} atmosphere surface pressure, and net flux at the top of the atmosphere ($F_{OLR}-F_{ASR}$) for three different atmosphere models. These models are the gray model without any correction (solid red line), the corrected gray model (dotted red line), and the radiative transfer adaptation (solid purple line).}
    \label{fig: Earth_CO2}
\end{figure*}

Figure~\ref{fig: Earth_CO2} demonstrates the thermal evolution of our Earth magma ocean simulations with a dense \ce{CO2} atmosphere. We find that solidification occurs after 0.5~Myrs when the gray model is used, in qualitative agreement with the solidification time of 0.8~Myrs reported by \citet{Elkins-Tanton2008}.  Table~\ref{Tab_Results_Earth} also illustrates general agreement between the final volatile budget and atmospheric pressure in \magmoc{2.0} compared to the results of \citet{Elkins-Tanton2008}. The agreement between both models in volatile reservoirs and solidification times with the gray atmosphere model, provides once again confidence that our model, despite its simplification,  captures the magma ocean accurately enough for comparison with more complex magma ocean models.

\begin{figure}
    \centering
    \includegraphics [width=0.49\textwidth]{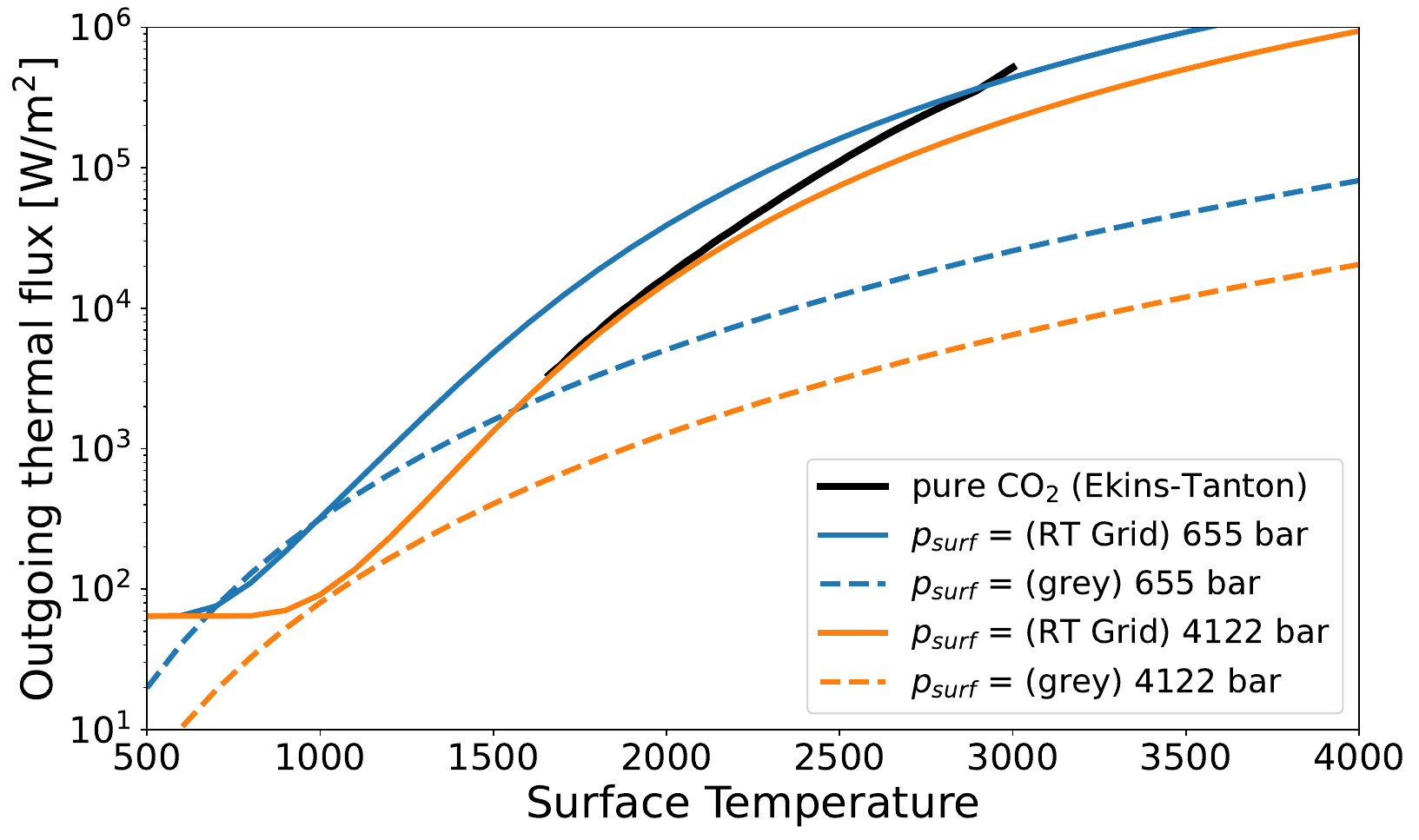}
   \caption{Pure \ce{CO2} Earth magma ocean simulation with 14.7 TO of initial \ce{CO2} mass following \citet{Elkins-Tanton2008} and using the RT atmosphere model (solid black line). Two RT thermal emission grid lines for $p_{surf}=655$~bar (blue solid line) and $p_{surf}=4122$~bar (orange solid line) are shown, respectively,  as specified in Sect.~\ref{sec: RT}. The corresponding thermal emission for the gray atmosphere model as used in \citet{Elkins-Tanton2008} is shown for comparison (dashed lines). The thermal emission deviates between the RT and gray atmosphere model by up to two orders of magnitude during the magma ocean stage in this case with $T_{surf}=1500$-~$3000$~K. }
    \label{fig: Pure_CO2Earth}
\end{figure}

However, significant differences in solidification time compared to the gray atmosphere model arise in simulations that use the corrected gray and RT atmosphere model. More precisely, the solidification time decreases by one order of magnitude  to approximately 20~000 - 30~000 years. This discrepancy stems from the fact that the nominal gray model neglects the vertical extent of the atmosphere. The large extension of a hot, dense \ce{CO2} atmosphere is evident from Fig.~\ref{fig: H2O_PT} and also supported by \citet[][Fig. 3]{Lichtenberg2021}. A pure \ce{CO2} atmosphere has a much steeper temperature gradient in the troposphere compared to \ce{H2O}, resulting in hotter atmosphere layers that contribute to the thermal emission on top of the atmosphere. Therefore, a particularly high thermal flux occurs on top of a \ce{CO2}-dominated atmosphere for the initial magma ocean stage with high surface temperatures ($T_{surf}\geq 2000$~K) and pressures ($p_{surf}>1$~bar), which are not captured with a gray model (Sect.~\ref{sec: grey}). 

Figure~\ref{fig: Pure_CO2Earth} provides a detailed illustration of the strong deviation in emission between the gray model and the full radiative transfer (RT) grid. The RT grid model yields OLR~$>10^5$~W/m${}^2$ for $T_{surf}>2500$~K, whereas the gray model yields for the same temperature and pressure range OLR~$<10^5$~W/m${}^2$. Figure~\ref{fig: Earth_CO2} also demonstrates how the magma ocean simulation with 14.7~TO initial \ce{CO2} mass evolves with the RT grid model (black line) evolves within the emission grid, build with full radiative transfer calculations. The comparison reveals that the OLR differs by more than one order of magnitude already at the start of the simulation (Fig.~\ref{fig: Earth_CO2}, panel right). 

Similarly, \citet{Lichtenberg2021} demonstrate that a magma ocean  with a pure \ce{CO2} atmosphere and full radiative transfer yields emits large amounts of flux ($OLR>10^5$~W/m${}^2$) for $T_{surf}>2500$~K and $p_{surf}=260$~bar. When we further compare our \ce{CO2} magma ocean simulation to that of \citet{Lichtenberg2021}, we likewise find that their magma ocean with a 200 - 300 bar dense \ce{CO2} atmosphere reaches surface temperatures between 1500-1600~K, which corresponds to our solidification criterion, after  10 000 years in the same order of magnitude than our \ce{CO2} magma ocean simulation.

While the gray atmosphere model fails to accurately capture the thermal evolution of a magma ocean with a dense \ce{CO2} atmosphere, the agreement between the RT grid simulation and the simulation with the corrected gray atmosphere model is much better (Fig.~\ref{fig: Pure_CO2Earth}). The latter yields a solidification time of the same order of magnitude (40 000 years). The results of the corrected gray model shows that our analytical approximation generally reproduce thermal emission even for a hot, vertically extended \ce{CO2} atmosphere (Sect.~\ref{sec: grey}).

 \subsection{Mixed CO2-H2O atmospheres - Different atmosphere models}
 \label{sec: Earth -Mixed CO2 scenarios}
\begin{table*}[h]
    \caption{\magmoc{2.0} Results for Earth: Comparison between \citet[Tab. 3, Earth ($\SI{2000}{\kilo\metre}$)]{Elkins-Tanton2008} and \magmoc{}2.0 at the end of their respective magma ocean stages. }
    \begin{tabular}{c|cc|cc|cc}		
        \noalign{\smallskip}
		\hline
		\noalign{\smallskip}
		Melt fraction [wt\%]
	 & \multicolumn{2}{c|}{Dry case: 0.05 \ce{H2O}, 0.01 \ce{CO2} } & \multicolumn{2}{c|}{Wet case: 0.5 \ce{H2O}, 0.1 \ce{CO2} }  & \multicolumn{2}{c}{Only \ce{CO2}: 0.6 \ce{CO2} } \\
		Initial mass$^{a}$ [TO] &\multicolumn{2}{c|}{\hspace{1.5cm} 1.02 \ce{H2O} 0.21 \ce{CO2} } & \multicolumn{2}{c|}{\hspace{1.5cm}  10.2 \ce{H2O} 2.1 \ce{CO2}  } & \multicolumn{2}{c}{\hspace{1.5 cm} 14.7 \ce{CO2}  }\\ 
		\noalign{\smallskip}
		\hline
		\noalign{\smallskip}
	Model* & Elkins-Tanton & \magmoc{2.0} & Elkins-Tanton & \magmoc{2.0}& Elkins-Tanton & \magmoc{2.0}  \\
		\noalign{\smallskip}
		\hline
		\noalign{\smallskip}
		\multicolumn{5}{l}{Fraction of initial volatile content outgassed into the atmosphere [$\si{\%]}$} \\
		\noalign{\smallskip}
		\hline
		\ce{H2O} & 70 & \textit{63}/\textbf{63} &91 & \textit{86}/\textbf{86} & 0 & 0 \\
		\ce{CO2} & 78 & \textit{81}/\textbf{81} &95  & \textit{95}/\textbf{95} & 97 & \textit{98}/\textbf{98} \\
		\noalign{\smallskip}
		\hline
		\noalign{\smallskip}
		\multicolumn{5}{l}{Final atmospheric pressure (sum of partial pressures of \ce{H2O} and \ce{CO2}) [bar]} \\
		\noalign{\smallskip}
		\hline
		\noalign{\smallskip}
		& 240 & \textit{238}/\textbf{238} & 3150  &\textit{2940}/\textbf{2940} & 3350 & \textit{3841}/\textbf{3843} \\
		\noalign{\smallskip}
		\hline
		\noalign{\smallskip}
		\multicolumn{5}{l}{Solidification time, for $k_{\ce{H2O}} = 0.01$ and $k_{\ce{CO2}} = 0.001$ [m${}^2$/kg] for gray model} [Myrs] \\
		\noalign{\smallskip}
		\hline
		\noalign{\smallskip}
		& 0.06 & \textit{0.113}/\textbf{0.61}  &2.4 & \textit{1.7}/\textbf{1.7} & 0.8 & \textit{0.51}/\textbf{0.023} \\
		\noalign{\smallskip}
		\hline
		\noalign{\smallskip}
		\multicolumn{5}{l}{Remaining Volatile content in the melt [$\si{wt\%]}$} \\
		\noalign{\smallskip}
		\hline
		\noalign{\smallskip}
		\ce{H2O} & 1.5 & \textit{1.6}/\textbf{1.6} &5.3 & \textit{5.3}/\textbf{5.3} & 0 & 0\\
		\ce{CO2} & 0.2 & \textit{0.2}/\textbf{0.2}  &0.7  & \textit{0.5}/\textbf{0.5} & 1.5 & \textit{1.6}/\textbf{1.6} \\
		\noalign{\smallskip}
		\hline
	\end{tabular}
    \newline Cursive: gray model, without the runaway greenhouse limit \,
    Bold: Full RT grid model.
     \newline *: In all cases, the outgassing laws of \citet{Elkins-Tanton2008} were used.
	\label{Tab_Results_Earth}
\end{table*}

To validate \magmoc{2.0} for a mixed \ce{H2O}-\ce{CO2} atmosphere, we follow again \citet{Barth2021} and compare our simulations first to results of the Earth simulations of \citet{Elkins-Tanton2008}. In \citet{Barth2021}, the pure \ce{H2O} atmosphere already showed promising agreement with \citet{Elkins-Tanton2008}.

We test here two scenarios: one with melt fractions $0.05$ and $0.01$ for \ce{H2O} and \ce{CO2}, respectively, which we call henceforth the "dry scenario", the other with melt fractions $0.5$ and $0.1$ for \ce{H2O} and \ce{CO2}, respectively, which we call henceforth the "wet scenario". With an initial magma ocean depth of 2000~km, these melt fractions correspond for the 'dry case' to an initial water mass 1.02~TO of \ce{H2O} and for the 'wet case' to an initial mass of 10.2~TO \ce{H2O}. The initial \ce{CO2} mass is 0.21 and 2.12 TO, respectively. These melt fractions (Table~\ref{Tab_Results_Earth}) also correspond to a 5:1 ratio between \ce{H2O} and \ce{CO2} that is assumed by \citet{Elkins-Tanton2008}. We further adopt the  same solubility laws used by \citet{Elkins-Tanton2008} that differ from the \ce{H2O} outgassing law used in the previous section, where instead the law by \citet{Schaefer2016} was used. See Sect.~\ref{sec: outgas} for an overview of outgassing laws used in this work.

\begin{figure}
    \centering
    \includegraphics [width=0.49\textwidth]{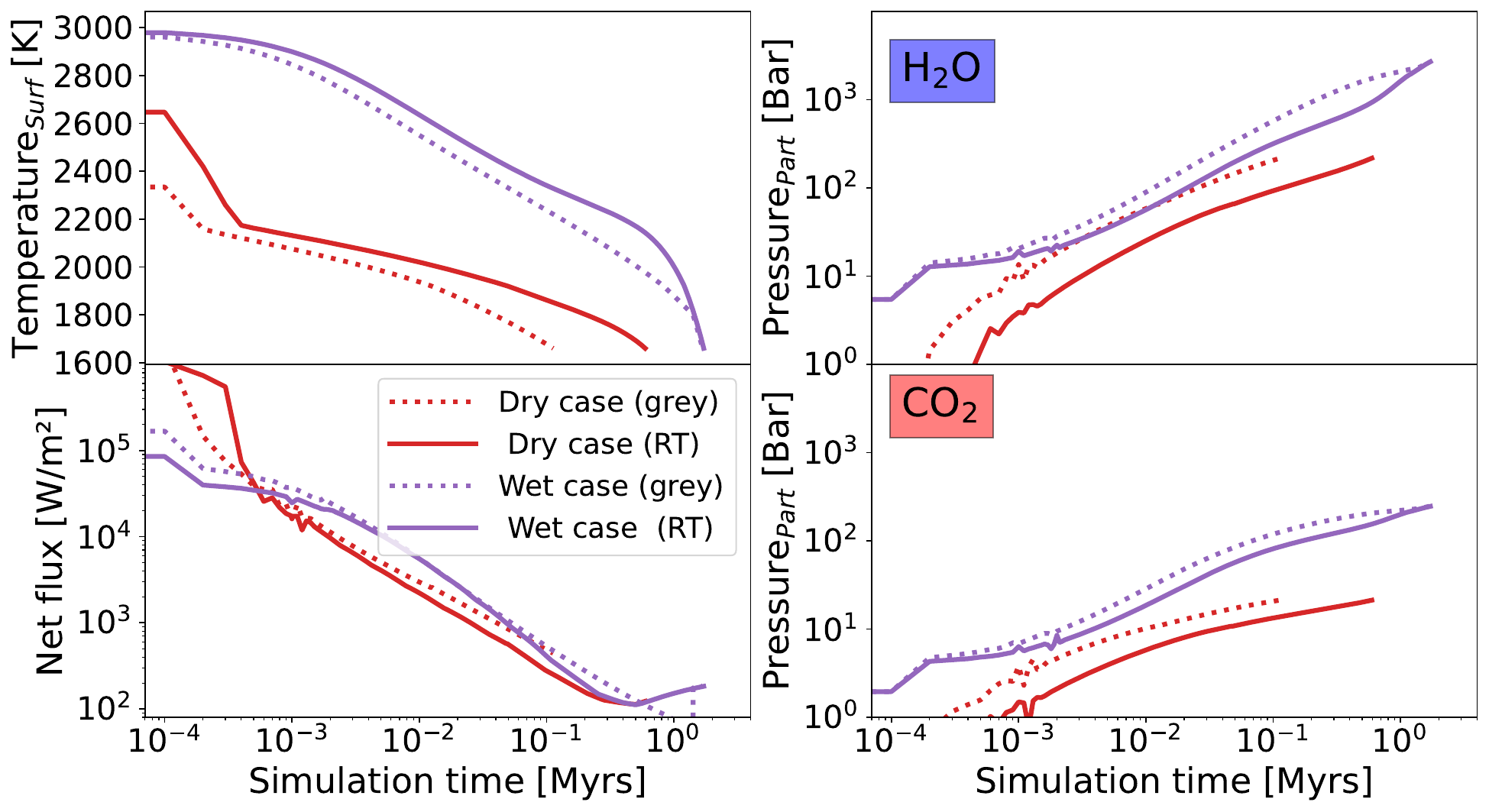}
   \caption{Earth magma ocean simulation set-up  for comparison with \citet{Elkins-Tanton2008} with the traditional gray model but without greenhouse limit and the full radiative transfer grid model. Scenarios are as listed in Table~\ref{Tab_Results_Earth}. The solid lines denote results of simulations with the gray atmosphere model. The dotted lines denote results of simulations with full radiative transfer. We not again, that we show net flux, which is equal ASR-OLR.}
    \label{fig: Elkins_all}
\end{figure}

First, we investigate the impact of the gray atmosphere and the full RT atmosphere model for mixed \ce{H2O}-\ce{CO2} outgassing. We emphasize, however, that we 'switched off' the runaway greenhouse radiation limit in the gray atmosphere model for these specific simulations. This choice is made because \citet{Elkins-Tanton2008} did not include this limiting factor for the OLR. In all other "gray" simulations in this work, we consider the radiation limit. Figure~\ref{fig: Elkins_all} (solid lines) and Table~\ref{Tab_Results_Earth} illustrate that our "gray" simulations without the runaway greenhouse radiation limit are in general agreement with the solidification times reported by \citet{Elkins-Tanton2008}: The magma ocean in the "wet case" solidifies after 1.7~Myrs, and in the "dry case" after only 0.1~Myrs. 

For the "wet case", the differences between simulations with the gray atmosphere model and the RT model do not significantly impact the magma ocean solidification times. However, the RT model simulation yields a hotter surface during the early evolution due to a lower initial net flux, resulting in less outgassing of volatiles initially. The hotter magma ocean persists until solidification at 1.7~Myrs, after which the gray and RT evolution tracks converge. This convergence occurs because the net flux in the RT atmosphere simulation is larger compared to the gray simulation in the end, thereby  offsetting initial differences.

For the "dry case", the magma ocean begins its evolution essentially without a significant atmosphere, when the outgassing laws of \citet{Elkins-Tanton2008} are used.  This dry, hot initial state once again leads to a hotter magma ocean in the simulation with the RT atmosphere model compared to the gray simulation. Unlike "in the wet case", the evolution tracks do not converge at the end. This is because the net flux of the RT simulation mostly remains below that of the gray simulation until solidification. Consequently, the magma ocean stage is prolonged with the RT atmosphere model compared to simulations using the gray model.   
 
While the lifetime of the Earth magma ocean lifetime may vary with different atmosphere models, this variability does not appear to affect the final volatile content. Simulations using the RT and gray atmosphere model show virtually identical quantities of volatiles in the end (Table~\ref{Tab_Results_Earth}).

We further compare the RT grid simulations for \ce{H2O}-\ce{CO2} atmospheres with simulations using the corrected gray atmosphere model. Our analysis reveals that in the mixed \ce{H2O}-\ce{CO2} set-up, the corrected gray atmosphere model underestimates the OLR in the initial evolution stage, during which surface temperatures are very high and surface pressures are particularly low ( $<10$~bar). This tendency is illustrated in Fig.~\ref{fig: Mixed_Emission} (left): The corrected gray atmosphere model consistently underestimates the OLR compared to the RT grid for high surface temperatures ($T_{surf}\geq 2000$~K) and low pressures ($p_{surf}\leq 26$~bar).

\begin{figure}
    \centering
    \includegraphics [width=0.49\textwidth]{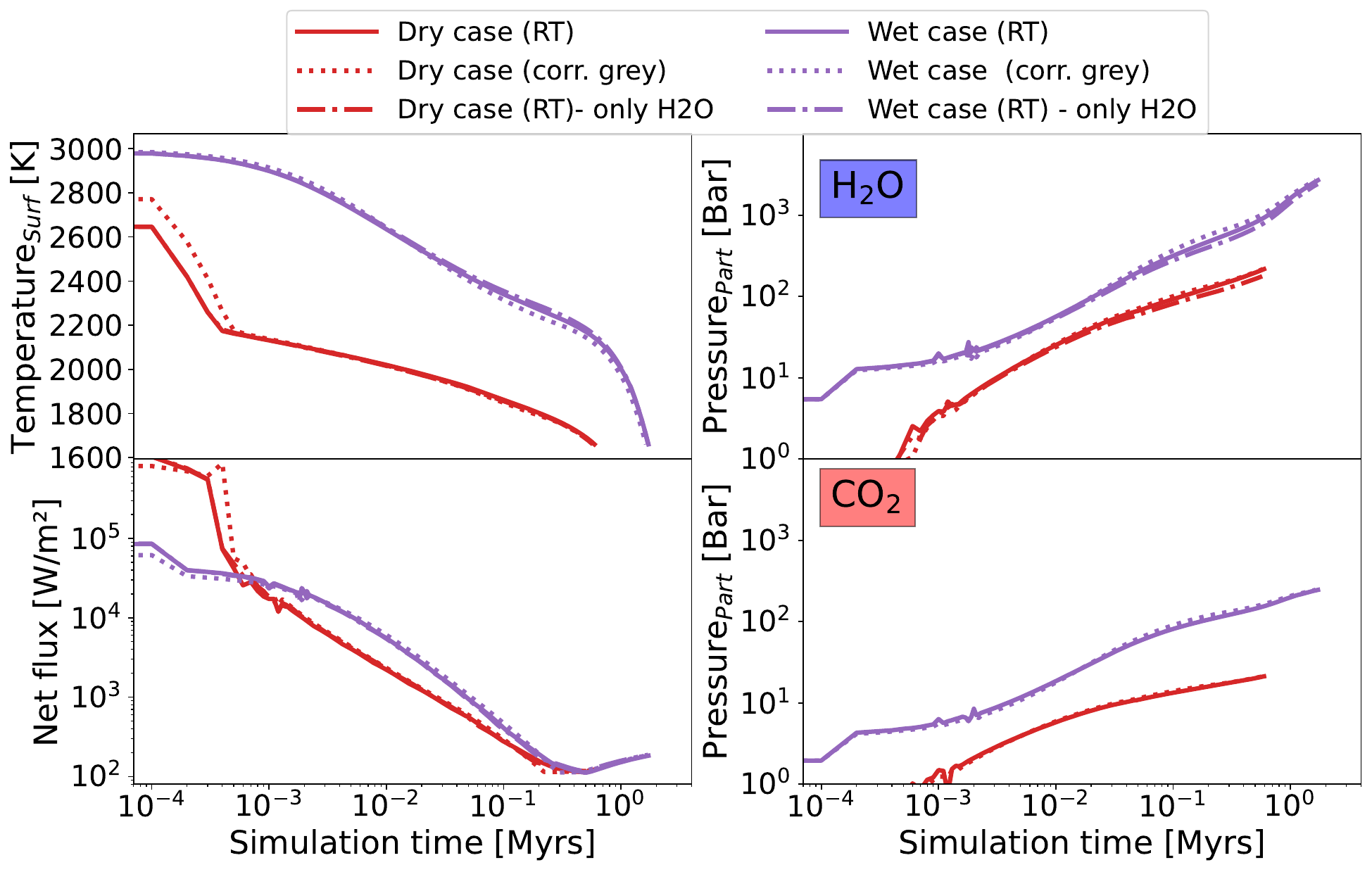}
   \caption{Earth magma ocean simulation for comparison with \citet{Elkins-Tanton2008} with the corrected gray and full radiative transfer atmosphere model. Scenarios are as listed in Table~\ref{Tab_Results_Earth}. The dotted lines denote results of simulations with the corrected grqy atmosphere model. Solid lines denote results of simulations with full radiative transfer. The dashed dotted line denote the results of simulations with full radiative transfer only with \ce{H2O} to assess the impact of \ce{CO2}.}
    \label{fig: Elkins_all_greycorr}
\end{figure}

However, these discrepancies in flux are limited to the first thousand years of simulation. More importantly, they do not lead to significant differences in later stages of the magma ocean evolution. Throughout the majority of the magma ocean evolution, both for the "dry" and "wet case", the surface temperature and atmosphere surface pressures agree well between the RT and corrected gray simulation.

Finally, we present RT atmosphere simulations for the "dry" and "wet case" that exclude \ce{CO2} (Fig.~\ref{fig: Elkins_all_greycorr} dashed dotted line). We find that the absence of \ce{CO2} has a minimal influence on the magma ocean evolution,  when the outgassing laws of \citet{Elkins-Tanton2008} are used. However, this set-up severely suppresses outgassing of \ce{CO2}, such that it remains a minor constituent throughout the entire magma ocean evolution. In the following, we will thus tackle the impact of alternative solubility laws.

In summary, we generally demonstrate with \magmoc{2.O} agreement in \ce{H2O} and \ce{CO2} distribution with \citet{Elkins-Tanton2008}, and also in solidification times, when the gray atmosphere model without runaway greenhouse limit is used. Different atmosphere models may yield variations in solidification time, confirming \citet{Niko2019}. Generally, we find that the differences between atmosphere models are largest during the initial stage and we confirm again that the runaway greenhouse radiation limit has a strong impact on the later magma oecean evolution and thus the solidification times.

\subsection{Mixed CO2-H2O atmospheres - Different outgassing laws}
 \label{sec: Earth -Mixed CO2 scenarios-Outgas}
More recent solubility laws \citep[e.g.,][]{Niko2019,Lichtenberg2021} suggest that \ce{CO2} should dominate the atmosphere during the initial magma ocean stage due its significantly lower solubility in the magma compared to \ce{H2O} (Sect.~\ref{sec: outgas}). To test the impact of such a scenario, we implement in \magmoc{2.0} the outgassing laws of \citet{Niko2019}. In the following, we will denote simulations using the outgassing laws of \citet{Elkins-Tanton2008} as "ET" and simulations using outgassing laws of \citet{Niko2019} as "Ni".

To facilitate a comparison of the impact of different outgassing laws with results from previous sections, we once again perform 'dry' and 'wet case' simulations (Table~\ref{tab: Niko_Comp}), maintaining an initial magma ocean depth of 2000~km. For the "dry case" simulation, volatile inventories are broadly consistent with the set-up of the simulations outlined in \citet{Niko2019}.

\begin{table}[]
\caption{Parameters for comparing outgassing laws.}
    \centering
    \begin{tabular}{l|c|c|c|c|}
    &\multicolumn{2}{|c|}{Dry case${}^*$} &  \multicolumn{2}{c|}{Wet case${}^*$}\\
    \hline
   Outgassing laws  &  ET & Ni & ET & Ni \\
   \hline
    initial \ce{H2O} mass [TO] &\multicolumn{2}{c|}{1.02} &  \multicolumn{2}{c|}{10.2}\\ 
    \hline
    initial \ce{CO2} mass [TO] & 0.21 & 0.48 &2.1 & 4.8 \\
    
    \end{tabular}
    \newline
     ${}^*$ An initial magma ocean depth of 2000~km is assumed for all cases.
    \label{tab: Niko_Comp}
\end{table}

\begin{figure}
    \centering
    \includegraphics [width=0.49\textwidth]{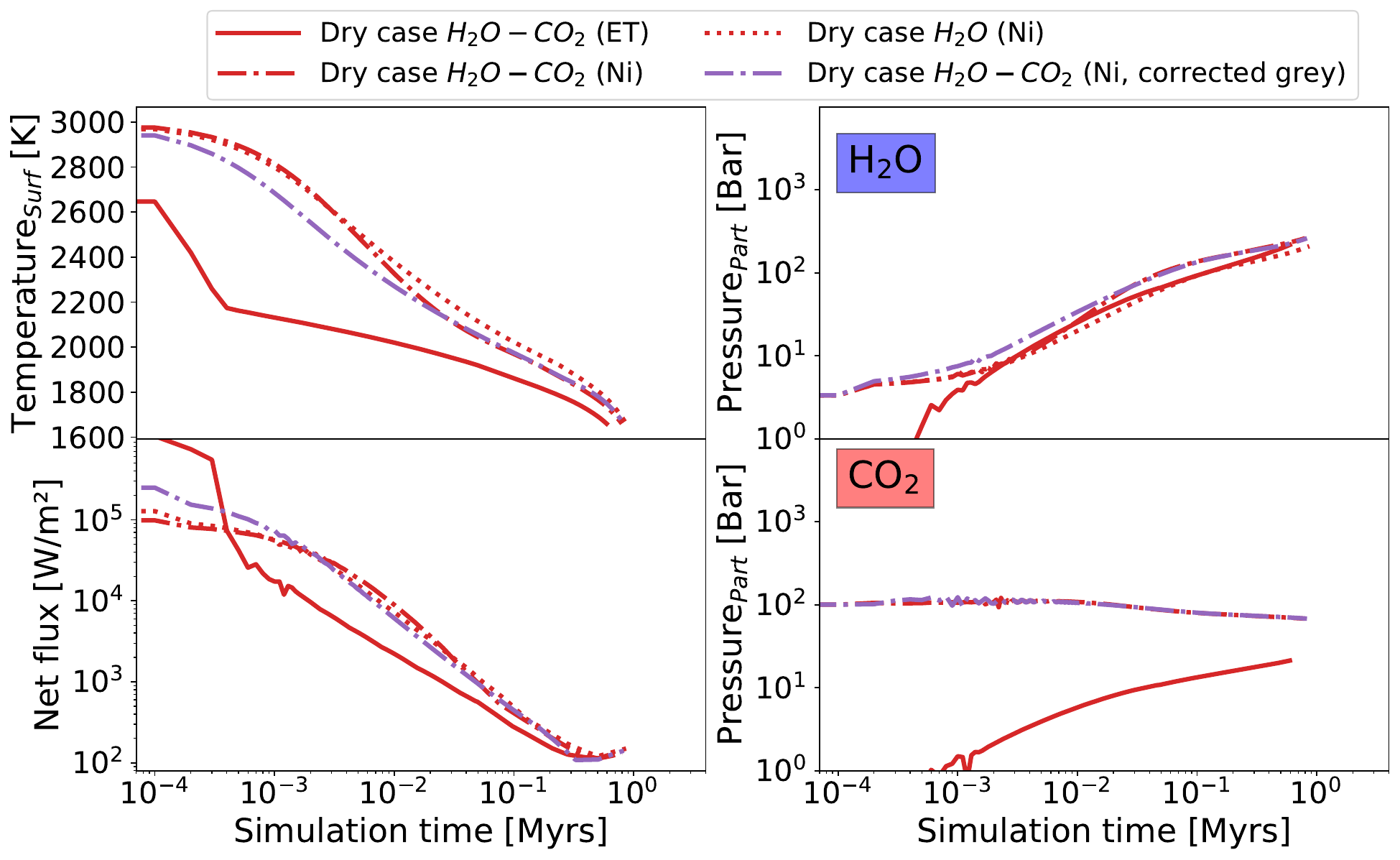}
   \caption{Earth magma ocean simulations with mixed \ce{H2O}-\ce{CO2} atmospheres to compare the outgassing laws of \citet{Elkins-Tanton2008} (ET) and \citet{Niko2019} (Ni). Here, the results of the 'dry case' scenarios are shown (see Table~\ref{tab: Niko_Comp}). For most scenarios the full radiative transfer atmosphere model is used (red). The solid lines denote the dry case scenario of \citet{Elkins-Tanton2008} with 0.21~TO \ce{CO2} and the same outgassing laws. The dashed-dotted lines denote results of a 'dry' simulation with the outgassing laws from \citet{Niko2019}. For the latter scenario, an additional simulation with the corrected gray atmosphere model is shown (purple line). The red dotted line denotes a simulation with only \ce{H2O}, using the outgassing laws of \citet{Niko2019} and the RT model.}
    \label{fig: Niko_comp}
\end{figure}

Figure~\ref{fig: Niko_comp} illustrates the impact of different outgassing laws on the 'dry case', where major differences are evident for \ce{CO2} surface pressures. In the ET-simulation, \ce{CO2} is outgassed relatively late and never becomes a dominant component of the atmosphere, consistent with results from the previous section. Conversely, in the Niko-simulations, a significant portion of \ce{CO2} is already outgassed at the onset of the magma ocean stage, in agreement with magma ocean simulations of \citet{Niko2019} and \citet{Lichtenberg2021}. Consequently, \ce{CO2} initially dominates the atmosphere, with substantial \ce{H2O} outgassing occurring only after 10 000 years of simulation time. Eventually \ce{H2O} becomes the dominant atmospheric species. 

Because the magma ocean evolution in the Ni-simulation begins in the 'dry case' with a thick atmosphere rather than a 'bare rock scenario', the magma ocean cooling is less efficient (Fig.~\ref{fig: Niko_comp} dashed-dotted red line). Consequently, the Ni-simulation reaches solidification later, at 0.8~Myrs, compared to the solidification time of 0.5~Myrs in the ET-simulation (Fig.~\ref{fig: Niko_comp} solid red line). 

\begin{figure}
    \centering
    \includegraphics [width=0.49\textwidth]{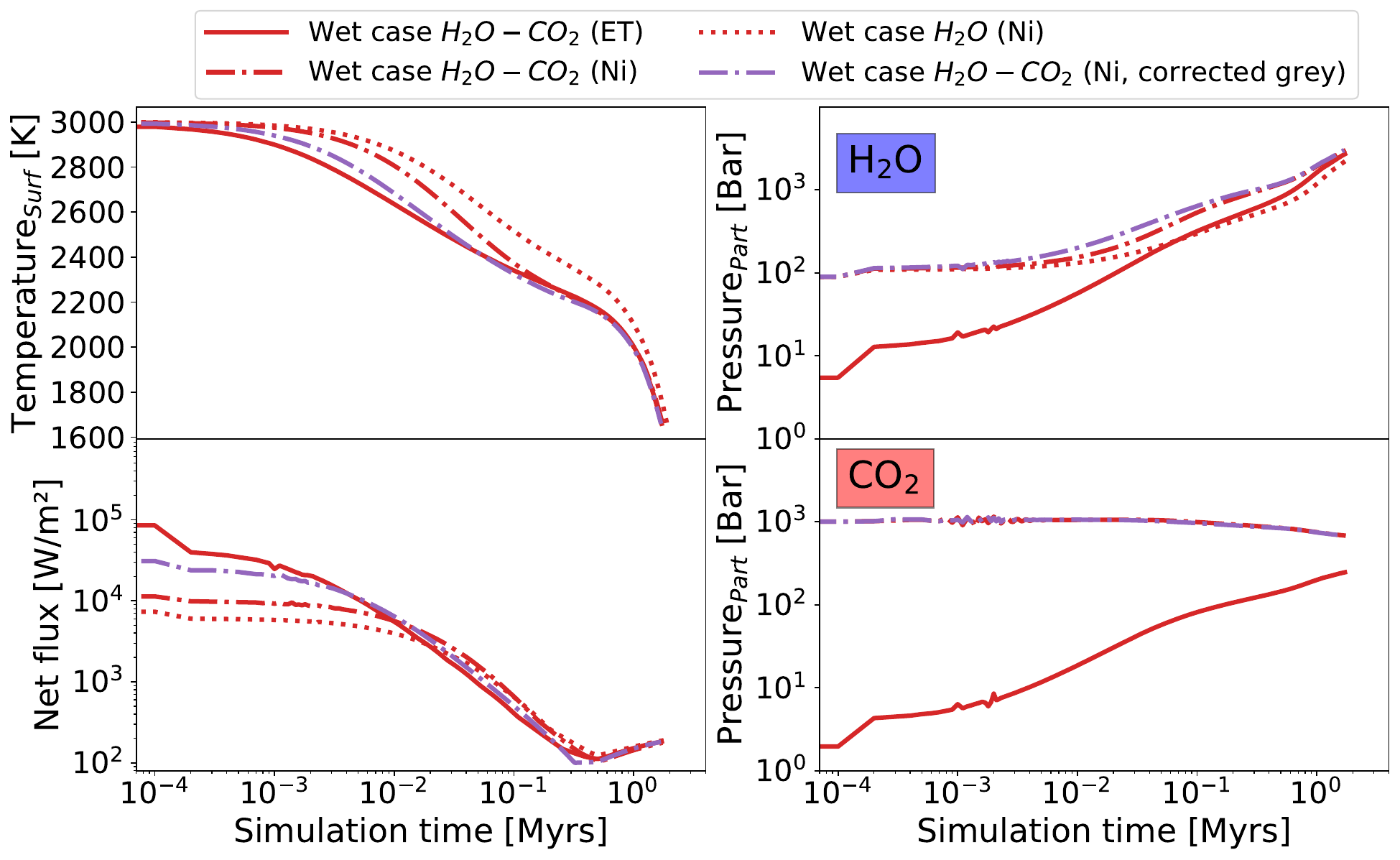}
   \caption{Earth magma ocean simulation that compares the mixed \ce{H2O}-\ce{CO2} magma oceans of \citet{Elkins-Tanton2008} (ET) and \citet{Niko2019} (Ni) and with a 'wet case' volatile budget (Table~\ref{tab: Niko_Comp}). For most scenarios the full radiative transfer atmosphere model is used (red). The solid lines denote ET-simulations. The dashed-dotted lines denote a 'wet' simulation with the outgassing laws from \citet{Niko2019}. For the latter scenario, an additional simulation with the corrected gray atmosphere model is shown (purple lines). The red dotted line denotes a simulation with only \ce{H2O} using the outgassing laws of \citet{Niko2019} and the RT atmosphere model.}
    \label{fig: Niko_compwet}
\end{figure}

We conduct a similar comparison between different outgassing laws and atmosphere models for a volatile-rich 'wet case' Earth scenario. Here, all initial masses are scaled from the 'dry case' by a factor of ten (Fig.~\ref{fig: Niko_compwet}, Table~\ref{tab: Niko_Comp}). In this scenario, the thermal evolution of the magma ocean across diverse scenarios is quite similar, converging at around 2~Myrs. Significant differences are again obtained for the \ce{H2O} and \ce{CO2} partial pressures. In the  ET-simulations, \ce{CO2} consistently remains a minor constituent in the atmosphere. Conversely, in the Ni-simulations, the atmospheric composition initially \ce{CO2}-dominated and eventually becomes \ce{H2O}-dominated.

Evaluating again the performance of the corrected gray atmosphere model against the RT model in the Ni-simulations, we find that the former overestimates the OLR of a hot, \ce{CO2}-dominated atmosphere by a factor of two to three   (Figures~\ref{fig: Niko_comp} and \ref{fig: Niko_compwet}, dashed-dotted purple line). This discrepancy has been already previously noted (Fig.~\ref{fig: Mixed_Emission}, right panel and  Fig.~\ref{fig: Earth_CO2}), where it was found that the corrected gray model can deviate from the RT model by up to half an order of magnitude for thick \ce{CO2} atmospheres with high surface temperatures ($T_{surf}> 2500$~K). In any case, also here we find that the surface temperature evolution tracks with the corrected gray atmosphere model eventually converge with those with the RT atmosphere. In the 'dry case', convergence occurs after 10,000 years of simulation time, while in the 'wet case', convergence occurs after 0.1~Myrs. 
Despite these discrepancies in surface temperatures, the outgassed \ce{H2O} and \ce{CO2} volatile content generally agrees throughout the entire magma ocean evolution between the Ni-simulations with the corrected gray and RT atmosphere model. 

Correctly capturing the amount of outgassed volatiles is crucial for assessing the impact of atmospheric erosion on magma oceans around M dwarf stars. For the Earth simulations, the corrected gray atmosphere model, benchmarked against the numerically more costly RT atmosphere model (Sect.~\ref{sec: Stability}), appears to adequately capture the volatile evolution in Earth magma oceans. This good performance also holds true with outgassing laws that result in atmospheric composition changing from \ce{CO2}-dominated to \ce{H2O}-dominated.

\subsection{Mixed CO2-H2O atmospheres - Impact of CO2}
\label{sec: Earth -Mixed CO2 scenarios- CO2}

In the ET-simulations, \ce{CO2} consistently remains a minor atmospheric constituent with a negligible influence on the magma ocean evolution. However, in simulations using the outgassing laws used of \citet{Niko2019}, \ce{CO2} is initially the dominant atmospheric species that can affect the initial magma ocean stage, as evidenced by the prolongation of the magma ocean stage in the Ni-simulations compared to the ET-simulations in the 'dry case'. To assess the impact of \ce{CO2} in the Ni-simulations more coherently, we conducted several simulations without \ce{CO2} (Figures~\ref{fig: Niko_comp} and \ref{fig: Niko_compwet}, red dotted curves). 

A detailed inspection of the thermal evolution of the 'dry case' shows that a mixed \ce{CO2}-\ce{H2O} atmosphere exhibits higher thermal emission during the first 0.3~Myrs of the evolution compared to a pure \ce{H2O} atmosphere. This difference arises because a \ce{CO2}-rich atmosphere emits more flux than an equivalent \ce{H2O}-atmosophere for high surface temperatures ($T_{surf}>2000$~K, see Sections~\ref{sec: Earth - pure CO2} and \ref{sec: grey}).  However, beyond 0.3 Myrs, the system reaches the runaway greenhouse limit that is set primarly determined by the water content in the atmosphere. 

As discussed in Sect.~\ref{sec: PT}, the addition of \ce{CO2} tends to cool the emitting atmosphere layers, thereby reducing the OLR limit and diminishing thermal cooling in a mixed atmosphere compared to a pure \ce{H2O} atmosphere. In other words, the addition of \ce{CO2} for a sufficiently oxidized mantle \citep{Ortenzi2020} tends to shorten the magma ocean lifetime outside of the runaway greenhouse limit and tends to extend the solidification time within it. For the "dry case", both effects counterbalance each other, resulting in a slightly longer solidification time for the mixed atmosphere (by few 10,000 years or a few percent) . 

The higher thermal emission of a mixed atmosphere outside of the runaway greenhouse limit is more pronounced when more volatiles are in the system as in the 'wet case' (Fig.~\ref{fig: Niko_compwet}, red dotted line). However, even in this scenario, the runaway greenhouse limit reached towards the end of the evolution results  after 2~Myrs in a convergence of solidification time compared to a magma ocean model with a pure \ce{H2O} atmosphere. Upon close examination, the mixed atmosphere exhibits a slightly shorter solidification time, albeit by only a few percent.

Our assessment of a minimal impact of \ce{CO2} on the solidification times of magma oceans disagrees with the results of \citet{Niko2019} who report an extension of magma ocean lifetime with additional \ce{CO2}. These authors used, however, a gray atmosphere model. As discussed in Sect.~\ref{sec: grey}, a gray atmosphere model consistently overestimates thermal emission for thick \ce{H2O} atmospheres (Fig.~\ref{fig: Grey_H2O}) and underestimates thermal emission for thick \ce{CO2} atmospheres (Fig.~\ref{fig: Grey_CO2}) because it does not adequately capture the thermal emission in a vertically extended atmosphere, as already outlined by \citet{Lichtenberg2021}.

\subsection{Mixed CO2-H2O atmospheres - Importance of initial conditions}

We compare, similarly to Table~\ref{Tab_Results_Earth}, the volatile budget at the end of the magma ocean between the ET and Ni-simulations (Table~\ref{Tab_Results_Earth_Niko}). The most notable difference is the very low remaining \ce{CO2} content in the latter cases. This discrepancy can be explained by the different initial conditions, driven by the different \ce{CO2} outgassing laws. 

In the model of \citet{Elkins-Tanton2008}, \ce{CO2} is mostly dissolved in the melt initially, resulting in a relatively large initial melt fraction $F_{\ce{CO2}}$ compared to the outgassing laws used by \citet{Niko2019}. As the magma ocean solidifies and the thickness of the magma ocean decreases, as a result significant enrichment of volatile mass fraction occurs, assuming no substantial sink terms. Only a very small fraction of \ce{CO2} is partitioned from the melt into the solid mantle (Table~\ref{Tab_geo}) and atmospheric erosion for Earth is negligible during the magma ocean evolution. Thus, the volatile mass fraction at the end of the magma ocean stage can only be larger than the initial value. 

Because \ce{CO2} outgassing is suppressed in the ET-simulations, initial values of $F_{\ce{CO2}}$ are relatively high already. Consequently, the magma ocean ends with large \ce{CO2} mass fractions of 0.2 - 0.7 wt\%. Conversely, the Ni-simulations begin with a substantially lower mass fraction in the magma ocean and consequently end the magma ocean stage with \ce{CO2} mass fractions of 0.003 - 0.3 wt\% , which is up to two orders of magnitudes lower compared to the ET-simulations. The differences are much smaller for \ce{H2O}, because the ET and Ni-outgassing laws assume both a high solubility of \ce{H2O}, resulting in similar initial values of $F_{\ce{H2O}}$.

We thus conclude that the initial conditions, which determine the initial volatile mass fraction in the magma ocean, are critical for determining the overall volatile budget at the end of the magma ocean phase and how much of the volatiles can be retained in the mantle.

\begin{table*}[ht]
    \caption{\magmoc{2.0} results for Earth to compare with \citet[Tab. 3, Earth ($\SI{2000}{\kilo\metre}$)]{Elkins-Tanton2008}.}
    \begin{tabular}{c|cc|cc}		
        \noalign{\smallskip}
		\hline
		\noalign{\smallskip}
		melt fraction$^{b}$ [wt\%]
	 & \multicolumn{2}{c|}{Dry case: 0.05 \ce{H2O} (0.01/$4\times 10^{-5}$)\{0.023\}${}^a$ \ce{CO2} } & \multicolumn{2}{c}{Wet case: 0.5 \ce{H2O} (0.1/$4\times 10^{-4}$)\{0.23\}${}^a$ \ce{CO2} }  \\
		Initial mass$^{a}$ [TO] &\multicolumn{2}{c|}{\hspace{1.5cm} 1.02 \ce{H2O} (0.21/0.48) \ce{CO2} } & \multicolumn{2}{c}{\hspace{1.5cm}  10.2 \ce{H2O} (2.1/4.8) \ce{CO2}  } \\ 
		\noalign{\smallskip}
		\hline
		\noalign{\smallskip}
	Model${}^{*}$ & Elkins-Tanton & \magmoc{2.0} (Ni) & Elkins-Tanton & \magmoc{2.0} (Nicolaou)  \\
		\noalign{\smallskip}
		\hline
		\noalign{\smallskip}
		\multicolumn{5}{l}{Fraction of initial volatile content outgassed into the atmosphere [$\si{\%]}$} \\
		\noalign{\smallskip}
		\hline
		\ce{H2O} & 70 & 75 &91 & 85\\
		\ce{CO2} & 78 & 99.84 &95  & 99.85 \\
		\noalign{\smallskip}
		\hline
		\noalign{\smallskip}
		\multicolumn{5}{l}{Final atmospheric pressure (sum of partial pressures of \ce{H2O} and \ce{CO2}) [bar]} \\
		\noalign{\smallskip}
		\hline
		\noalign{\smallskip}
		& 240 & 261 & 3150  & 3687 \\
		\noalign{\smallskip}
		\hline
		\noalign{\smallskip}
		\multicolumn{5}{l}{Solidification time, for $k_{\ce{H2O}} = 0.01$ and $k_{\ce{CO2}} = 0.001$ [m${}^2$/kg] for gray model [Myrs]} \\
		\noalign{\smallskip}
		\hline
		\noalign{\smallskip}
		& 0.06 & 0.8  &2.4 & 1.8\\
		\noalign{\smallskip}
		\hline
		\noalign{\smallskip}
		\multicolumn{5}{l}{Volatile content of liquids remaining [$\si{wt\%]}$} \\
		\noalign{\smallskip}
		\hline
		\noalign{\smallskip}
		\ce{H2O} & 1.5 & 1.1 &5.3 & 5.9 \\
		\ce{CO2} & 0.2 & 0.003  &0.7  & 0.03 \\
		\noalign{\smallskip}
		\hline
	\end{tabular}
	\newline
    *: Using the full RT grid model and the outgassing laws of \citet{Niko2019}.\newline
    ${}^a$ The \ce{CO2} content differs between the ET and Nicolaou-simulations. Thus, the content is given in brackets divided by slashes as ET/Nicolaou).\newline
	${}^b$: In the ET-simulations, all \ce{CO2} is assumed to be in the melt initially. In the Nicolaou (Ni)-simulations, however, we already start with a substantially outgassed \ce{CO2} atmosphere. For better comparison, we thus give here in curly brackets for the Nicolaou-simulation also the initial melt fraction assuming that all \ce{CO2} is dissolved.  
	\label{Tab_Results_Earth_Niko}
\end{table*}

\subsection{Mixed CO2-H2O atmospheres - Outgassing feedback}
\label{sec: feedback}

We confirm here that the evolution of the atmospheric \ce{CO2} and \ce{H2O} exhibit a feedback effect, when the mean molecular weight of the atmosphere changes during the magma ocean evolution \citep{Bower2019}. A change in atmospheric composition occurs during the magma ocean evolution because \ce{CO2} is less soluble than \ce{H2O} and thus dominates outgassing at the beginning of the magma ocean stage, as captured by the outgassing laws of \citet{Niko2019} (Sect.~\ref{sec: Earth -Mixed CO2 scenarios-Outgas}). The model of \citet{Elkins-Tanton2008} avoids this effect by suppressing \ce{CO2} outgassing and hence these simulations show no such feedback.

\begin{figure}
    \centering
    \includegraphics [width=0.49\textwidth]{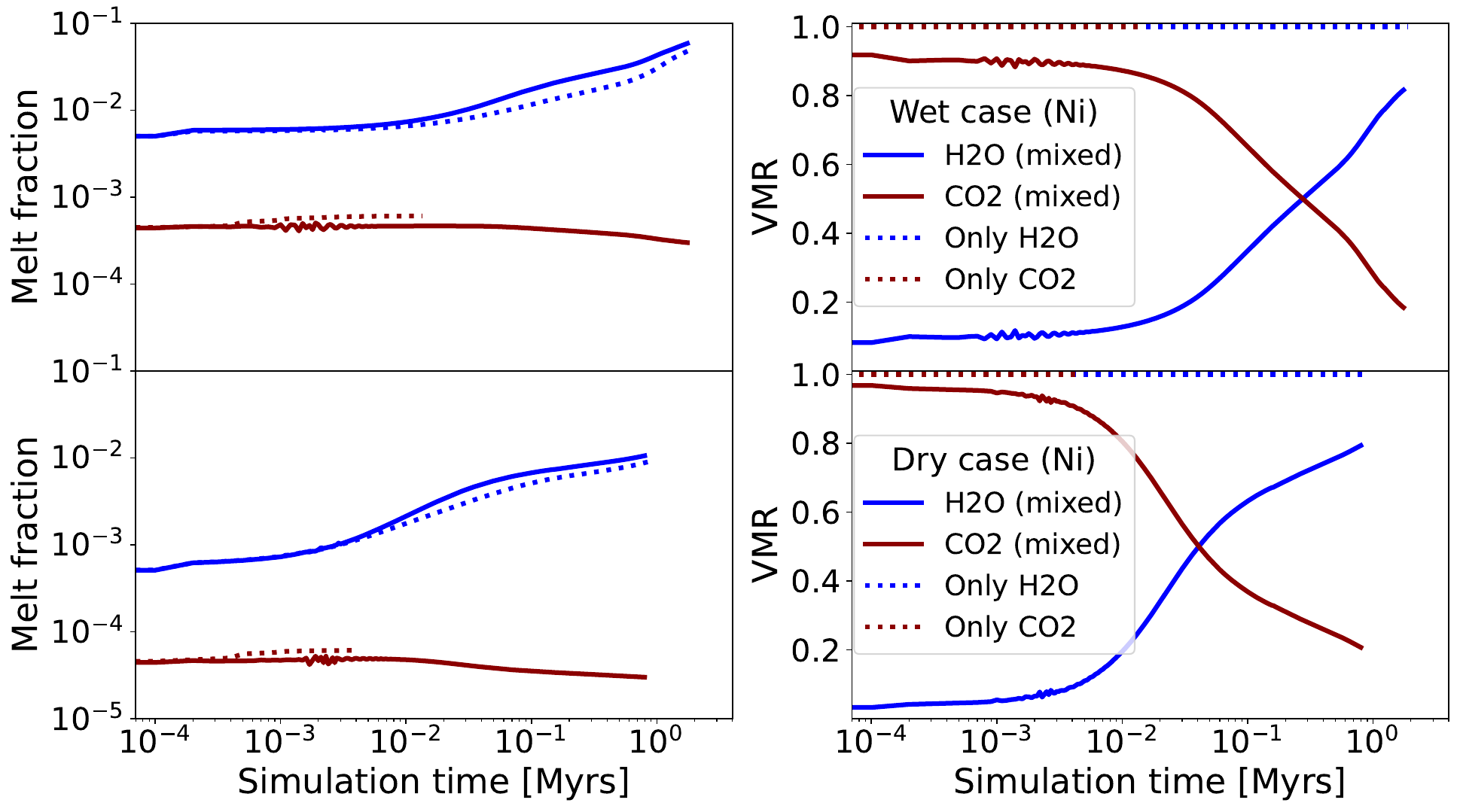}
   \caption{ Volatile content in the magma ocean and the atmosphere for an Earth magma ocean simulation with the outgassing laws of \citet{Niko2019}, using the RT grid  and for the 'dry case' (top panels) and 'wet case' (bottom panels). See Table~\ref{tab: Niko_Comp} for the respective volatile content of the simulations. Shown are melt fraction (left) and  volume mixing ratio (VMR, right) for \ce{H2O} (in blue) and \ce{CO2} (in dark red) for the mixed case and the cases with only \ce{H2O} (blue dotted) and only \ce{CO2} (red dotted lines), respectively.}
    \label{fig: Niko_volatile}
\end{figure}

Figure~\ref{fig: Niko_volatile} provides more detailed insights about the outgassing feedback in the Nicolaou (short: Niko) simulations. In the mixed volatile set-up (solid lines), the atmospheric composition undergoes drastic changes from \ce{CO2}-dominated (molecular weight of \ce{CO2}: $\mu=44$~g/mol) to \ce{H2O}-dominated (molecular weight of \ce{H2O}: $\mu=18$~g/mol), as also reported by \citet{Niko2019,Bower2019}. Consequently, the mean molecular weight of the atmosphere decreases as the magma ocean solidifies. Additionally, the melt fraction of \ce{CO2} decreases as \ce{H2O} begins to accumulate in the atmosphere (Fig.~\ref{fig: Niko_volatile}, left panel, red dotted lines). In the absence of \ce{H2O}, \ce{CO2} outgassing would continue to increase steadily as the magma ocean solidifies. In contrast, the melt fraction of \ce{H2O}, and thus the partial pressure of \ce{H2O}, increases in the mixed atmosphere case. In the absence of \ce{CO2}, the melt fraction still increases as the mantle solidifies, albeit not as strongly (Fig.~\ref{fig: Niko_volatile}, left panel, blue dotted line). 

The feedback between atmosphere composition and outgassing affects the volatile distribution at the end of the magma ocean. For the dry case, the final \ce{CO2} partial pressure decreases from $p_{\ce{CO2}}=100$~bar in a pure \ce{CO2} atmosphere to $p_{\ce{CO2}}=67$~bar in a mixed atmosphere as \ce{H2O} becomes the dominant species. The interplay between atmospheric composition and outgassing results in higher \ce{H2O} outgassing ($p_{\ce{H2O}}=260$~bar) compared to the scenario with no \ce{CO2} in the system, where 'only' $p_{\ce{H2O}}=$~206~bar is outgassed. The changes in \ce{H2O} and \ce{CO2} pressures for the 'dry case' are in first order agreement with \citet[see][ their Fig. 7a]{Bower2019} for a similar volatile content. \citet{Niko2019} did not account for the outgassing feedback effect and thus report an increase in both, \ce{CO2} and \ce{H2O} partial pressures as the magma ocean solidifies.  

We further find for the 'wet case' magma ocean, that volatile outgassing is modified similarly to the 'dry case'.  We also emphasize again that a drastic change in atmosphere composition and thus in outgassing does not occur when the outgassing laws of \citet{Elkins-Tanton2008} are used, where \ce{H2O} is always the dominant volatile in the atmosphere.

We conclude that \magmoc{2.0} adequately accounts for outgassing feedback between \ce{H2O} and \ce{CO2}, and that the corrected gray and RT atmosphere model capture thermal emission in vertically extended thick atmosphere.  Even more, \magmoc{2.0} is well-equipped to also tackle planets in the habitable zone around M dwarf stars, where atmospheric erosion of \ce{H2O} mainly by XUV photolysis \citep[see e.g.,][]{Luger2015} has the potential to further change the composition of a mixed \ce{H2O}-\ce{CO2} atmosphere during the magma ocean stage.

\FloatBarrier
\section{Numerical stability and runtime}
\label{sec: Stability}
The code \magmoc{2.0} is designed to efficiently explore diverse planetary and stellar scenarios. To achieve this goal, the code solves various sets of ordinary differential equations (ODEs), including equations that describe outgassing (Sect.~\ref{sec: outgas}). In this appendix, we provide details on numerical stability, convergence tests, and the typical wall clock times of the simulations discussed earlier. We focus here on the albedo=0.75 simulations.

One parameter in these specific ODEs is the decrease in liquid magma ocean $\frac{d M^{liq}}{dt}$ as the solidification radius increases $\frac{d r_s}{dt}$, which undergoes, however, phase state transitions for the critical mantle melt fraction $\psi_c=0.4$ \citep[][Sections 2.1.1 and 2.1.2]{Barth2021}. To tackle numerical instabilities introduced by non-continuous changes of melt fractions $F_{\ce{CO2}}(t)$ and $F_{\ce{H2O}}(t)$ in time $t$, we carefully monitor the volatile masses in the different reservoirs of the coupled magma ocean-atmosphere system ($M_i^{cystal}+M_i^{liq}+M_i^{atm}$) to ensure that their sum is equal to $M_i^{\mathrm{moa}}$ within reasonable limits (less than 5\% differences). $M_i^{\mathrm{moa}}$ also has to be initially equal to the prescribed initial volatile mass, $M_i^{\mathrm{ini}}$. To avoid numerical 'leakage of mass', we add source/sink terms $\Delta M_{\ce{H2O}}^{Corr}$  and $\Delta M_{\ce{CO2}}^{Corr}$ to the partial derivative of $ M_{\ce{H2O}}^{\mathrm{moa}}$ of the form

\begin{align}
 \Delta M_{\ce{H2O}}^{Corr} &= \frac{M_{\ce{H2O}}^{\mathrm{ini}}-   M_{\ce{H2O}}^{\mathrm{moa}}}{\Delta t_{curr}} \nonumber\\
  \Delta M_{\ce{CO2}}^{Corr} &= \frac{M_{\ce{CO2}}^{\mathrm{ini}}-   M_{\ce{CO2}}^{\mathrm{moa}}}{\Delta t_{curr}},
\end{align}
where $\Delta t_{curr}$ is the current time step during the magma ocean evolution
 that ensures that the volatile mass budget during runtime does not deviate more than 5\% from the initial mass budget. We note that $\Delta M_{\ce{H2O}}^{Corr}$ and $\Delta M_{\ce{CO2}}$ are rates in units of [kg/s] to counterbalance numerical mass loss.

 That is, the differentiation of the mass balance equations for volatile $i$ is modified during simulation run time such that:
 \begin{align}
		\frac{d M_i^{\mathrm{moa}}}{d t}+ \Delta M_{i}^{Corr}&= \frac{d M_i^{\mathrm{crystal}}}{d t} +  \frac{ d M_i^{\mathrm{liq}}}{d t} + \frac{d M_i^{\mathrm{atm}}}{d t}.
\label{eq: mass balance plus numetrics}
\end{align}
In terms of the substitution framework (Table~\ref{tab: Placeholders}), we add $\Delta M_{i}^{Corr}$ to $a'_1$ (for \ce{H2O}) and $a'_2$ (for \ce{CO2}), respectively. To avoid overcompensation, we limit the correction term to values smaller than 1\% of the total volatile budget. Despite these measures, sometimes noticeable fluctuations in the outgassed volatiles can occur due to fluctuations in $F_i$.

\begin{figure}
    \centering
    \includegraphics [width=0.49\textwidth]{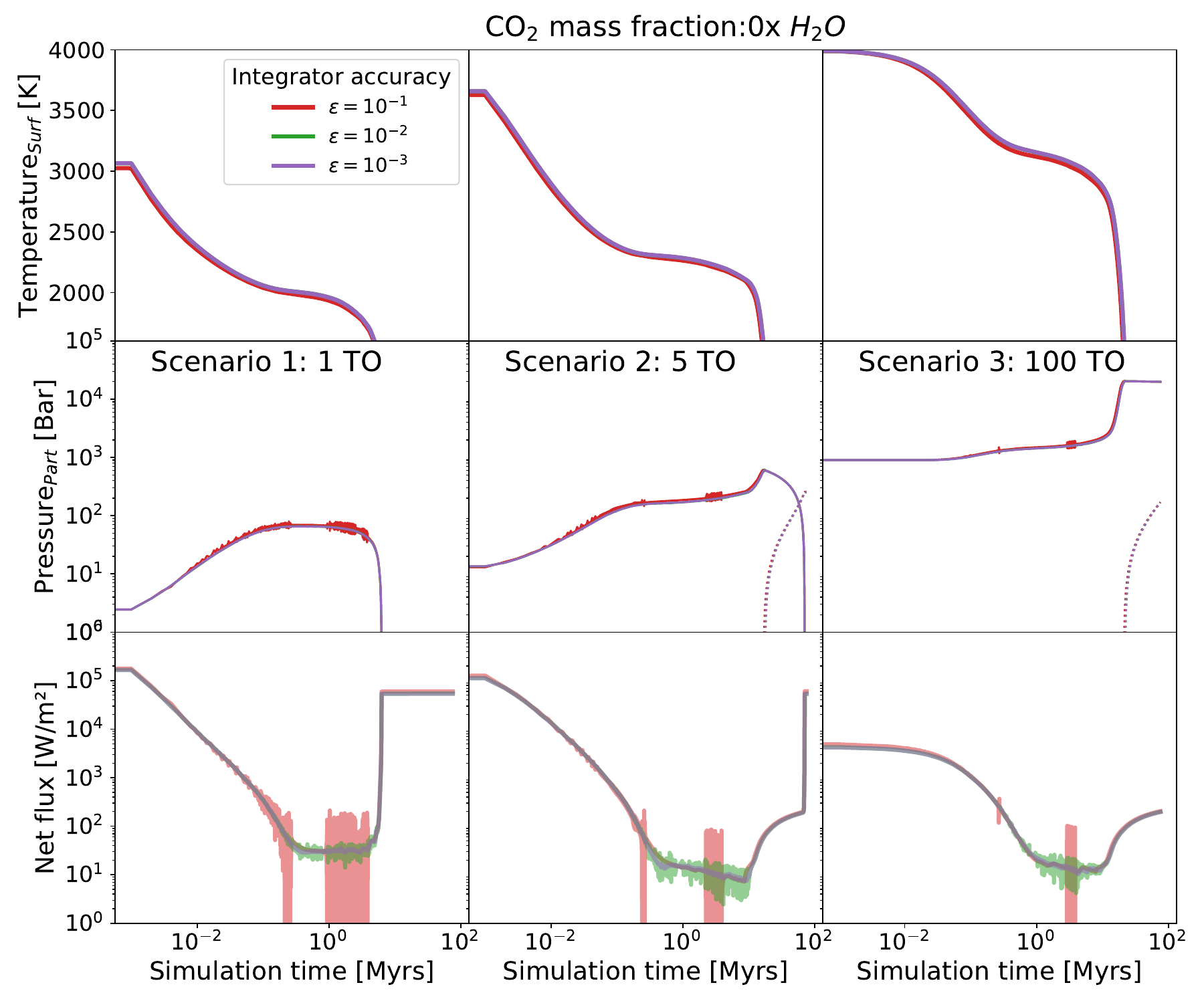}
   \caption{TRAPPIST-1 g magma ocean evolution with different relative accuracies $\epsilon$ in the integration during runtime for no additional \ce{CO2} in the system (red: $\epsilon=10^{-1}$, green: $\epsilon=10^{-2}$, purple: $\epsilon=10^{-3}$). Top panels show surface temperatures, middle panels show volatile partial pressures (solid: \ce{H2O}, dotted: \ce{O2}), bottom panels show net flux for initial water masses of 1 TO, 5 TO, and 100 TO from left to right.}
    \label{fig: Stab 0 CO2}
\end{figure}

\begin{figure}[ht]
    \centering
    \includegraphics [width=0.49\textwidth]{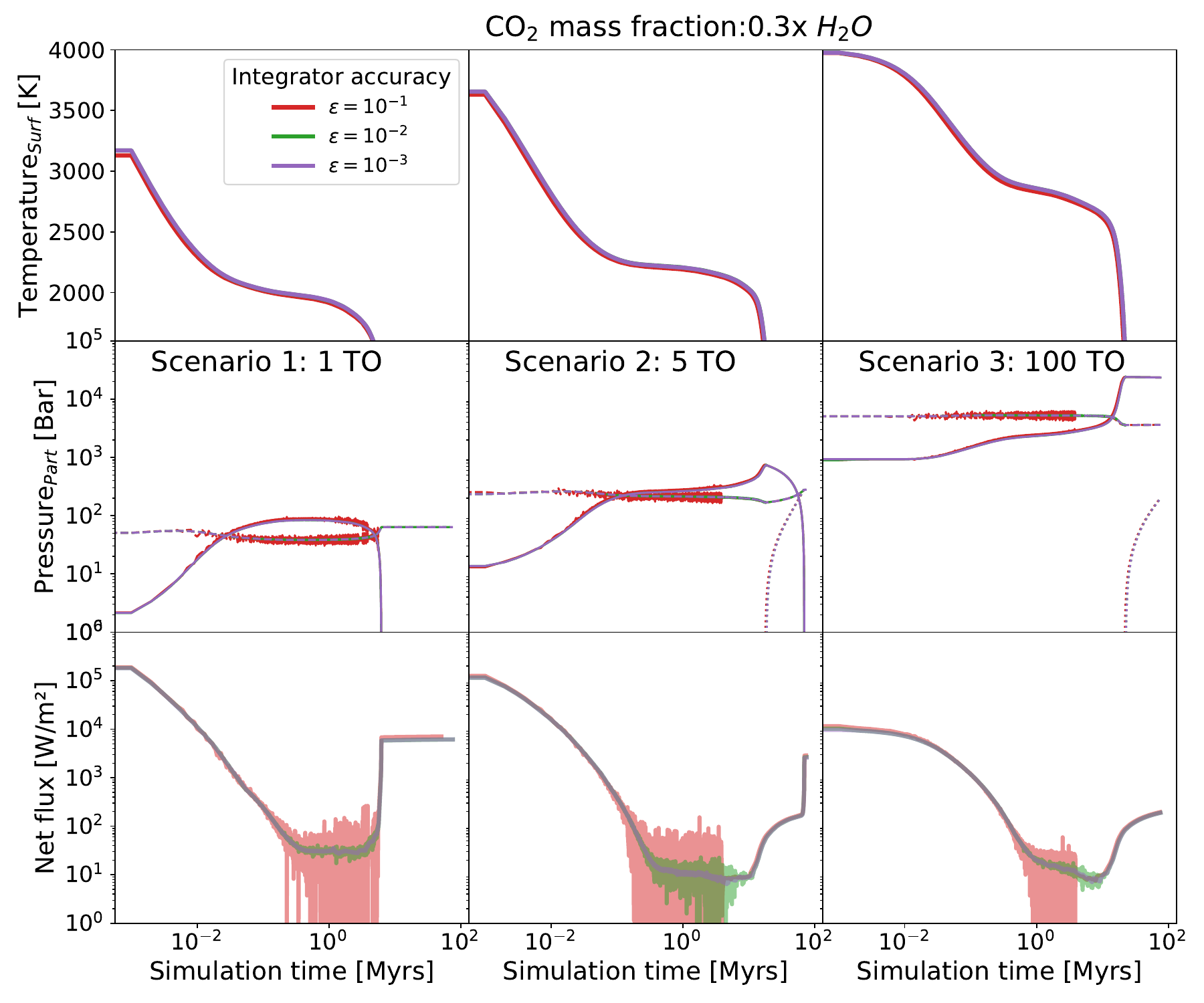}
   \caption{TRAPPIST-1 g magma ocean evolution with different relative accuracies $\epsilon$ in the integration during runtime for additional \ce{CO2}, the mass of which is scaled by a factor of $0.3$ with the initial \ce{H2O} masses in the system (red: $\epsilon=10^{-1}$, green: $\epsilon=10^{-2}$, purple: $\epsilon=10^{-3}$). Top panels show surface temperatures, middle panels show volatile partial pressures (solid: \ce{H2O}, dashed: \ce{CO2} dotted: \ce{O2}), bottom panels show net flux for initial water masses of 1 TO, 5 TO, and 100 TO from left to right.}
    \label{fig: Stab 0.3 CO2}
\end{figure}

\begin{figure}[ht]
    \centering
    \includegraphics [width=0.49\textwidth]{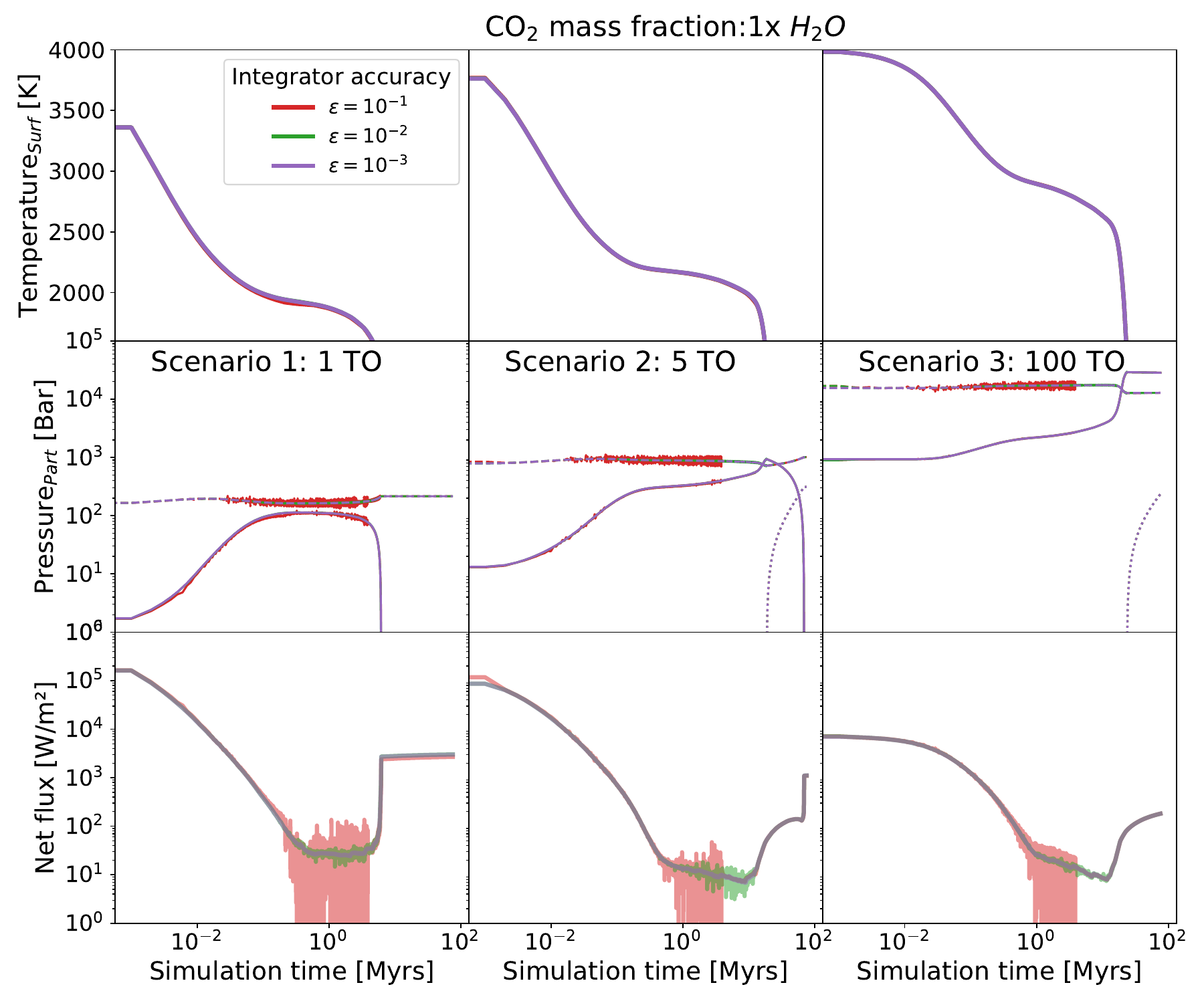}
   \caption{TRAPPIST-1 g magma ocean evolution with different relative accuracies $\epsilon$ in the integration during runtime for additional \ce{Co2}, the mass of which is scaled by a factor of $1$ with the initial \ce{H2O} masses in the system (red: $\epsilon=10^{-1}$, green: $\epsilon=10^{-2}$, purple: $\epsilon=10^{-3}$). Top panels show surface temperatures, middle panels show volatile partial pressures (solid: \ce{H2O}, dashed: \ce{CO2} dotted: \ce{O2}), bottom panels show net flux for initial water masses of 1 TO, 5 TO, and 100 TO from left to right.}
    \label{fig: Stab 1 CO2}
\end{figure}

We demonstrate such fluctuations for the TRAPPIST-1 g magma ocean evolution tracks for 1 TO, 5 TO, and 100 TO initial mass \ce{H2O} with \ce{CO2} initial mass scaled by \ce{H2O} mass with $0x,0.3x$ and $1x$ the inital \ce{H2O} mass simulated with different relative accuracies $\epsilon$ of the Runge-Kutta integrator (Figs.~\ref{fig: Stab 0 CO2}, \ref{fig: Stab 0.3 CO2}, \ref{fig: Stab 1 CO2}).

It is evident that atmospheric pressures and net flux can strongly fluctuate for the lowest integrator accuracy of $\epsilon=10^{-1}$. In all investigated cases, however, these fluctuations have a negligible impact on the surface temperature and overall volatile evolution. This is confirmed by comparison with simulations of higher numerical accuracy. More precisely, a relative accuracy of $\epsilon=10^{-3}$ in the Runge-Kutta integrator results in stable simulations that can be used to benchmark a specific evolution track in case of numerical issues.

We further show the run times of the 18 TRAPPIST-1 g simulations with 1 TO, 5 TO, and 100 TO initial mass of \ce{H2O} and an equivalent quantity of \ce{CO2} mass with different numerical accuracies (Table~\ref{tab: runtime}). The simulations were carried out on a single CPU on a AMD Ryzen Threadripper PRO 5955WX with 16 cores. The magma ocean simulations using the full RT atmosphere model may require up to 28 mins to complete. Simulations with the corrected gray atmosphere model, benchmarked with full radiative transfer calculations, are much more efficient, with runtimes of 1 minute and less, depending on numerical accuracy. The difference in runtime between $\epsilon=10^{-3}$ and $10^{-1}$ is just 15 to 20 seconds. Thus, with the corrected gray atmosphere model, the simulations can be carried out with highest numerical stability without excessive numerical costs. \magmoc{2.0} as part of the \vplanet{}-software framework is thus ideally suited to explore the magma ocean evolution for rocky exoplanets in the habitable zone of their host stars, spanning a wide range in fundamental parameters and  exploring different processes to account for the large diversities in exoplanet evolution. 

\begin{table}[ht]
\centering
\caption{Run times of TRAPPIST-1 g simulations with initial \ce{CO2}  mass equal to the initial \ce{H2O} mass.}
\begin{tabular}{ |p{2cm} | c| c | c| } 
\hline
   Initial water mass [TO] & 1 & 2 & 100\\
  \hline
\multicolumn{4}{|c|}{RT atmosphere model}  \\ 
  \hline
  $\epsilon =10^{-1}$ &6 min 11 s & 6 min 4 s  &  6 min 1 s \\
$\epsilon =10^{-2}$ &10 min 22 s & 11 min 25 s  &  11 min 15 s \\
$\epsilon =10^{-3}$ &27 min 41 s & 22 min 54 s  & 22 min 40 s \\
  \hline
  \multicolumn{4}{|c|}{Corrected gray atmosphere model}  \\ 
    \hline
      $\epsilon =10^{-1}$ & 0 min 40 s &  0 min 41 s  & 0 min 41 s  \\
$\epsilon =10^{-2}$ & 0 min 45 s & 0 min 48 s  & 0 min 47 s \\
$\epsilon =10^{-3}$ &  1 min 3 s & 1 min 2 s  &  1 min 2 \\
\hline
  
\end{tabular}
\newline On a state-of-the-art multicore machine with different numerical accuracies $\epsilon$. 
\label{tab: runtime}
\end{table}
\section{Chemical composition} \label{sec chem comp}

Since stars and their planetary accretion disks are formed by the collapse of the same interstellar dust cloud, the composition of a star can be used as a first estimate for the upper limit of the composition of the accretion disk.
However, the stellar composition of TRAPPIST-1 has not yet been determined. We therefore derive elemental abundances for the main planet-forming elements from the stellar metallicity. For this, large-scale astronomical surveys play a crucial role in providing the necessary data for understanding the chemical composition of stars. One such surveys is the Galactic Archaeology with HERMES (GALAH) project. The GALAH survey focuses on determining the detailed chemical composition of Milky Way stars, contributing to a better understanding of Galactic chemical evolution. In this study, the third release of the GALAH survey by \citet{Buder2021} serves as the main resource, offering a rich dataset that facilitates the calculation of metallicity-dependent compositional variations. However, the GALAH survey has limitations in providing data for all chemical elements of interest to our study. To address this gap, the Hypatia catalog was incorporated specifically for elements such as Nitrogen and Sulfur. The Hypatia Catalog, compiled from 84 literature sources, presents spectroscopic abundance data for 50 elements across 3058 stars in the solar neighborhood (within 150 pc of the Sun). Employing a binning strategy with intervals set at [Fe/H] = 0.05, stars were systematically organized to facilitate thorough data analysis. The mean metallicity and chemical composition for each bin was computed, providing a representative value within that specific metallicity range. For TRAPPIST-1, we use a metallicity value of 0.04$\pm$0.08 \citep{Gillon2017}, which leads to the predicted stellar composition reported in Section \ref{sec atm int model}. 

\citet{Bitsch2020} suggested a stoichiometric model to obtain a first-order estimate on the compositional variation of planetary building material depending on the local temperature within an accretion disk. In this approach, the gas within the accretion disk is assumed to have achieved the state of chemical
equilibrium before condensation, with the complete set of molecules preexisting in the gas. Consequently, the relative abundance of molecules can be calculated
stochiometrically and based on their condensation temperature \citet{Lodders2003}.

For a first estimate on the temperature profile within the accretion disk, we employ a simple power law \citet{Williams2011}:
\begin{equation}
    T_{disk}(r) = T_S \left( \frac{r}{x R_i} \right)^{-3/4},
\end{equation}
where $T_S$ represents the sublimation temperature and is set to 1500 K at the inner edge of the disk $R_i$ (following observations by \citealp{Dullemond2010}). $x$ is a free scaling parameter which we set here to $x=2$ for a good agreement of our profile compared to \citet{Jorge2022}. Using this estimate for the disk temperature profile allows us to determine the composition of the planetary building blocks at different distances of the star. For this, we extended the model of \citet{Bitsch2020} to include nitrogen, 
aluminum and calcium species. The full set of molecules, their condensation temperatures and stoichiometric calculations are listed in table \ref{table stochio}. Due to the overabundance of hydrogen, the elemental abundances are expressed as X/H.

\begin{table}[]
    \centering
       \caption{50\% condensation temperatures and volume mixing ration for different molecules in the accretion disk. 
    }
    \begin{tabular}{c|c|c}
Molecule & T$_c$ [K] & volume mixing ratio\\
\hline
CO &20 &0.45 $\cdot$ C/H\\
CH$_4$ &30 &0.45 $\cdot$ C/H\\
CO$_2$ &70 &0.1 $\cdot$ C/H\\
NH$_3$ &123& N/H\\
H$_2$O &150 &
O/H - (3 $\cdot$ MgSiO$_3$/H + \\
&&4 $\cdot$ Mg$_2$SiO$_4$/H + CO/H + 2 $\cdot$ CO$_2$/H +\\ 
&&3 $\cdot$ Fe$_2$O$_3$/H + 4 $\cdot$ Fe$_3$O$_4$/H)\\
Fe$_3$O$_4$ &371 &16 $\cdot$ (Fe/H - S/H)\\
FeS &664 &S/H\\
MgSiO$_3$ &1316& Mg/H - 2$\cdot$(Mg/H - Si/H)\\
Fe$_2$O$_3$ &1328 &0.25 $\cdot$ (Fe/H - S/H)\\
Mg$_2$SiO$_4$ &1336 &Mg/H - Si/H\\
CaAl$_{12}$O$_{19}$ &1529& (Al/H - Ca/H)/11\\
Ca$_2$Al$_2$SiO$_7$ &1659& 0.5$\cdot$(Ca/H - CaAl$_{12}$O$_{19}$)   
    \end{tabular}
 
    \label{table stochio}
\end{table}

\section{Interior structure models} \label{sec int struct}

\begin{figure}
    \centering
    \includegraphics[width=0.9\linewidth]{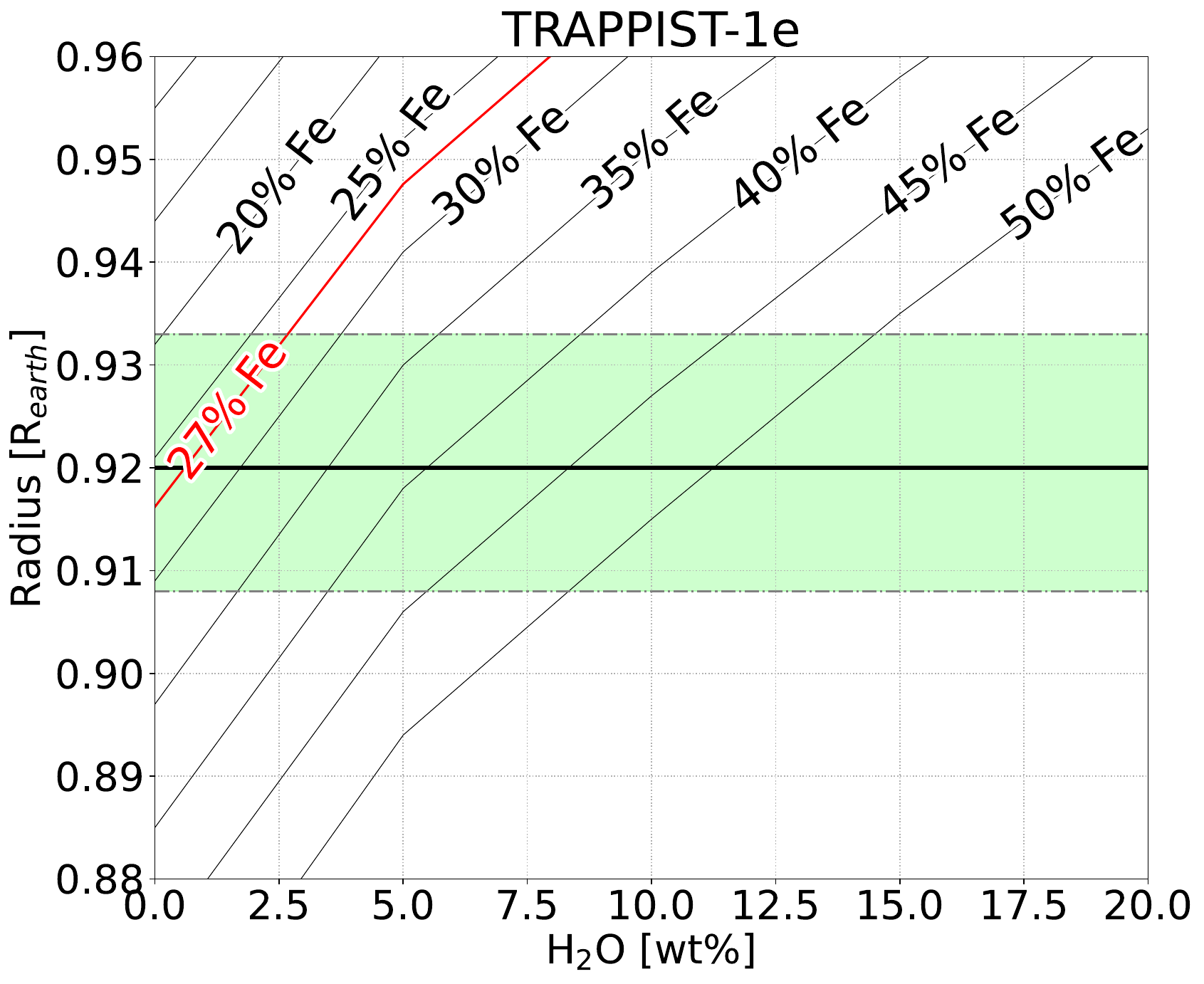}
     \includegraphics[width=0.9\linewidth]{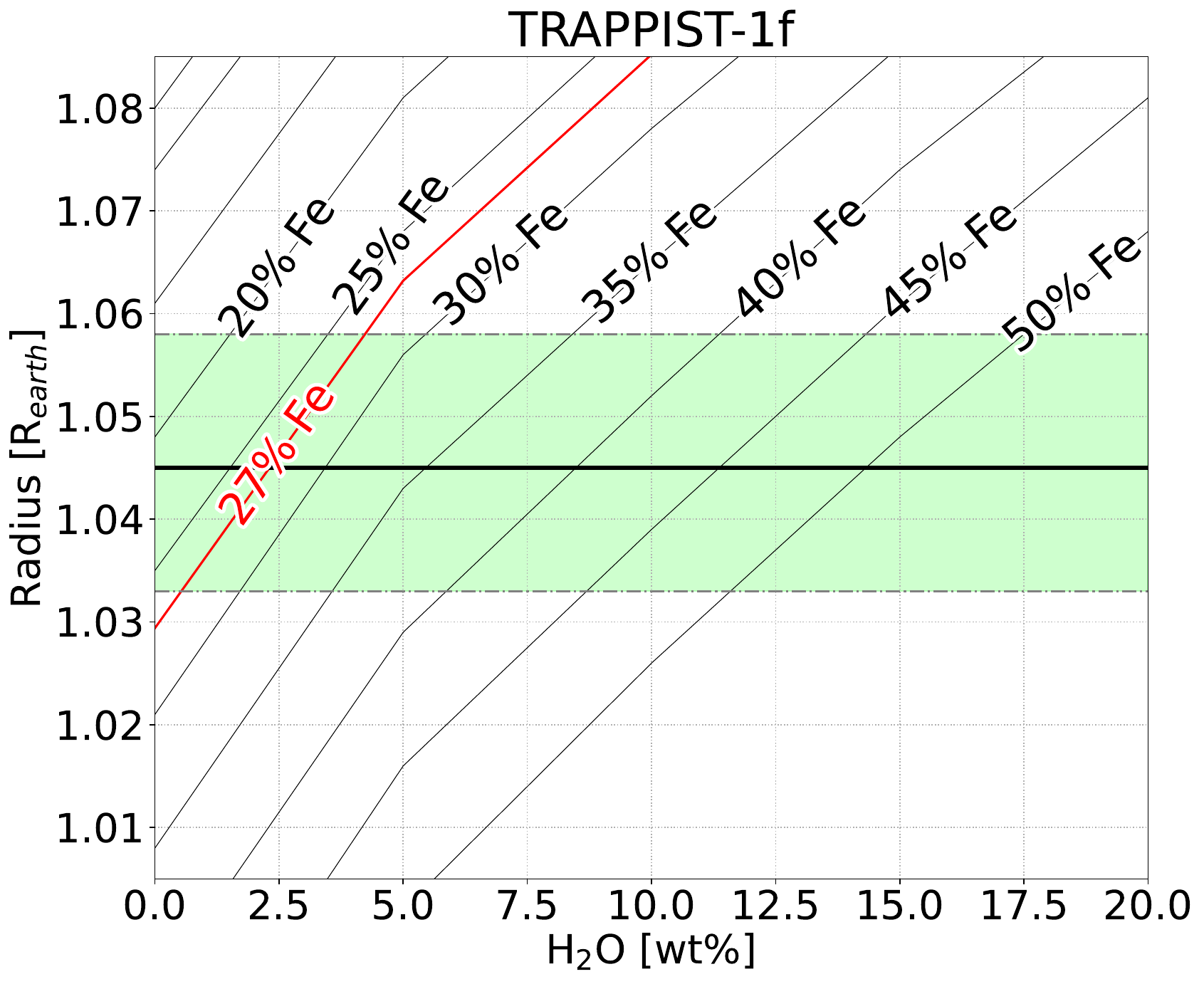}
        \includegraphics[width=0.9\linewidth]{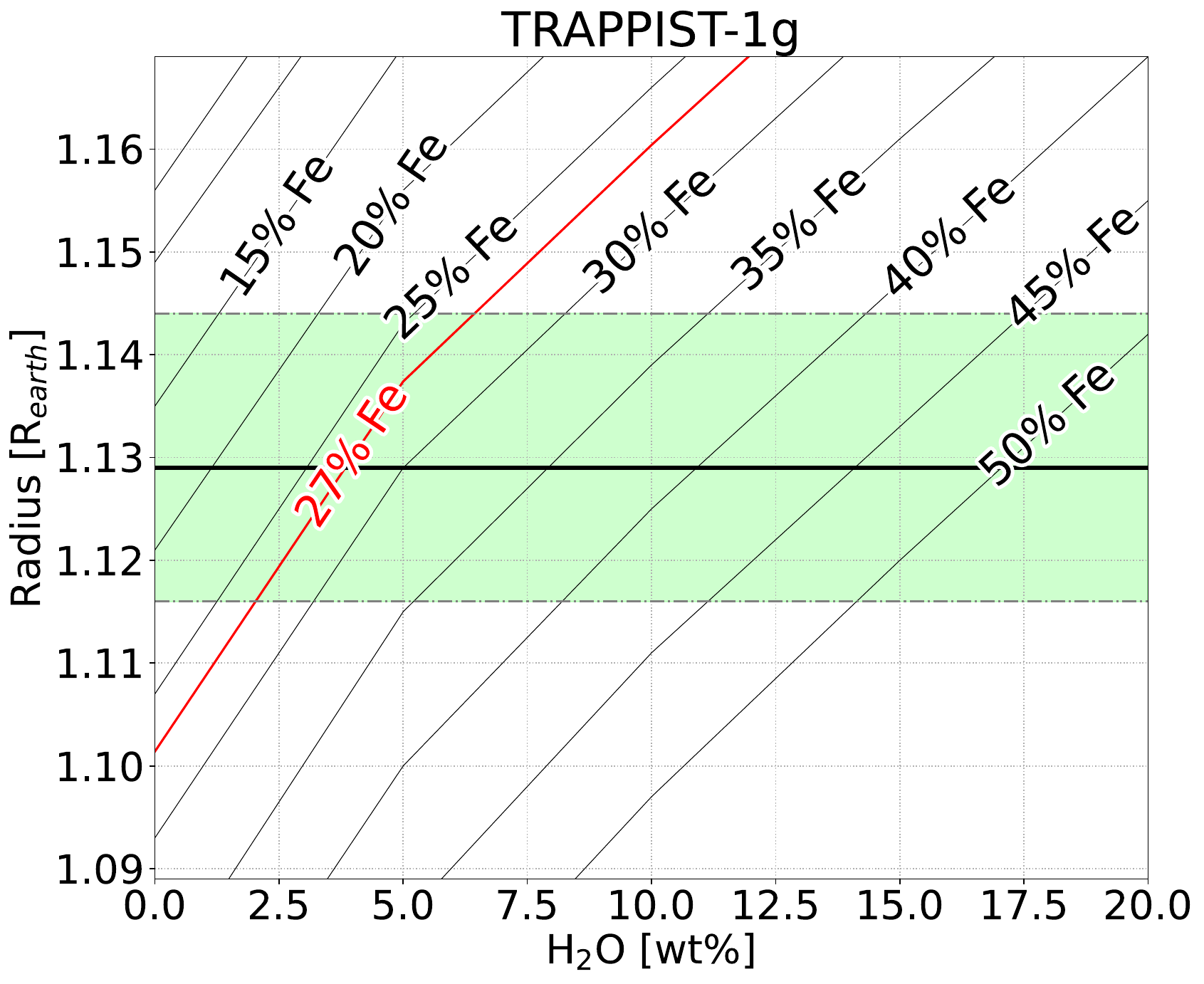}
    \caption{Ranges of possible water fractions of TRAPPIST-1e, f and g for different iron mass fractions within error bars of the mass-radius values of \citet{Agol2021}.}
    \label{fig:H2O_content_1e,f,g}
\end{figure}

We integrate the compositional information in our interior-structure models for all three planets. It is important to note here, that while we directly employ the refractory elements condensing out of the accretion disk in our interior-structure model, we allow for the core-mass fraction (considering here for simplicity a pure iron core) as well as the volatile fraction (considering here for simplicity only water, since it is the most abundant volatile molecule) to vary to explore the range of potential interior structures of TRAPPIST-1e, f and g. For a first justification of our model, we applied our interior structure model without any volatile material to the three innermost planets, and obtain modeled planet radii consistent with observations (see Section \ref{sec atm int model}). Figure~\ref{fig:H2O_content_1e,f,g} displays the full range of water content within error bars of planet radii and mass measurements.

The interior structure model follows \citet{Noack2016}, where we dissect the planet into 1000 shells, which we fill with core, mantle or ice/water material for given planet mass as well as core and water mass fractions. We assume an adiabatic temperature profile from surface to the center of the planet, while adopting a temperature increase at the core-mantle boundary following \citet{Stixrude2014} and \citet{Noack2020}. The surface temperature is set at 300 K, which means that we consider here for simplicity at the surface a liquid water layer. High-pressure ice forms for all three planets for water fractions starting at a few wt-\%.

For the core and water thermodynamic properties (such as density or heat capacity), we employ equations of state as outlined in \citet{Noack2016}. For the silicate mantle, we improved the approach of that study by developing lookup-tables for all relevant thermodynamic properties for silicate mantles of varying composition calculated with Perple\_X \citep{Connolly2009} and interpolated for our predicted mantle composition as well as local temperature and pressure conditions in each shell. We use a different resolution for our look-up tables for low, intermediate and high pressures, where we vary temperature in steps of 25 K between 200 and 5000 K and pressure in steps of 1250 bar for pressures below 25 GPa, 1.9 GPa for pressures between 25 and 400 GPa, and 5.5 GPa for pressures above 400 GPa going up to 1500 GPa (which ensures coverage of the entire pressure range encountered in the planetary interior in this study).

\end{appendix}

\end{document}